\documentclass[11pt, a4paper]{article}
\pdfoutput=1
\usepackage{jheppub}
\input{epsf.tex}
\usepackage{graphics} 
\usepackage{epsfig}
\usepackage{amsfonts}
\usepackage{amsmath}
\usepackage{transparent, colortbl}
\usepackage{booktabs}
\usepackage{multirow}
\usepackage{anysize}
\usepackage{array}
\usepackage{latexsym, amscd, amsthm}
\usepackage{pdfsync}
\usepackage{epsf}
\usepackage[all]{xy}
\usepackage[usenames,dvipsnames]{xcolor}
\usepackage{enumerate}
\usepackage{float}
\restylefloat{table}
\usepackage{upgreek}
\usepackage{subcaption}
\usepackage{array}
\usepackage{multirow}
\usepackage{graphicx,color}
\DeclareSymbolFont{extraup}{U}{zavm}{m}{n}
\DeclareMathSymbol{\varheart}{\mathalpha}{extraup}{86}
\DeclareMathSymbol{\vardiamond}{\mathalpha}{extraup}{87}

\usepackage{fancyhdr}

\makeatletter
\def\CT@@do@color{%
  \global\let\CT@do@color\relax
        \@tempdima\wd\z@
        \advance\@tempdima\@tempdimb
        \advance\@tempdima\@tempdimc
        \kern-\@tempdimb
\transparent{0.6}%
        \leaders\vrule
                \hskip\@tempdima\@plus  1fill
        \kern-\@tempdimc
        \hskip-\wd\z@ \@plus -1fill }
\makeatother

\makeatletter
\newcommand{\thickhline}{%
    \noalign {\ifnum 0=`}\fi \hrule height 1pt
    \futurelet \reserved@a \@xhline
}
\newcolumntype{"}{@{\hskip\tabcolsep\vrule width 1pt\hskip\tabcolsep}}
\makeatother

\newtheorem{Theorem}{Theorem}[section]
\newtheorem{Lemma}[Theorem]{Lemma}
\newtheorem{Proposition}[Theorem]{Proposition}
\theoremstyle{definition}
\newtheorem{Definition}[Theorem]{Definition}
\theoremstyle{remark}

\newcommand{\Hom}{{\rm Hom}}

\newcommand{\D}{\mathbb{D}}

\newcommand{\im}{\mathrm{im}}

\newcommand{\bp}{\begin{Proposition}}
\newcommand{\ep}{\end{Proposition}}
\newcommand{\bl}{\begin{Lemma}}
\newcommand{\el}{\end{Lemma}}
\newcommand{\bt}{\begin{Theorem}}
\newcommand{\et}{\end{Theorem}}
\newcommand{\bd}{\begin{Definition}}
\newcommand{\ed}{\end{Definition}}
\newcommand{\End}{\mathrm{End}}

\newcommand{\Mat}{\mathrm{Mat}}
\newcommand{\ev}{\mathrm{ev}}
\newcommand{\eqdef}{\stackrel{{\rm def.}}{=}}

\newcommand{\cinf}{{{\cal \cC}^\infty(M,\R)}}

\DeclareFontFamily{U}{rsf}{}
\DeclareFontShape{U}{rsf}{m}{n}{<5> <6> rsfs5 <7> <8> <9> rsfs7 <10-> rsfs10}{}
\DeclareMathAlphabet\Scr{U}{rsf}{m}{n}

\newcommand{\KA}{K\"{a}hler-Atiyah~}

\def\cU{\mathcal{U}}

\def\cW{\mathcal{W}}

\def\hV{{\hat V}}
\def\hU{{\hat U}}

\def\N{\mathbb{N}}
\def\Z{\mathbb{Z}}
\def\C{\mathbb{C}}
\def\R{\mathbb{R}}

\def\rk{{\rm rk}}

\def\dd{\mathrm{d}}

\def\AdS{\mathrm{AdS}}
\def\flat{\mathrm{flat}}

\def\tb{{\tilde b}}

\def\Stab{\mathrm{Stab}}

\def\Int{\mathrm{Int}}
\def\Fr{\mathrm{Fr}}

\def\fr{\mathrm{fr}}
\def\D{\mathbf{D}}
\def\U{\mathbf{U}}

\def\q{\mathfrak{q}}
\def\hG{\hat{G}}

\def\hzeta{{\hat \zeta}}
\def\txi{{\tilde \xi}}
\def\tb{{\tilde b}}
\def\tV{{\tilde V}}
\def\tW{{\tilde W}}
\def\tbeta{{\tilde \beta}}

\newcommand{\be}{\begin{equation*}}
\newcommand{\ee}{\end{equation*}}
\newcommand{\ben}{\begin{equation}}
\newcommand{\een}{\end{equation}}
\newcommand{\beqa}{\begin{eqnarray*}}
\newcommand{\eeqa}{\end{eqnarray*}}
\newcommand{\beqan}{\begin{eqnarray}}
\newcommand{\eeqan}{\end{eqnarray}}

\newcommand{\nn}{\nonumber}

\newcommand{\id}{\mathrm{id}}

\def\cR{{\mathcal R}}

\def\fG{\Gamma}
\def\cC{{\mathcal C}}
\def\cB{\Scr B}

\def\cZ{{\cal Z}}

\def\Cl{\mathrm{Cl}}

\def\Grn{\mathrm{Grn}}
\def\Trivf{\mathrm{Trivf}}
\def\cK{\mathrm{\cal K}}

\def\Spin{\mathrm{Spin}}

\def\SO{\mathrm{SO}}
\def\O{\mathrm{O}}
\def\Hol{\mathrm{Hol}}
\def\cD{\mathcal{D}}

\def\cA{\mathcal{A}}

\def\cP{\mathcal{P}}
\def\cN{\mathcal{N}}
\def\cG{\mathcal{G}}

\def\cC{\mathcal{C}}

\def\SU{\mathrm{SU}}

\def\G_2{\mathrm{G_2}}

\def\cS{\mathcal{S}}
\def\cZ{\mathcal{Z}}
\def\cV{\mathcal{V}}

\def\fC{\mathfrak{C}}
\def\fK{\mathfrak{K}}
\def\fF{\mathfrak{F}}
\def\fI{\mathfrak{I}}
\def\fD{\mathfrak{D}}

\def\fP{\mathfrak{P}}
\def\fS{\mathfrak{S}}
\def\fA{\mathfrak{A}}

\def\cA{\mathcal{A}}

\def\mbD{\mathbb{D}}

\def\mF{\mathbf{F}}

\def\op{\mathrm{op}}

\newcommand{\twopartdef}[4]
{
	\left\{
		\begin{array}{ll}
			#1 & \mbox{if } #2 \\
			#3 & \mbox{if } #4
		\end{array}
	\right.
}

\title{The landscape of G-structures in eight-manifold compactifications of M-theory}

\author{Elena Mirela Babalic$^{1,2}$, Calin Iuliu Lazaroiu$^3$ }

\affiliation{
   $^1$ Department of Theoretical Physics, National
  Institute of Physics and Nuclear Engineering, Str. Reactorului
  no.30, P.O.BOX MG-6, Postcode 077125, Bucharest-Magurele, Romania  \\
 $^2$ Department of Physics, University
  of Craiova, 13 Al. I. Cuza Str., Craiova  200585, Romania\\
 $^3$ Center for Geometry and Physics, Institute for Basic Science, Pohang 790-784, Republic of Korea
}

\emailAdd{mbabalic@theory.nipne.ro, calin@ibs.re.kr} 

\abstract{We consider spaces of ``virtual'' constrained generalized
  Killing spinors, i.e.  spaces of Majorana spinors which correspond
  to ``off-shell'' $s$-extended supersymmetry in compactifications of
  eleven-dimensional supergravity based on eight-manifolds $M$. Such
  spaces naturally induce two stratifications of $M$, called the
  chirality and stabilizer stratification.  For the case $s=2$, we
  describe the former using the canonical Whitney stratification of a
  three-dimensional semi-algebraic set $\cR$. We also show that the
  stabilizer stratification coincides with the rank stratification of
  a cosmooth generalized distribution $\cD_0$ and describe it
  explicitly using the Whitney stratification of a four-dimensional
  semi-algebraic set $\fP$.  The stabilizer groups along the strata
  are isomorphic with $\SU(2)$, $\SU(3)$, $\G_2$ or $\SU(4)$, where
  $\SU(2)$ corresponds to the open stratum, which is generically
  non-empty. We also determine the rank stratification of a larger
  generalized distribution $\cD$ which turns out to be integrable in
  the case of compactifications down to $\AdS_3$.}

\begin{document}

\maketitle 

\pagebreak

\vskip .6in

\section*{Introduction}

General compactifications of M-theory on eight-manifolds provide a rich
class of geometries which are of physical interest due to their relation
to F-theory \cite{GranaShahbaziZambon, Bonetti, Shahbazi}. They can serve to test 
ideas such as exceptional generalized geometry \cite{Hull, Waldram1, Waldram2,
  Waldram3, Waldram4, Waldram5, Baraglia} since eight is the
first dimension for which the problem of ``dual gravitons''
\cite{CurtrightDG, HullDG1, WestDG, HullDG2, DGnogo1, DGnogo2}
appears.  Given these aspects, it is rather surprising that current
understanding of such backgrounds is quite limited.  The notable
exception is the class of compactifications down to 3-dimensional
Minkowski space, which were studied intensively following the
seminal work of \cite{BeckerSpin7} (for the $\cN=1$ case) and
\cite{BeckerCY} (for the $\cN=2$ case). Such backgrounds are obtained
by constraining the internal part of the supersymmetry generators to
be Majorana-Weyl rather than merely Majorana.  As expected from no-go
theorems (first used within this setting in \cite{MartelliSparks}),
such Minkowski compactifications cannot support a flux at the
classical level. However, they can support {\em small} fluxes at the
quantum level, which are suppressed by inverse powers of the size of
the compactification manifold. Since such fluxes are difficult to
control beyond leading order \cite{BeckerConstantin, Constantin}, a
natural idea is to consider instead compactifications down to $\AdS_3$
spaces.

As pointed out in \cite{MartelliSparks}, compactifications of M-theory
down to $\AdS_3$ {\em do} support classical fluxes, which are
therefore not suppressed. This happens because the internal parts of
the supersymmetry generators are no longer required to be
Majorana-Weyl. This seemingly innocuous extension leads to a
surprisingly intricate geometry, as already apparent in the case of
$\cN=1$ unbroken supersymmetry \cite{MartelliSparks, Tsimpis}, which
can be described using the theory of singular foliations
\cite{g2,g2s}. By comparison, little is known\footnote{Such
  backgrounds were considered in \cite{ga2} using a nine-dimensional
  formalism and were also discussed in \cite{Palti} with similar
  methods, but without carefully studying the corresponding geometry
  of the eight-manifold. Certain compactifications down to
  three-dimensional Minkowski space but with torsion-full $\SU(4)$
  structure were studied in \cite[Section 3]{TsimpisSU4}.} about
$\cN=2$ compactifications down to $\AdS_3$. In this paper, we consider
certain aspects of the geometry of $\cN=2$ eight-dimensional
backgrounds by working directly in eight dimensions.  Namely, we solve
the question of classifying the stratified reductions of structure
group which arise on the internal eight-manifold $M$, showing that the
full picture is considerably richer than has been previously
presumed. Pointwise positions of internal supersymmetry generators as
well as their stabilizer groups are described by stratifications of
the internal space $M$: the first by the {\em chirality
  stratification} and the second by the {\em stabilizer
  stratification}.  Unlike the case $\cN=1$, the two stratifications
need not coincide. We find that these stratifications can be described
explicitly using the preimages through certain smooth maps
$b:M\rightarrow \R^3$ and $B:M\rightarrow \R^4$ of the connected
refinements of the canonical Whitney stratifications \cite{Whitney, Gibson} of
semi-algebraic \cite{BCR, AK, BPR} subsets $\cR\subset \R^3$ and
$\fP\subset \R^4$, where $\cR$ is obtained from $\fP$ by projection on
the three-dimensional space corresponding to the first three
coordinates of $\R^4$.  The maps $b$ and $B$ are constructed from
bilinears in the internal supersymmetry generators, while the
semi-algebraic set $\fP$ can be described explicitly using algebraic
constraints implied by the Fierz identities. This gives a geometric
picture of such backgrounds which shows how they can be approached
using the theory of stratified manifolds. We classify the stabilizer
groups for each stratum, thus giving a complete description of the
``stratified G-structure'' which arises in such backgrounds. In
particular, we find that a generic eight-manifold $M$ of this type
contains an open stratum on which the structure group reduces to
$\SU(2)$. In a certain sense, this stratum is the ``largest'', but it
was not considered previously. We also classify the amount of
supersymmetry preserved by an M2-brane transverse to $M$ along each
stratum.

Since the classification results mentioned above are independent of
the precise form of the supersymmetry equations, they hold more
generally than the case of compactifications down to $\AdS_3$. To
highlight this, we develop the formalism required to describe the
``topological part'' of the conditions for supersymmetry,
characterizing those finite-dimensional spaces of globally-defined
Majorana spinors which can be spanned by solutions of constrained
generalized Killing equations (so-called ``virtual CGK spaces''). We
show that such spaces must obey a local non-degeneracy condition which
puts them in bijection with trivial sub-bundles of the bundle of
Majorana spinors, endowed with a trivial flat connection. This
formulation clarifies some aspects of the mathematical description of
so-called ``off-shell  supersymmetric'' backgrounds.

The paper is organized as follows. Section \ref{sec:basics} gives the
general description of virtual CGK spaces $\cK$ and of the chirality
and stabilizer stratifications which they induce on $M$ and shows how
this framework arises in the case of compactifications down to
$\AdS_3$.  We also treat the case $\cN=1$ as a warm-up, pointing out
its differences with the case $\cN=2$. The rest of the paper is
devoted to the detailed study of the latter case.  Section
\ref{sec:cD} discusses the scalar and one-form bilinears which can be
constructed using a basis of $\cK$ when $\dim \cK=2$ and introduces two
cosmooth generalized distributions $\cD$ and $\cD_0$ (where
$\cD_0\subset \cD$) which are naturally associated to the one-form
bilinears. The rank stratification of $\cD_0$ turns out to coincide
with the stabilizer stratification, thus providing a way to identify
the latter.  In the case of compactifications to $\AdS_3$, the
distribution $\cD_0$ need not be integrable, but one can show that the
larger distribution $\cD$ integrates to a singular foliation in the
sense of Haefliger (topologically, this is a Haefliger structure
\cite{Haefliger} which may be non-regular). Section \ref{sec:spinors}
discusses the chirality stratification, giving its explicit
description in terms of a convex three-dimensional semi-algebraic body
$\cR$ and a smooth map $b:M\rightarrow \R^3$ whose image is contained
in $\cR$. Section \ref{sec:alg} discusses the algebraic constraints on
zero- and one-form spinor bilinears which are induced by Fierz
identities, showing how they can be described using a four-dimensional
semi-algebraic set $\fP$. In the same section, we discuss the
geometry of $\fP$ and of its boundary, its canonical Whitney
stratification and the preimage of $\partial\cR$ inside $\partial \fP$
through the map which projects on the first three coordinates. Section
\ref{sec:degen} shows that the rank stratifications of $\cD$ and
$\cD_0$ (where the latter coincides with the stabilizer
stratification) are different coarsenings of the $B$-preimage of the
connected refinement of the canonical Whitney stratification of $\fP$,
where $B$ is a map from $M$ to $\R^4$ with image contained in $\fP$.
The two coarsenings are given explicitly, leading to the
classification of stabilizer groups.  In the same section, we show how
the chirality stratification fits into this picture, while in Section
\ref{sec:concl} we conclude.  The appendices contain various proofs as
well as other technical details. The main results of this paper are
Theorems 1, 2, 3 and 4 which can be found in Subsections
\ref{sec:Theorem1}, \ref{sec:Theorems2and3} and
\ref{sec:Gstratif}. For ease of reference, various results are summarized in tables and
figures. The notations and conventions used in
the paper are explained in Appendix \ref{app:notations}.

\section{Virtual CGK spaces}
\label{sec:basics}

The eight-manifold $M$ can be used in various ways to construct a
supersymmetric background $\mathbf{M}$ of eleven-dimensional
supergravity \cite{sugra11}, for example by taking $\mathbf{M}$ to be
foliated in eight-manifolds with typical leaf $M$ or by taking it to
be a (warped) product between $M$ and some non-compact 3-manifold $N$
endowed with a metric of Minkowski signature. In such backgrounds,
supersymmetry generators can be constructed starting from
globally-defined solutions $\xi\in \Gamma(M,S)$ of equations of the
type:
\ben
\label{par_eq}
\mbD\xi = 0~~,~~Q\xi = 0~~,
\een 
which we shall call {\em constrained generalized Killing (CGK) spinor
  equations}.  Here $\mbD:\Gamma(M,S)\rightarrow \Omega^1(M,S)$ is a
connection on the bundle $S$ of Majorana spinors over $M$ and $Q\in
\Gamma(M,\End(S))$ is a globally-defined endomorphism of $S$. Such
equations encode the condition that a supersymmetry transformation
whose generator has $\xi$ as its ``internal part'' preserves the
background.  The explicit forms of $\mbD$ and $Q$ depend on the
precise background under consideration and will generally involve the
metric of $M$ as well as various differential forms defined on $M$. We
let $\cK(\mbD,Q)$ denote the (finite-dimensional) space of solutions
to \eqref{par_eq}.

\paragraph{Definition} 
A finite-dimensional subspace $\cK$ of $\Gamma(M,S)$ is called a {\em
  virtual CGK space} if there exists a connection $\mbD$ on $S$ and a
globally-defined endomorphism $Q\in \Gamma(M,\End(S))$ such that $\cK=
\cK(\mbD,Q)$.

\paragraph{Definition.} 
A virtual CGK space $\cK$ is called {\em $\cB$-compatible} if there
exists a $\cB$-compatible connection $\mbD$ on $S$ and a global
endomorphism $Q\in \Gamma(M,\End(S))$ such that $\cK=\cK(\mbD,Q)$.

\paragraph{First remarks.}  
The physics literature of flux compactifications sometimes makes a
distinction between\footnote{The topological conditions are sometimes
  called ``algebraic conditions'' \cite{MinasianGG2, GranaOrsi} while
  the supersymmetry conditions are called ``differential conditions'',
  but this terminology is inaccurate for our purpose. In this paper,
  we are interested in supersymmetry conditions for the ``internal
  part'' of spinors, hence the equations on the internal manifold $M$
  will generally have both a differential and an algebraic part as in
  \eqref{par_eq}. On the other hand, existence of a certain number of
  globally-defined independent spinors is clearly a topological,
  rather than algebraic, condition.}:
\begin{enumerate}[(a)]
\itemsep 0.0em
\item The {\em topological condition} for supersymmetry, namely that
  the given background admits a number $s$ of independent
  and globally-defined spinors $\xi_1,\ldots, \xi_s$ of the desired
  type;
\item The {\em algebro-differential conditions} for supersymmetry,
  namely that the spinors at (a) satisfy an equation of the form 
  \eqref{par_eq}.
\end{enumerate}

\noindent To clarify this, let $\xi_1,\ldots, \xi_s\in \Gamma(M,S)$ be $s$
globally defined Majorana spinors on $M$. Recall that $\Gamma(M,S)$
has a canonical structure of module over $\cinf$. Since the latter is an
$\R$-algebra, this also endows $\Gamma(M,S)$ with a structure of
(infinite-dimensional) vector space over $\R$.

\paragraph{Definition.} 
The globally-defined spinors $\xi_1,\ldots, \xi_s$ are called {\em
  weakly linearly independent} if they are linearly independent over
the field $\R$ of real numbers, i.e. linearly independent as elements
of the infinite-dimensional $\R$-vector space $\Gamma(M,S)$ of smooth
globally-defined sections of $S$. They are called {\em strongly
  linearly independent} if they are linearly independent over $\cinf$, i.e. linearly
independent as elements of the $\cinf$-module $\Gamma(M,S)$.

\

\noindent Weak linear independence of $\xi_1,\ldots, \xi_s$ means that the
relation:
\be
c_1\xi_1(p)+\ldots +c_s\xi_s(p)=0~~~\forall p\in M~~,
\ee
where $c_1,\ldots, c_s$ are real {\em constants}, implies
$c_1=\ldots=c_s=0$. Strong linear independence
means that the relation:
\be
c_1(p)\xi_1(p)+\ldots +c_s(p)\xi_s(p)=0~~~\forall p\in M~~,
\ee
where $c_1,\ldots, c_s\in \cinf$ are smooth real-valued {\em
  functions} defined on $M$, implies $c_1(p)=\ldots=c_s(p)=0$ for all
$p\in M$.  Since constant real-valued functions are smooth, it is
clear that strong linear independence implies weak linear
independence. It is also clear that strong linear independence amounts
to the condition that $\xi_1(p),\ldots, \xi_s(p)$ are linearly
independent inside the vector space $S_p$ for all $p\in M$.  As we
show below, condition (b) implies that the independence condition at
(a) should be understood as {\em strong} linear independence.

The supersymmetry equations \eqref{par_eq} do not specify precise
choices of globally-defined spinors but only a subspace $\cK$ of
$\Gamma(M,S)$, namely the space $\cK(\mbD,Q)$ of all globally-defined
solutions of \eqref{par_eq}. Hence we need a formulation of the strong
linear independence condition which does not rely on choosing a basis
for $\cK$. Since this is a pointwise condition, it can be formulated
in a frame-free manner using the evaluation map. This leads to the
notion of locally non-degenerate subspaces of $\Gamma(M,S)$. As we
show below, a subspace $\cK$ of $\Gamma(M,S)$ is a virtual CGK space iff it
obeys this non-degeneracy condition. When $\cK$ is $\cB$-compatible,
the freedom to change an orthonormal basis of $\cK$ is related to the
R-symmetry of supersymmetric effective actions built using such
backgrounds.

\paragraph{Remark.} 
The fact that some subspace $\cK\subset \Gamma(M,S)$ is a virtual CGK
space does {\em not} mean that $\cK$ consists of internal parts of
supersymmetry generators for any specific background of
eleven-dimensional supergravity built on $M$.  To know whether this is
the case, one has to study which pairs $(\mbD,Q)$ can arise in a given
class of backgrounds. The notion of virtual CGK space encodes the
``topological part'' of the supersymmetry conditions, which is much
weaker than the full supersymmetry conditions in a given background or
class of backgrounds.

\subsection{Locally non-degenerate subspaces of $\Gamma(M,S)$}

For any $p\in M$, let $\ev_p:\Gamma(M,S)\rightarrow S_p$ be the
evaluation map at $p$:
\be
\ev_p(\xi)\eqdef \xi(p)~~,~~\forall \xi\in \Gamma(M,S)~~.
\ee
Notice that $\ev_p$ is $\R$-linear and surjective. Any subspace
$\cK\subset \Gamma(M,S)$ induces a generalized linear sub-bundle
$\ev_\ast(\cK)\eqdef \sqcup_{p\in M}\ev_p(\cK)$ of $S$, which is
smooth in the sense of \cite{Drager}.

\paragraph{Definition.} 
A subspace $\cK\subset \Gamma(M,S)$ is {\em locally non-degenerate} if the 
restriction $\ev_p|_\cK:\cK\rightarrow S_p$ is injective for all $p\in M$. 

\

\noindent The local non-degeneracy condition means that any element
$\xi\in \cK$ is either the zero section of $S$ or a section of $S$
which does not vanish anywhere on $M$.  A locally non-degenerate
subspace $\cK$ of $\Gamma(M,S)$ has finite dimension $s\eqdef \dim \cK
\leq \rk S=16$. In this case, it is easy to see that $\ev_\ast(\cK)$
is an ordinary sub-bundle of $S$ which is topologically trivial,
because any basis $\xi_1,\ldots,\xi_s$ of $\cK$ obviously forms a
frame of $K$. Let $\Grn_s(M,S)$ denote the set of locally
non-degenerate $s$-dimensional subspaces of $\Gamma(M,S)$; notice that
$\Grn_s(M,S)$ can be viewed as an infinite-dimensional manifold. Let
$\Trivf_s(M,S)$ denote the set of pairs $(K,\D)$, where $K$ is a
trivial (in the sense of globally trivializable) smooth rank $s$
sub-bundle of $S$ and $\D$ is a trivial flat connection on $K$.

\paragraph{Remark.} 
Given a trivial rank $s$ sub-bundle $K$ of $S$ and a point $p\in M$,
trivial flat connections on $K$ can be identified (using parallel
transport) with bundle isomorphisms
$\varphi_p:K\stackrel{\sim}{\rightarrow}M\times S_p$, so
$\Trivf_s(M,S)$ can be identified with the set of all pairs
$(K,\varphi_p)$. Notice that this identification depends 
on the choice of $p\in M$ and hence
it is natural only if we work with pointed manifolds $(M,p)$. A
natural description which does not require the choice of a base point
is given below.

\paragraph{Proposition.} 
There exists a natural bijection
$\Phi_s:\Grn_s(M,S)\stackrel{\sim}{\longrightarrow}\Trivf_s(M,S)$,
whose inverse is given by $\Phi_s^{-1}(K,\D)=\Gamma_{\flat}(K,\D)$,
where:
\be
\Gamma_\flat(K,\D)\eqdef \{\xi \in \Gamma(M,K)|\D\xi=0\}~~
\ee
is the space of all $\D$-flat sections of $K$.

\

\noindent{\bf Proof.} 
Let $\Pi_1(M)$ be the first homotopy groupoid of $M$ and $A(K)$ be the
isomorphism groupoid of $K$ (the groupoid whose objects are the points
of $M$ and whose Hom-set from $p$ to $q$ is the set of linear
isomorphisms from $K_p$ to $K_q$).  The map which assigns the pair
$(p,q)$ to curves starting at $p$ and ending at $q$ induces a functor
$E:\Pi_1(M)\rightarrow M\times M$, where $M\times M$ is the trivial
groupoid whose objects are the points of $M$. Given $\cK\in
\Grn_s(M,S)$, the rank $s$ bundle $K\eqdef \ev_\ast(\cK)$ is trivial,
as pointed out above. The corestriction:
\ben
\label{epdef}
e_p\eqdef
  \ev_p|_\cK^{K_p}:\cK\rightarrow K_p~~ 
\een
of $\ev_p|_\cK$ to its image is
  bijective for all $p\in M$.  Given $p,q\in M$, consider the
  bijection:
\ben
\label{Upq}
\U_{pq}\eqdef e_q\circ e_p^{-1}:K_p\stackrel{\sim}{\rightarrow}K_q~~. 
\een
This satisfies:
\be
\U_{qr}\circ \U_{pq}=\U_{pr}~~\mathrm{and}~~~\U_{pp}=\id_{K_p}~~,~~\forall p,q,r\in M~~
\ee
and hence defines a functor $\U:M\times M\rightarrow A(K)$ whose image
is a trivial sub-groupoid of $A(K)$ (the Hom-sets of the image being
singleton sets). There exists a unique flat connection $\D$ on $K$
whose holonomy functor $\Hol_\D$ (the functor which associates to
every morphism of the groupoid $\Pi_1(M)$ the parallel transport of
$\D$ along curves belonging to that homotopy class) coincides with the
composition $\U\circ E:\Pi_1(M)\rightarrow A(K)$.  This flat
connection is trivial since the image of $\Hol_\D=\U\circ E$ (which
coincides with the image of $\U$) is a trivial groupoid. This
construction gives a natural map $\Phi_s:\Grn_s(M,S)\rightarrow
\Trivf_s(M,S)$ given by $\Phi_s(\cK)=(K,\D)$. Relation \eqref{Upq}
implies that any $\xi\in \cK$ satisfies:
\ben
\label{xiflat}
\xi(q)=\U_{pq}(\xi(p))~~,~~\forall p,q\in M~~,
\een
which implies $\D\xi=0$. Hence $\cK$ is contained in the space $\Gamma_\flat(K,\D)$. 
Since $\dim \Gamma_{\flat}(K,\D)=\rk
K=s=\dim \cK$, we must have $\cK=\Gamma_\flat(K,\D)$. This shows that
$\cK$ is uniquely determined by $(K,\D)$ and hence that $\Phi_s$ is
injective. Consider now a pair $(K,\D)\in \Trivf_s(M,S)$ and set $
\cK\eqdef \Gamma_\flat(K,\D)$. We have $\dim \cK=\rk K=s$. The map
$\ev_p|_\cK$ is injective with image equal to $K_p$. Thus $\cK$ is
locally non-degenerate and $K=\ev_\ast(\cK)$. Since $\D$ is a trivial
flat connection, its parallel transport along curves from $p$ to $q$
depends only on $p$ and $q$, being given by \eqref{Upq}. Thus
$(K,\D)=\Phi_s(\cK)$, which shows that $\Phi_s$ is
surjective. $\blacksquare$

\subsection{$\cB$-compatible locally non-degenerate subspaces of $\Gamma(M,S)$}

\paragraph{Definition.} 
A locally nondegenerate subspace $\cK\subset \Gamma(M,S)$ is $\cB$-{\em compatible} if 
the following condition is satisfied: 
\ben
\label{KE}
\cB(\xi,\xi')=\mathrm{constant~on~}M~~,~~\forall \xi,\xi'\in \cK~.
\een

\

\noindent Any $\cB$-compatible locally nondegenerate subspace $\cK$ is
endowed with a Euclidean metric $\cB_0:\cK\times \cK\rightarrow \R$
which is defined through $\cB(\xi,\xi')=\cB_0(\xi,\xi')1_M$, where
$1_M\in \cinf$ is the constant function equal to one on $M$. For
simplicity, we will not distinguish notationally between $\cB_0$ and
the $\cinf$-valued bilinear form $\cB|_{\cK\otimes \cK}=\cB_0
1_M$. Condition \eqref{KE} is an invariant way of saying that $\cK$
admits a basis $\xi_1,\ldots, \xi_s$ having the property that the
scalar products $\cB_p(\xi_i(p),\xi_j(p))$ are independent of the
point $p\in M$ for all $i,j=1\ldots s$.  Using the Gram-Schmidt
algorithm for $\cB_0$, it is easy to see that this amounts to the
condition that $\cK$ admits a basis which is everywhere orthonormal in
the sense $\cB_p(\xi_i(p),\xi_j(p))=\delta_{ij}$ for all $i,j=1\ldots
s$ and all $p\in M$.

Let $\Grn_s(M,S,\cB)$ be the subset of $\Grn_s(M,S)$ consisting of
$\cB$-compatible locally nondegenerate subspaces of dimension $s$ and
$\Trivf_s(M,S,\cB)$ be the subset of $\Trivf_s(M,S)$ consisting of
those pairs $(K,\D)\in \Trivf_s(M,S)$ for which $\D$ is a
$\cB$-compatible connection.

\paragraph{Corollary.} 
$\Phi_s$ restricts to a bijection between $\Grn_s(M,S,\cB)$ and $\Trivf_s(M,S,\cB)$. 

\

\noindent{\bf Proof.} 
Let $\cK\in \Grn_s(M,S)$ and $(K,\D)\eqdef \Phi_s(\cK)$.  Condition \eqref{KE} is equivalent
with:
\ben
\label{KE1}
\cB_q\circ (e_q\otimes e_q)=\cB_p\circ (e_p\otimes e_p)~,~\forall ~p,q\in M~~,
\een
where the map $e_p$ was defined in \eqref{epdef}.  Since
$e_p:\cK\rightarrow K_p$ is bijective for all $p$, the relation
$e_q=\U_{pq}\circ e_p$ (which follows from \eqref{Upq}) shows that
\eqref{KE1} is equivalent with the condition:
\ben
\label{Uisom}
\cB_q|_{K_q}\circ (\U_{pq}\otimes \U_{pq})=\cB_p|_{K_p}~~,
\een
which amounts to the requirement that $\U_{pq}$ be an isometry from
$(K_p,\cB_p|_{K_p})$ to $(K_q,\cB_q|_{K_q})$ for all $p,q\in M$. In
turn, this is equivalent with the requirement that the trivial flat
connection $\D$ be $\cB$-compatible. $\blacksquare$

Let $\cK\in \Grn_s(M,S)$ and $(K,\D)\eqdef \Phi_s(\cK)$. The following
statement is obvious in view of the above:

\paragraph{Proposition.} 
Let $\xi_1,\ldots, \xi_s\in \cK$ and $\Xi\eqdef
(\xi_1,\ldots,\xi_s)$. Then:
\begin{enumerate}
\itemsep 0,0em
\item $\Xi$ is a basis of $\cK$ iff it is a $\D$-flat global frame of
  $K$.
\item When $\cK$ is $\cB$-compatible, $\Xi$ is an orthonormal basis of
  $\cK$ iff it is an everywhere-orthonormal $\D$-flat global frame of
  $K$.
\end{enumerate}

\subsection{Relation to virtual CGK spaces}

Let $\cK(\mbD,Q)$ denote the space of solutions to
\eqref{par_eq} and $s\eqdef \dim \cK(\mbD,Q)$.  

\paragraph{Proposition.} 
$\cK(\mbD,Q)$ is a locally non-degenerate subspace of $\Gamma(M,S)$.

\

\noindent{\bf Proof.}
For ease of notation, let $\cK\eqdef \cK(\mbD,Q)$. Let $\cP_{pq}(M)$
denote the set of curves in $M$ starting at $p$ and ending at $q$.
For any $\gamma\in \cP_{pq}(M)$, let:
\be
U_{pq}(\gamma):S_p\stackrel{\sim}{\rightarrow} S_q
\ee
denote the parallel transport of $\mbD$ along $\gamma$. Since the
connection $\mbD$ need not be flat, the isomorphisms $U_{pq}(\gamma)$
may depend on $\gamma$ and not only on its homotopy class. For any
$\xi\in \cK$, the first equation in \eqref{par_eq} implies:
\ben
\label{pt}
\xi(q)=U_{pq}(\gamma)\xi(p)~~,~~\forall p,q\in M~~,~~\forall \gamma\in \cP_{pq}(M)~~.
\een
When $\xi\in \ker (\ev_p)$ (i.e. $\xi(p)=0$), relation \eqref{pt}
gives $\xi(q)=0$ for all $q\in M$ and hence $\xi=0$.  This shows that
the restriction $\ev_p|_{\cK}:\cK\rightarrow S_p$ is injective for all
$p\in M$ and thus that $\cK$ is a locally non-degenerate subspace of
$\Gamma(M,S)$.  $\blacksquare$

\

\noindent Let $(K,\D)\eqdef \Phi_s(\cK(\mbD,Q))$. 

\paragraph{Proposition.} 
The bundle $K$ is $\D$-invariant, thus:
\ben
\label{Kinv}
\mbD(\Gamma(M,K))\subset \Omega^1(M,K)~~.
\een
Furthermore, the restriction of $\mbD$ to $K$ is a trivial flat connection on
$K$ which coincides with $\D$:
\ben
\label{Dind}
\mbD \xi=\D\xi~~,~~\forall \xi\in \Gamma(M,K)~~.
\een

\noindent{\bf Proof.}
Defining $e_p$ as in \eqref{epdef}, relation \eqref{pt} implies:
\ben
\label{Ures}
U_{pq}(\gamma)|_{K_p}=\ev_q\circ e_p^{-1}~~,
\een
showing that $U_{pq}(\gamma)(K_p)=K_q$ for all $p,q\in M$ and
$\gamma\in \cP_{pq}(M)$. This means that $\mbD$ preserves the bundle
$K$, i.e. relation \eqref{Kinv} holds.  Corestricting $U_{pq}$ to its
codomain, \eqref{Ures} gives the parallel transport of the connection
$\mbD_0$ induced by $\mbD$ on the sub-bundle $K$:
\be
U_{pq}(\gamma)|_{K_p}^{K_q}=e_q\circ e_p^{-1}=\U_{pq}~~,
\ee
where in the last line we used formula \eqref{Upq} for the parallel
transport $\U_{pq}$ of the trivial flat connection $\D$ of $K$. This
shows that $\D$ coincides with the restriction of $\mbD$ to
$K$. $\blacksquare$

\paragraph{Remark.}
Let us fix $p\in M$. Using relations \eqref{pt}, it is
  easy to see that $K_p$ can be written as:
\ben
\label{Kp}
K_p=\cap_{\gamma\in \cP_{pp}(M)} \ker (U_{pp}(\gamma)-\id_{K_p})\cap 
\cap_{q\in M, \gamma\in \cP_{pq}(M)} \ker (U_{pq}(\gamma)^{-1}\circ Q_q\circ U_{pq}(\gamma))~~.
\een
Given $\xi(p)\in K_p$, the element $U_{pq}(\gamma)\xi(p)\in S_q$ is
independent of the choice of $\gamma\in \cP_{pq}(M)$ and $\xi$ can be
recovered using \eqref{pt}.  Thus \eqref{par_eq} is equivalent with
the condition $\xi(p)\in K_p$, where $K_p$ is given by \eqref{Kp}.  

\paragraph{Proposition.} 
Assume that $\mbD$ is $\cB$-compatible. Then $\cK(\mbD,Q)$ is a
$\cB$-compatible locally non-degenerate subspace of $\Gamma(M,S)$.

\

\noindent{\bf Proof.}  When $\mbD$ is $\cB$-compatible, its parallel
transport satisfies:
\be
\cB_q(U_{pq}(\gamma)\otimes U_{pq}(\gamma))=\cB_p~~,~~\forall p,q\in M~~,~~\forall \gamma\in \cP_{pq}(M)~~.
\ee
Restricting this to $K_p$ shows that $\U_{pq}\eqdef U_{pq}(\gamma)|_{K_p}$ is
an isometry from $(K_p,\cB_p)$ to $(K_q,\cB_q)$ for all $p,q\in M$,
i.e. relation \eqref{Uisom} is satisfied. This implies the conclusion
since \eqref{Uisom} is equivalent with \eqref{KE}. $\blacksquare$

\paragraph{Proposition.} Let $\cK$ be an $s$-dimensional subspace of $\Gamma(M,S)$. 
Then the following statements are equivalent: 
\begin{enumerate}[(a)]
\item $\cK$ is a virtual CGK space.
\item $\cK$ is locally non-degenerate.
\end{enumerate}

\

\noindent{\bf Proof.}
The implication $(a)\Rightarrow (b)$ was proved before. To prove the
inverse implication, let $\cK\in \Grn_s(M,S)$ and $(K,\D)\eqdef
\Phi_s(\cK)$.  Choosing a complement $K'$ of $K$ inside $S$ gives a
direct sum decomposition:
\be
S=K\oplus K'~~.
\ee 
We have $\cK=\Gamma_\flat(K,\D)\subset \Gamma(M,K)$ and hence
$\D\xi=0$ for all $\xi\in \cK$.  Let $Q\in \Gamma(M,\End(S))$ denote
the projector of $S$ onto $K'$ parallel to $K$.  Then $K=\ker Q$ and
hence $Q\xi=0$ for any $\xi\in \cK$. Let $D'$ be any connection on
$K'$. Then the direct sum $\mbD\eqdef \D\oplus D'$ is a connection on
$S$ which satisfies $\mbD\xi=0$ for all $\xi\in \cK$. It follows that
we have $\cK\subset \cK(\mbD,Q)$.  To show the inverse inclusion, let
$\xi\in \cK(\mbD,Q)$. Then $Q\xi=0$ and hence $\xi\in
\Gamma(M,K)$. The equation $\mbD\xi=0$ is thus equivalent with
$\D\xi=0$.  It follows that we have $\xi\in \Gamma_\flat(K,\D)=\cK$
and hence $\cK(\mbD,Q)\subset \cK$.  $\blacksquare$

\paragraph{Proposition.} Let $\cK$ be an $s$-dimensional subspace of $\Gamma(M,S)$. 
Then the following statements are equivalent: 
\begin{enumerate}[(a)]
\item $\cK$ is a $\cB$-compatible virtual CGK space.
\item $\cK$ is a $\cB$-compatible locally non-degenerate subspace of $\Gamma(M,S)$.
\end{enumerate}

\

\noindent{\bf Proof.} The implication $(a)\Rightarrow (b)$ was proved
before. For the inverse implication, let $\cK\in \Grn_s(M,S,\cB)$ and 
$(K,\D)\eqdef \Phi_s(\cK)$. Let $K^\perp$ be the $\cB$-orthocomplement
of $K$ inside $S$ and $Q$ the $\cB$-orthoprojector on $K^\perp$. Let
$D'$ be any $\cB$-compatible connection on $K^\perp$ and let
$\mbD=\D\oplus D'$.  The same argument as in the proof of the previous
proposition shows that we have $\cK=\cK(\mbD,Q)$. Since $\cK$ is
$\cB$-compatible, the connection $\D$ is $\cB$-compatible and hence
$\mbD$ is $\cB$-compatible as well. $\blacksquare$

\subsection{The chirality stratification}

Let:
\be
P_\pm\eqdef \frac{1}{2}(1\pm \gamma(\nu))\in \Gamma(M,\Hom(S,S^\pm))~~
\ee
be the $\cB$-orthogonal projectors of $S$ onto $S^\pm$ and let
$(K,\D)=\Phi_s(\cK)$ for some locally non-degenerate subspace
$\cK\subset \Gamma(M,S)$.

\paragraph{Definition.} 
The {\em chiral projections} of $K$ are the smooth generalized sub-bundles of
$S^\pm$ defined through:
\be
K_\pm\eqdef P_\pm K\subset S^\pm~~.
\ee
The {\em chiral rank functions} $r_\pm$ of $K$ are the rank functions
of $K_\pm$:
\be
r_\pm\eqdef \rk K_\pm:M\rightarrow \N~~.
\ee

\

\noindent Notice that $r_\pm$ are lower semicontinuous and that they
satisfy:
\ben
\label{rineq}
r_\pm \leq s~~,~~r_++r_- \geq s~~,
\een
where the last inequality follows from the fact that $K$ is a
sub-bundle of the generalized bundle $K_+\oplus K_-$.

\paragraph{Definition.} 
The {\em chiral slices} of $K$ are the
following cosmooth generalized sub-bundles of $K$:
\be
K^\pm\eqdef S^\pm \cap K~~.
\ee

\

\noindent The identity $S^\pm=\ker P_\mp$ implies $K^\pm=\ker
(P_\mp|_K)$, hence we have exact sequences of generalized sub-bundles
of $S$:
\be
0\rightarrow K^\mp \hookrightarrow K\stackrel{P_\pm|_K}{\longrightarrow} K_\pm\rightarrow 0~~,
\ee
which give the relations: 
\ben
\label{rhopm}
\sigma_\pm\eqdef \rk  K^\pm=s-r_\mp~~.
\een

\paragraph{Definition.} 
We say that $p\in M$ is a {\em $K$-special point} if
$(r_-(p),r_+(p))\neq (s,s)$. The {\em $K$-special locus} is the
following subset of $M$:
\ben
\cS\eqdef \{p\in M|p~\mathrm{is~a~}K\rm{-special~point}\}~.
\een
The open complement: 
\be
\cG\eqdef M\setminus \cS=\{p\in M|r_-(p)=r_+(p)=s\}
\ee
will be called the {\em non-special locus} of $K$; its elements are
the {\em non-special points}. The special locus admits a
stratification induced by the chiral rank functions:
\be
\cS=\sqcup_{\tiny \begin{array}{c}0\leq k,l\leq s\\k+l\geq s\\(k,l)\neq (s,s)\end{array}}\cS_{kl}~~,
\ee
where:
\be
\cS_{kl}\eqdef \{p\in \cS|r_-(p)=k ~\&~ r_+(p)=l\}~~.
\ee

\paragraph{Definition.} 
The {\em chirality stratification} of $M$ induced by $\cK$ is the decomposition: 
\be
M=\cG\sqcup \sqcup_{\tiny \begin{array}{c}0\leq k,l\leq s\\k+l\geq s\\(k,l)\neq (s,s)\end{array}}\cS_{kl} ~~.
\ee

\subsection{The stabilizer stratification}

For any $p\in M$, consider the natural representation of the group
$\Spin(T_pM,g_p)\simeq \Spin(8)$ on $S_p$. 

\paragraph{Definition.} 
The {\em stabilizer group of $K$ at $p$} is the 
closed subgroup of $\Spin(T_pM,g_p)$ consisting of those elements
which act trivially on the subspace $K_p\subset S_p$:
\ben
H_p\eqdef \{h\in \Spin(T_pM,g_p)|~h u=u~~\forall u\in K_p\}~~.
\een

\paragraph{Definition.} 
Let $\cK$ be an $s$-dimensional locally-nondegenerate subspace of
$\Gamma(M,S)$.  The {\em stabilizer stratification of $M$ induced by
  $\cK$} is the stratification of $M$ given by the isomorphism type of
$H_p$. Two points $p,q\in M$ belong to the same stratum of this
stratification iff $H_p$ and $H_q$ are isomorphic.

\paragraph{Remark.} 
Given a frame $(\xi_1,\ldots, \xi_s)$ of $\cK$, the group
$H_p$ coincides with the common stabilizer of $\xi_i(p)$:
\be
\!\!\!H_p=\Stab_{\Spin(T_pM,g_p)}(\xi_1(p),\ldots, \xi_s(p))=
\{h\in \Spin(T_pM,g_p)|~h\xi_i(p)=\xi_i(p)~~\forall i=1\ldots s\}~~.
\ee
When $\cK$ is $\cB$-compatible, we can formulate this as follows. Let
$V_p^{(s)}(S_p,\cB_p)$ be the Stiefel manifold of orthonormal
$s$-frames of the Euclidean space $(S_p,\cB_p)$ and $V^{(s)}(S,\cB)$
be the fiber bundle over $M$ having $V_p^{(s)}(S_p,\cB_p)$ as its
fiber at $p$.  Since the action of $\Spin(T_pM,S_p)$ on $S_p$
preserves $\cB_p$, it induces an action on $V^{(s)}(S_p,\cB_p)$:
\ben
\label{Spin8Stiefel}
(u_1,\ldots u_s)\rightarrow (h u_1,\ldots, h u_s)~~,
~~\forall  h\in \Spin(T_pM,g_p)~~,~~
\forall (u_1,\ldots, u_s)\in V^{(s)}(S_p,\cB_p)~~.
\een
An orthonormal basis $\Xi\eqdef (\xi_1,\ldots, \xi_s)$ of $\cK$
can be viewed as a smooth section of the fiber bundle
$V^{(s)}(S,\cB)$.  Then $H_p$ coincides with the stabilizer of the
value $\Xi(p)$ of this section under the action
\eqref{Spin8Stiefel}. The Stiefel manifold $V^{(s)}(S_p,\cB_p)$ has a
stratification by the isomorphism type of stabilizers inside
$\Spin(T_pM,g_p)$.  Similarly, there is a stratification
$\Sigma^{(s)}$ of the total space of $V^{(s)}(S,\cB)$ by the
isomorphism type of stabilizers. Since $H_p$ is independent of the
choice of $\Xi$, the $\Xi$-preimage of the stratification
$\Sigma^{(s)}$ is independent of $\Xi$ and coincides with the
stabilizer stratification of $M$ induced by $\cK$. A similar
formulation exists when $\cK$ is not $\cB$-compatible, if one
replaces $V^{(s)}(S,\cB)$ by the bundle $V^{(s)}(S)$ whose fiber at
$p\in M$ is the Stiefel manifold $V^{(s)}(S_p)$ of all $s$-frames of
the fiber $S_p$.

Assuming $\rk K\geq 1$, let $\q_p:\Spin(T_pM,g_p)\rightarrow
\SO(T_pM,g_p)$ denote the double covering morphism. The image
$G_p\eqdef \mathfrak{q_p}(H_p)$ is a subgroup of $\SO(T_pM,g_p)$. The
$\q_p$-preimage of the unit element $\id_{T_pM}$ of $\SO(T_pM,g_p)$ is
a two-point set which consists of the unit element of
$\Spin(T_pM,g_p)$ and another element which we denote by
$\epsilon_p$. The latter acts on $S_p$ as minus the identity and hence
it cannot be contained in $H_p$. It follows that the restriction of
$\q_p$ to $H_p$ is injective and hence it gives an isomorphism from
$H_p$ to $G_p$. Thus the stabilizer stratification coincides with the
stratification of $M$ by the isomorphism type of $G_p\simeq H_p$.  Let
$T$ be a stratum of the connected refinement of this stratification and
let $G_T$ denote the isomorphism type of $G_p\simeq H_p$ for $p\in T$.
Endow $T$ with the topology induced from $M$.  The restriction
$\Fr_+(M)|_T$ of the oriented frame bundle $\Fr_+(M)$ of $M$ is a
principal $\SO(8)$ bundle (in the sense of general topology) defined
over the connected topological space $T$.  Picking specific
$G_p$-orbits inside the fibers $\Fr_p(M)$ for $p\in T$ specifies a
$G_T$-reduction of structure group of $\Fr(M)|_T$ and such reductions
for all connected strata $T$ fit together into a ``stratified
G-structure'' defined on $M$.

\paragraph{Remark.} 
In the Physics literature, what we call a stratified G-structure is
sometimes called a ``local G-structure''. In Mathematics, the word
``local'' refers to a structure or property which is defined/which
holds for all points of some {\em open} subset of a topological
space. Since most strata of the stabilizer stratification are not open
subsets of $M$, it is clear that a stratified G-structure cannot be a
local G-structure in the sense used in Mathematics.

\subsection{The case of compactifications to $\AdS_3$}

As an example, consider compactifications down to an $\AdS_3$ space of
cosmological constant $\Lambda=-8\kappa^2$, where $\kappa$ is a
positive parameter. In this case, the eleven-dimensional background
$\mathbf{M}$ is diffeomorphic with $N\times M$, where $N$ is an
oriented 3-manifold diffeomorphic with $\R^3$ and carrying the
$\AdS_3$ metric $g_3$. The metric on $\mathbf{M}$ is taken to be a
warped product:
\beqan
\label{warpedprod}
\dd \mathbf{s}^2  & = & e^{2\Delta} \dd s^2~~~{\rm where}~~~\dd s^2=\dd s^2_3+ g_{mn} \dd x^m \dd x^n~~.
\eeqan
The warp factor $\Delta$ is a smooth real-valued function defined on
$M$ while $\dd s_3^2$ is the squared length element of the $\AdS_3$
metric $g_3$. The Ansatz for the field strength $\mathbf{G}$ of
eleven-dimensional supergravity is:
\ben
\label{Gansatz}
\mathbf{G} = \nu_3\wedge \mathbf{f}+\mF~~,~~~~\mathrm{with}~~ 
\mF\eqdef e^{3\Delta}F~~,~~\mathbf{f}\eqdef e^{3\Delta} f~~
\een
where $f\in \Omega^1(M)$, $F\in \Omega^4(M)$ and $\nu_3$ is the volume
form of $(N,g_3)$. The Ansatz for the supersymmetry generator is:
\ben
\label{eta_ansatz}
\boldsymbol{\eta}=e^{\frac{\Delta}{2}}\sum_{i=1}^s{\zeta_i\otimes \xi_i}~~,
\een
where $\xi_i\in \Gamma(M,S)$ are Majorana spinors of spin $1/2$ on the
internal space $(M,g)$ and $\zeta_i$ are Majorana spinors on $(N,g_3)$
which satisfy the Killing equation with positive Killing
constant\footnote{With our conventions (see Appendix
  \ref{app:notations}), gamma matrices in signature $(-1,2)$ can be
  taken to be real, for example $\gamma_0=i\sigma_2,
  \gamma_1=\sigma_1, \gamma_2=\sigma_3$ where $\sigma_k$ are the Pauli
  matrices.  In the Mathematics convention for Clifford algebras,
  $\gamma_k$ are replaced by ${\hat \gamma}_k=i\gamma_k$. A Killing
  Majorana spinor on $\AdS_3$ satisfies $\nabla_k\zeta=\lambda\gamma_k
  \zeta$, with a real Killing constant $\lambda=\pm \kappa$. In the
  Mathematics convention, this corresponds to $\nabla_k\zeta={\hat
    \lambda}{\hat \gamma}_k\zeta$, with imaginary ${\hat
    \lambda}=-i\lambda=\mp i\kappa$; these are known as ``imaginary
  Killing spinors''. In the Ansatz, we choose
  $\lambda=+\kappa$.}. Assuming that $\zeta_i$ are Killing spinor on
the $\AdS_3$ space $(N,g_3)$, the supersymmetry condition is satisfied if
$\xi_i$ satisfies \eqref{par_eq}, where:
\be
\mbD_X=\nabla_X^S+\frac{1}{4}\gamma(X\lrcorner F)+\frac{1}{4}\gamma((X_\sharp\wedge f) \nu) +\kappa \gamma(X\lrcorner \nu)~~,~~X\in \Gamma(M,TM)
\ee
and:
\be
Q=\frac{1}{2}\gamma(\dd \Delta)-\frac{1}{6}\gamma(\iota_f\nu)-\frac{1}{12}\gamma(F)-\kappa\gamma(\nu)~~.
\ee 
Here $\nabla^S$ is the connection induced on $S$ by the Levi-Civita
connection of $(M,g)$, while $\nu$ is the volume form of
$(M,g)$. Neither $Q$ nor the connection $\mbD$ preserve the chirality
decomposition $S=S^+\oplus S^-$ of $S$ when $\kappa\neq 0$:
\be
\mbD(S^\pm)\not\subseteq T^\ast M\otimes S^\pm~~,~~Q(S^\pm)\not\subseteq S^\pm~~.
\ee
It is not hard to check \cite{ga1} that $\mbD$ is $\cB$-compatible:
\ben
\label{flatness}
\dd \cB(\xi',\xi'')=\cB(\mbD\xi', \xi'')+\cB(\xi',\mbD\xi'')~~,
~~\forall \xi',\xi''\in \Gamma(M,S)~~.
\een
This implies that any $\xi,\xi'\in \cK(\mbD,Q)$ satisfy
$\cB(\xi,\xi')=\mathrm{constant}$, i.e. $\cK$ is a $\cB$-compatible
flat subspace of $\Gamma(M,S)$.  The restriction $\D=\mbD|_K$ is a
$\cB$-compatible trivial flat connection on $\cK(\mbD,Q)$.

\paragraph{Remarks.}
\begin{enumerate}
\itemsep 0.0em
\item An equivalent formulation of the Ansatz \eqref{eta_ansatz} is
  that the supersymmetry generators of the background span the space
  $\cK_3\otimes \cK$, where $\cK_3$ is the two-dimensional space of
  real Killing spinors on $\AdS_3$ with positive Killing
  constant. Then $\xi_i$ in the Ansatz can be taken to form an
  orthonormal basis of $\cK$, while $\zeta_i$ are {\em
    arbitrary} elements of $\cK_3$, so that the Ansatz describes the
  general element of $\cK_3\otimes \cK$. Notice that one does not gain
  anything by decomposing $\xi_i$ into their positive and negative
  chirality parts in the Ansatz since $\mbD$ and $Q$ do not preserve
  the sub-bundles $S^\pm$ and hence $\cK$ need not equal the direct
  sum of the intersections $\cK\cap \Gamma(M,S^+)$ and $\cK\cap
  \Gamma(M,S^-)$.
\item The amount $\cN$ of supersymmetry preserved by the background
  may be larger than $s$ in the limit $\Lambda=0$, when $\AdS_3$
  reduces to the three-dimensional Minkowski space. In that limit, the
  results of \cite{MartelliSparks,g2s} imply that all fluxes must
  vanish, thus $F=f=\kappa=0$ and that $\dd\Delta=0$, which imply
  $\mbD=\nabla^S$ and $Q=0$, hence both $\mbD$ and $Q$ preserve the
  sub-bundles $S^+$ and $S^-$ of $S$. A discussion of this
  phenomenon for the case $s=1$ (which gives $\cN=1$ for $\Lambda<0$
  and $\cN=2$ for $\Lambda=0$) can be found in \cite[Appendix
    B.1]{g2s}. 
\end{enumerate}
\subsection{A toy example: the case $s=1$}

Let us illustrate the discussion above with the case $s=1$. Then $\cK$
is a one-dimensional locally non-degenerate subspace of $\Gamma(M,S)$
while $(K,\D)$ is a trivial flat line sub-bundle of $S$.  Assume that
$\cK$ is $\cB$-compatible. Then a $\cB$-compatible frame of $\cK$ is
given by a single Majorana spinor $\xi\in \Gamma(M,S)$ which is
everywhere of norm one; the same spinor gives a global normalized
frame of $K$. The chiral projections $K_\pm$ are the generalized
sub-bundles of $S$ generated by the positive and negative chirality
parts $\xi^\pm\eqdef P_\pm\xi$ of $\xi$. The chiral rank functions are
given by:
\be
r_\pm(p)=\dim \langle \xi^\pm(p)\rangle=\twopartdef{0~}{\xi(p)\in S_p^\mp}{1~}{\xi(p)\not \in S_p^\mp}~.
\ee
The chiral slices are:
\be
K^\pm(p)=\twopartdef{0}{\xi(p)\not\in S_p^\pm}{\langle \xi(p)\rangle\simeq \R~}{\xi(p)\in S_p^\pm}~~.
\ee
Since $\xi(p)$ is everywhere non-vanishing, we have $r_++r_-\geq 1$,
thus the allowed values are $(r_-(p),r_+(p))\in \{(0,1),(1,0),
(1,1)\}$. Hence the chirality stratification takes the form:
\be
M=\cU \sqcup \cW_-\sqcup \cW_+
\ee
where: 
\beqa
&& \cU\equiv \cG\eqdef \{p\in M|r_-(p)=r_+(p)=1\}=\{p\in M|\xi(p)\not \in S_p^+\sqcup S_p^-\}~~\\
&&\cW_-\equiv \cS_{10}\eqdef \{p\in M|r_-(p)=1~,~r_+(p)=0\}=\{p\in M|\xi(p)\in S_p^-\}~~\\
&&\cW_+\equiv \cS_{01}\eqdef \{p\in M|r_-(p)=0~,~r_+(p)=1\}=\{p\in M|\xi(p)\in S_p^+\}~~.
\eeqa
Thus $\cU$ is the non-chiral locus while $\cS_{10}$ and $\cS_{01}$
are the negative and positive chirality loci of \cite{g2s}. The union
of the latter is the chiral locus $\cW=\cW_-\sqcup
\cW_+=\cS_{10}\sqcup \cS_{01}$ of loc. cit. In this case, the
stabilizer stratification is a coarsening of the chirality
stratification, namely we have $H_p\simeq \Spin(7)$ for $p\in \cW$ and
$H_p\simeq \G_2$ for $p\in \cU$. The stabilizer stratification
coincides with the rank stratification of the cosmooth generalized
distribution $\cD\eqdef \ker V$, where
$V\eqdef_U\cB(\xi,\gamma_a\xi)e^a$ is the one-form bilinear
constructed from $\xi$, where the expressions are given on an open subset $U\subset M$ 
 which supports a local coframe $(e^a)$. 
Namely, we have $\dim \cD(p)=7$ for $p\in \cU$
and $\dim \cD(p)=8$ for $p\in \cW$. The group $G_p=\q_p(H_p)$ is a
subgroup of $\SO(\cD(p),g_p)$ for any $p\in M$.

\begin{figure}[H]
\begin{center}
\includegraphics[width=0.32\linewidth]{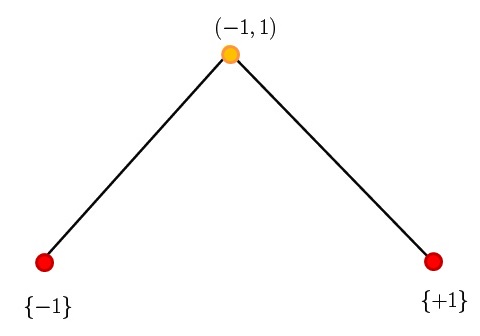}
\caption{Hasse diagram of the incidence poset (see Appendix
  \ref{app:stratif}) of the connected refinement of the Whitney
  stratification of the interval $[-1,1]$. The $b$-preimages of the
  strata represented by red and yellow dots correspond to the $\Spin(7)$
  and $G_2$ loci of $M$.}
\label{fig:WhitneyInterval}
\end{center}
\end{figure}

Let $b\eqdef_U\cB(\xi,\gamma(\nu)\xi)\in \cC^\infty(\R)$ denote the
scalar bilinear constructed from $\xi$. It was shown in
\cite{MartelliSparks,ga1} that the Fierz identities for $\xi$ imply
$1-b^2=||V||^2$ and hence $b^2\leq 1$. Thus the image of the map $b$
is contained within the interval $[-1,1]$.  This interval is a
semi-algebraic set given by the single polynomial inequality $b^2\leq 1
$ for a variable $b\in \R$. Its canonical Whitney stratification has a
0-dimensional stratum given by the two-point set $\{-1,1\}$ and a
one-dimensional stratum given by the open interval $(-1,1)$. The
connected refinement of the Whitney stratification has two connected
0-dimensional strata given by the one-point sets $\{+1\}$ and $\{-1\}$
and a connected 1-dimensional stratum given by the open interval
$(-1,1)$. The Hasse diagram of the incidence poset of this
stratification is depicted in Figure \ref{fig:WhitneyInterval}.
\noindent It was shown in \cite{g2s} that the rank/stabilizer
stratification coincides with the $b$-preimage of the canonical Whitney
stratification of $[-1,1]$:
\be
\cW=b^{-1}(\{-1, 1\})~~,~~\cU= b^{-1}((-1,1))~~.
\ee
On the other hand, the chirality stratification coincides with the
$b$-preimage of the connected refinement of the Whitney
stratification:
\be
\cW_\pm=b^{-1}(\{\pm 1\})~~,~~\cU=b^{-1}((-1,1))~~.
\ee
It was also shown in \cite{g2s} that, for compactifications down to
$\AdS_3$, the supersymmetry conditions \eqref{par_eq} imply that the
singular distribution $\cD$ integrates to a singular foliation in the
sense of Haefliger \cite{Haefliger}.

\paragraph{Remark.} 
The compactifications studied in \cite{BeckerSpin7} correspond to the
case $M=\cW_+$.

\

\noindent As we shall see in the next sections, the situation is much
more complicated when $s=2$. In that case (assuming that $\cK$ is
$\cB$-compatible):
\begin{enumerate}
\itemsep 0.0em
\item The chirality and stabilizer stratifications do not agree, in
  the sense that neither of them is a refinement of the other.
\item There exists a cosmooth singular distribution $\cD$ (determined
  by the intersection of the kernels of three one-form valued spinor
  bilinears $V_1,V_2$ and $V_3$) which integrates to a Haefliger
  foliation in the $\AdS_3$ case. The rank stratification of $\cD$
  does not agree with the chirality stratification or with the
  stabilizer stratification.
\item The stabilizer stratification coincides with the rank
  stratification of a cosmooth singular sub-distribution $\cD_0\subset
  \cD$ (given by the intersection of $\cD$ with the kernel of a fourth
  one-form spinor bilinear $W$), but $\cD_0$ need not be integrable in
  the case of compactifications down to $\AdS_3$. The group
  $G_p=\q_p(H_p)$ is a subgroup of $\SO(\cD_0(p),g_p)$ (and hence also
  a subgroup of $\SO(\cD(p),g_p)$) for any $p\in M$.
\item The chirality stratification coincides with the $b$-preimage of
  the connected refinement of the Whitney stratification of a
  three-dimensional semi-algebraic set $\cR$, where $b\in
  \cC^\infty(M,\cR)$ is a map constructed using scalar spinor
  bilinears defined by an orthonormal basis of $\cK$. We have $\im b\subset \cR$.
\item The stabilizer stratification and the rank stratification of
  $\cD$ are different coarsenings of the $B$-preimage of the connected
  refinement of the canonical Whitney stratification of a
  four-dimensional semi-algebraic set $\fP$, where $B:M\rightarrow
  \R^4$ is another map constructed using an orthonormal basis of $\cK$. We have
  $\im B\subset \fP$.
\end{enumerate}

\section{The generalized distributions $\cD$ and $\cD_0$ in the case $s=2$}
\label{sec:cD}

Throughout this section, $\cK$ denotes a $\cB$-compatible locally non-degenerate subspace of $\Gamma(M,S)$. 

\subsection{Functions and one-forms defined by a basis of $\cK$}
An orthonormal basis $(\xi_1,\xi_2)$ of $\cK$ induces three smooth
functions $b_i \in \cinf$  $(i=1,2,3$), namely:
\ben
\label{forms0}
b_1 =_U\cB(\xi_1,\gamma(\nu)\xi_1)~,~b_2 =_U\cB(\xi_2,\gamma(\nu)\xi_2)~,~b_3=_U\cB(\xi_1,\gamma(\nu)\xi_2)~~.
\een
It will be convenient to work with the combinations:
\ben
\label{forms0pm3}
b_\pm\eqdef \frac{1}{2}(b_1\pm b_2)~.
\een
Also consider the one-forms $V_i,V_3,W\in \Omega^1(M)$ (with $i=1,2$) given by:
\ben
\label{forms1}
V_i =_U \cB(\xi_i,\gamma_a\xi_i)e^a~~ ,~~V_3 \eqdef_U \cB(\xi_1,\gamma_a\xi_2)e^a~~,~~W\eqdef_U \cB(\xi_1,\gamma_a \gamma(\nu)\xi_2) e^a~~,
\een
where the relations hold in any local coframe $(e^a)$ defined above an
open subset $U\subset M$. It will be convenient to work with the linear
combinations:
\ben
\label{forms1pm3}
V_\pm\eqdef\frac{1}{2}(V_1\pm V_2)~~,~~V_3^\pm=\frac{1}{2}(V_3\pm W)~~.
\een
We have: 
\be
V_1=V_++V_-~~,~~V_2=V_+-V_-~~,~~V_3=V_3^++V_3^-~~,~~W=V_3^+-V_3^-~~.
\ee
Decomposing $\xi_i$ into their positive and negative chirality parts gives:
\ben
\label{VWspinors}
V_1=_U2\cB(\xi_1^-,\gamma_a\xi_1^+)e^a~~,~~V_2=_U2\cB(\xi_2^-,\gamma_a\xi_2^+)e^a~~,~~V_3^\pm=_U\cB(\xi_1^\mp,\gamma_a\xi_2^\pm)e^a~~.
\een

\subsection{The distributions $\cD$ and $\cD_0$}

The 1-forms $V_1,V_2,V_3$ generate a smooth generalized sub-bundle
$\cV$ (in the sense of \cite{Drager}) of the cotangent bundle of $M$,
which is also generated by $V_+,V_-,V_3$. Let:
\be
\cD\eqdef \ker V_1\cap \ker V_2\cap\ker V_3=\ker V_+\cap \ker V_-\cap \ker V_3
\ee 
denote the polar of $\cV$, which is a cosmooth generalized
distribution on $M$, i.e.  a cosmooth generalized sub-bundle of $TM$
in the sense of \cite{Drager}. Its orthogonal complement $\cD^\perp$
inside $TM$ is a smooth generalized sub-bundle of $TM$ which is
spanned by the three vector fields obtained from $V_+,V_-,V_3$ by
applying the musical isomorphism. Notice that $\cD$ contains the
cosmooth generalized distribution:
\be
\cD_0\eqdef \ker V_+\cap \ker V_-\cap \ker V_3^+\cap \ker V_3^-=\cD\cap \ker W\subset \cD~~.
\ee

\paragraph{Remark.} 
When considering compactifications to $\AdS_3$, one can show that the
supersymmetry conditions imply that $\cD$ is an integrable
distribution (namely, it integrates to a singular foliation in the
sense of Haefliger) while $\cD_0$ may fail to be integrable. This is
one reason for considering the generalized distribution $\cD$.

\subsection{Behavior under changes of orthonormal basis of $\cK$}
\label{sec:rot}

An orthonormal basis $(\xi'_1,\xi'_2)$ of $\cK$ having the same
orientation as $(\xi_1,\xi_2)$ has the form:
\ben
\label{basis_rot}
\begin{split}
&\xi'_1=\cos \left(\frac{u}{2}\right)\xi_1+\sin \left(\frac{u}{2}\right)\xi_2~,\\
&\xi'_2=-\sin \left(\frac{u}{2}\right)\xi_1+\cos \left(\frac{u}{2}\right)\xi_2~
\end{split}
\een
(where $u\in \R$) and defines the following 0-forms and 1-forms,
where $i=1,2$:
\ben
\begin{split}
&b'_i=_U\cB(\xi'_i,\gamma(\nu)\xi'_i)~~,~~b'_3=_U\cB(\xi'_1,\gamma(\nu)\xi'_2)~~\\
&V'_i =_U \cB(\xi'_i,\gamma_a\xi'_i)e^a~ ,~V'_3 \eqdef_U \cB(\xi'_1,\gamma_a\xi'_2)e^a~~,~~W'\eqdef_U \cB(\xi'_1,\gamma_a \gamma(\nu)\xi'_2) e^a~.
\end{split}
\een
Substituting \eqref{basis_rot} into these expressions, we find that
$b_+,V_+$ and $W$ are invariant while each of the pairs $b_-,b_3$ and
$V_-,V_3$ transforms in the fundamental representation of $\SO(2)$:
\ben
\label{Vrot}
\begin{split}
& b'_+=b_+~~,~~V'_+=V_+~~,~~W'=W\\
&b'_-=\cos(u) b_-+\sin (u)b_3~~~~,~~V'_-=\cos(u) V_-+\sin (u) V_3~~\\
&b'_3=-\sin (u) b_-+\cos (u) b_3~~,~~V'_3=-\sin (u) V_-+\cos (u) V_3~~.
\end{split}
\een
The improper rotation: 
\ben
\label{improt}
\left[\begin{array}{c}\xi'_1\\\xi'_2\end{array}\right]=
\left[\begin{array}{cc}0& 1\\1 & 0\end{array}\right]
\left[\begin{array}{c}\xi_1\\\xi_2\end{array}\right]
\een
which permutes $\xi_1$ and $\xi_2$ induces permutations
$b_1\leftrightarrow b_2$ and $V_1\leftrightarrow V_2$ while $V_3, b_3$
remain unchanged and $W$ changes sign (to arrive at these conclusions,
one uses the relations $\gamma(\nu)^t=\gamma(\nu)$,
$\gamma_a^t=\gamma_a$ and the fact that $\gamma(\nu)$ anticommutes
with $\gamma_a$). Hence \eqref{improt} induces the transformations:
\ben
\label{imptf}
\begin{split}
& b_+\rightarrow b_+~~~,~~V_+\rightarrow V_+~~~,~~W\rightarrow -W~\\
& b_-\rightarrow - b_-~,~~V_-\rightarrow -V_- ~\\
& b_3\rightarrow b_3~~~~,~~~V_3\rightarrow V_3~~.\\
\end{split}
\een
It follows that $b_+$ and $V_+$ depend only on $\cK$ while $W$ depends
on $\cK$ and on a choice of orientation of $\cK$. On the other hand,
$b_-$ and $V_-$ change sign while $b_3$ and $V_3$ are invariant under
a change of orientation of $\cK$. It also follows from the above that
$\cD$ and $\cD_0$ depend only on the space $\cK$ and do not depend on
the choice of basis $(\xi_1,\xi_2)$ for $\cK$.

\subsection{The rank stratification of $\cD$} 
\label{subsec:distrib}

The compact manifold $M$ decomposes into a disjoint union according to
the rank of $\cD$:
\ben
\label{Mdec}
M=\cU\sqcup \cW~~,
\een
where the open set: 
\be
\cU\eqdef \{p\in M|\rk \cD(p)=5\}=\{p\in M|V_+(p),V_-(p),V_3(p)~\mathrm{are~linearly~independent}\}
\ee
will be called the {\em generic locus} while its closed complement: 
\be
\cW\eqdef \{p\in M|\rk \cD(p)>5\}=\{p\in M|V_+(p),V_-(p),V_3(p)~\mathrm{are~linearly~dependent}\}
\ee
will be called the {\em degeneration locus}. The latter admits a stratification according to the corank of $\cD(p)$:
\be
\cW=\sqcup_{k=0}^2 \cW_k~~,
\ee
whose locally closed strata are given by:
\ben
\cW_k\eqdef \{p\in \cW|\dim \cV_p =k\}=\{p \in \cW| \rk \cD(p)=8-k\}~~.
\een
Combining everything gives the {\em rank stratification of $\cD$}:
\ben
\label{Mdecfull}
M=\cU\sqcup \cW_2\sqcup \cW_1\sqcup \cW_0~~.
\een

\paragraph{Definition.} 
$\cK$ is called {\em generic} if $\cU\neq \emptyset$ and {\em
  non-generic} otherwise.

\

\noindent Notice that $\cK$ is non-generic iff $\rk\cD(p)\geq 6$ for
all $p\in M$, i.e. iff $V_1(p),V_2(p)$ and $V_3(p)$ are linearly
dependent for all $p\in M$.

\paragraph{Remark.} 
For any $p\in M$, let $A_p\in \Hom(\R^3,T_p^\ast M)$ denote the linear
map which takes the canonical basis $\epsilon_i$ of $\R^3$ into
$V_i(p)$:
\be
A_p(\epsilon_i)=V_i(p)~~,~~\forall i=1\ldots 3~~.
\ee
This defines a smooth section $A\in
\Gamma(M,\Hom(\underline{\R}^3,T^\ast M))$, where
$\underline{\R}^3=M\times \R^3$ is the trivial rank 3 vector bundle
over $M$.  Each space $\Hom(\R^3,T_p^\ast M)\simeq \Mat(3,8,\R)$
admits a Whitney stratification (the so-called canonical
stratification \cite{Thom, Koriyama}) whose strata are the Stiefel
manifolds $V^{(k)}(T_p^\ast M)=\{A\in \Hom(\R^3,T_p^\ast M)|\rk
A=k\}\simeq \{{\hat A}\in \Mat(3,8,\R)|\rk {\hat A}=k\}$, where
$k=0,1,2,3$. This induces a stratification of the total space of the
bundle $\Hom(\underline{\R}^3,T^\ast M)$, whose preimage through the
section $A$ is the stratification \eqref{Mdecfull}. The preimage of
the stratum defined by $\rk A=3$ is the set $\cU$ while the preimages
of the strata defined by $\rk A=k$ with $k=0,1,2$ are the sets
$\cW_k$.

\subsection{The rank stratification of $\cD_0$}

The generalized distribution $\cD_0$ induces a decomposition:
\ben
\label{Mdec0}
M=\cU_0 \sqcup \cZ~~,
\een
where:
\be
\cU_0\eqdef \{p\in M|\rk \cD_0(p)=4\}=\{p\in M|V_+(p),V_-(p),V_3(p), W(p)~\mathrm{are~linearly~independent}\}~~
\ee
is an open subset of $M$ while:
\be
\cZ\eqdef \{p\in M|\rk \cD_0(p)>4\}=\{p\in M|V_+(p),V_-(p),V_3(p), W(p)~\mathrm{are~linearly~dependent}\}
\ee
is closed. The latter stratifies according to the corank of $\cD_0$:
\be
\cZ=\sqcup_{k=0}^3 \cZ_k~~,
\ee
with locally closed strata given by:
\ben
\cZ_k\eqdef \{p \in \cZ| \rk \cD_0(p)=8-k\}~~.
\een
We shall see later\footnote{This follows from the algebraic
  constraints satisfied by $V_i$ and $W$ --- see Theorem 4 of
  Subsection \ref{sec:Gstratif}.} that we always have:
\be
\cU_0=\cU~~\mathrm{and}~~\cZ_3=\emptyset~~,
\ee
so in particular $\rk\cD_0(p)$ can never equal five. 
We thus obtain the {\em rank stratification of $\cD_0$}:
\ben
\label{Mdecfull0}
M=\cU\sqcup \cZ_2\sqcup \cZ_1\sqcup \cZ_0~~.
\een

\subsection{Constraints on the stabilizer stratification}

Since the action of $\Spin(T_pM,g_p)$ on $S_p$ commutes
with $\gamma_p(\nu_p)$, relations \eqref{forms1} imply:
\ben
\label{Hinclusion}
H_p\subset \Stab_{\Spin(T_pM,g_p)}(V_+(p),V_-(p),V_3(p),W(p))~~,
\een
where $\Spin(T_pM,g_p)$ acts on $T_p^\ast M$ by the dual of the vector
representation.  The action of $\Spin(T_pM,g_p)$ on $T_p^\ast M$ is
obtained from that of $\SO(T_pM,g_p)$ by pre-composing with the
covering morphism $\q_p:\Spin(T_pM,g_p)\rightarrow
\SO(T_pM,g_p)$. Hence \eqref{Hinclusion} implies:
\ben
\label{Ginclusion}
G_p\subset \Stab_{\SO(T_pM,g_p)}(V_+(p),V_-(p),V_3(p),W(p))\simeq \SO(\cD_0(p),g_p)~~.
\een
In particular, we have:
\ben
\label{Ginclusion2}
G_p\subset \Stab_{\SO(T_pM,g_p)}(V_+(p),V_-(p),V_3(p))\simeq \SO(\cD(p),g_p)~~.
\een

\section{The chirality stratification for $s=2$}
\label{sec:spinors}

\begin{figure}[H]
\begin{center}
\includegraphics[width=0.5\linewidth]{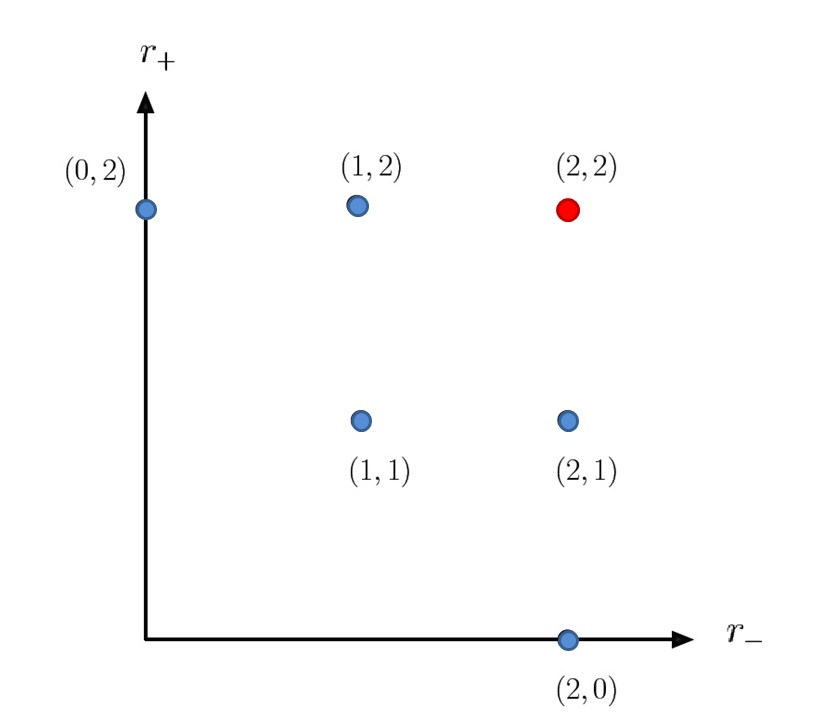}
\caption{Allowed values for the pair $(r_-(p),r_+(p))$. The values
  corresponding to $K$-special points are shown in blue, while the
  remaining value is shown as a red dot.}
\label{fig:ranks}
\end{center}
\end{figure}

Let $\cK$ be a two-dimensional $\cB$-compatible locally-nondegenerate
subspace of $\Gamma(M,S)$ and $(K,\D)$ be the associated trivial flat
sub-bundle of $S$. Relations \eqref{rineq} imply (see Figure
\ref{fig:ranks}):
\ben
(r_-(p),r_+(p))\in \{(0,2),(2,0),(1,1), (1,2),(2,1), (2,2)\}~~,~~\forall p\in M~~.
\een
\noindent A point $p\in M$ is $K$-special if $(r_-(p),r_+(p))\neq
(2,2)$ (the blue dots in Figure \ref{fig:ranks}). The special locus
decomposes as:
\be
\cS=\cS_{12}\sqcup \cS_{21}\sqcup \cS_{11}\sqcup \cS_{02}\sqcup \cS_{20}~~,
\ee
where $~\cS_{kl}=\{p\in M|r_-(p)=k,~r_+(p)=l\}$, while the chirality stratification is given by: 
\be
M=\cG\sqcup \cS_{12}\sqcup \cS_{21}\sqcup \cS_{11}\sqcup \cS_{02}\sqcup \cS_{20}~~,
\ee
where $\cG$ is the non-special locus.

\subsection{The semi-algebraic body $\cR$}

\begin{figure}[H]
\centering
\begin{subfigure}{0.49\textwidth}
\centering
\includegraphics[width=0.59\linewidth]{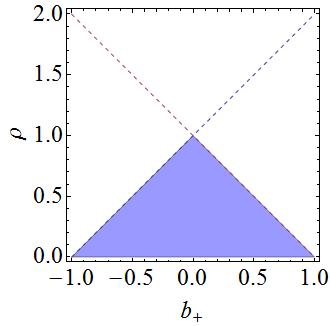}
\vskip 0.1in
\caption{The region $\Delta$ in the $(b_+,\rho)$ plane.}
\end{subfigure}
\begin{subfigure}{0.49\textwidth}
\centering
\includegraphics[width=0.65\linewidth]{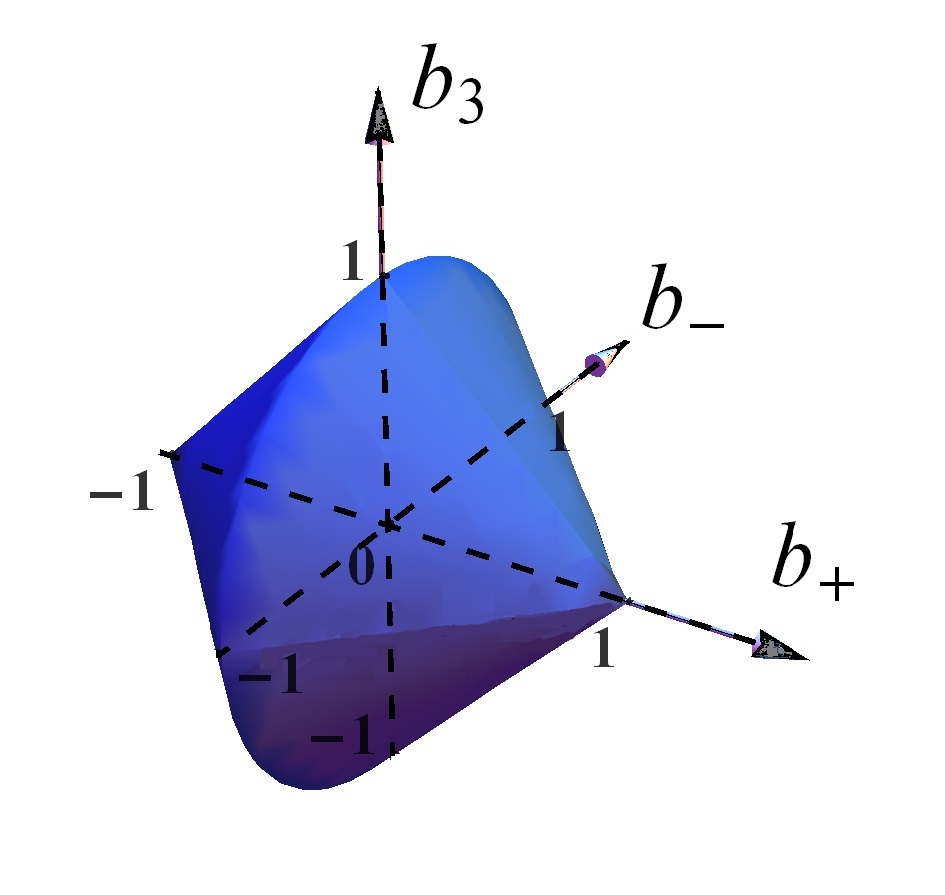}
\caption{The body $\cR$ is the solid of revolution obtained by
  rotating $\Delta$ around its hypotenuse, which lies on the $b_+$
  axis; it is the union of two compact right-angled cones whose bases
  coincide. }
\end{subfigure}
\caption{The region $\Delta$ (blue) and the body $\cR$. }
\label{fig:DeltaR}
\end{figure}

Consider the compact convex body (see Figure \ref{fig:DeltaR}):
\ben
\label{Rdef}
\boxed{\cR=\{(b_+,b_-,b_3)\in [-1,1]^3~|~\sqrt{b_-^2+b_3^2}\leq 1-|b_+|\}}~~,
\een
which is contained in the three-dimensional compact unit ball. Setting: 
\be
\rho\eqdef \sqrt{b_-^2+b_3^2}\in [0,1]~~,
\ee
one finds that $\cR$ is the solid of revolution obtained by rotating
the following isosceles right triangle around its hypothenuse:
\ben
\label{Deltadef}
\Delta\eqdef \{(b_+,\rho)\in [-1,1]\times [0,1]~|~\rho\leq 1-|b_+|\}~~.
\een

\begin{figure}[H]
\centering
\includegraphics[width=0.37\linewidth]{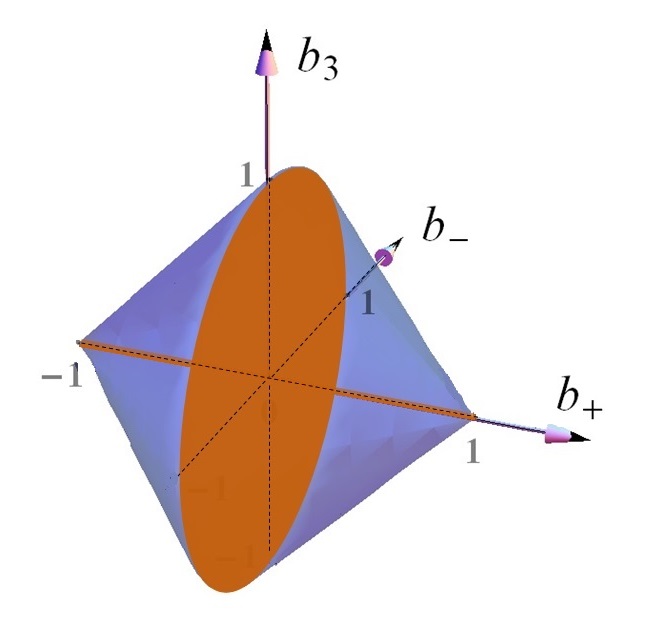}
\caption{The axis $I$ and the median disk $D$, depicted in orange.}
\label{fig:axis_disk}
\end{figure}

\noindent The compact interval:
\ben
I\eqdef \{(b_+,0,0)|b_+\in [-1,1]\}=\{b\in \cR|b_-=b_3=0\}
\een
will be called the {\em axis} of $\cR$ while the compact disk:
\ben
D\eqdef \{(0,b_-,b_3)|b_-^2+b_3^2\leq 1\}=\{b\in \cR|b_+=0\}
\een
will be called the {\em median disk} of $\cR$. The boundary $\partial
D$ of the median disk will be called the {\em median circle} (see
Figure \ref{fig:axis_disk}). 

\noindent Notice that $\cR$ is a semi-algebraic set, since it can be
described by polynomial inequalities:
\be
\cR=\{(b_+,b_-,b_3)\in \R^3|b_+^2+b_-^2+b_3^2\leq 1 ~\&~ b_-^2+b_3^2\leq \frac{1}{4}(1+b_-^2+b_3^2-b_+^2)^2\}~~.
\ee
Hence both $\cR$ and its frontier $\partial \cR$ (which is again a
semi-algebraic set) admit \cite{Whitney} canonical\footnote{Recall
  that the canonical Whitney stratification of a semi-algebraic set is
  the coarsest stratification which satisfies the frontier conditions
  as well as Whitney's regularity condition (b). The strata of this
  stratification need not be connected. The general algorithm through which
  such stratifications can be obtained is due to \cite{Whitney} and is 
  discussed in detail in \cite{Rannou}.} stratifications by
semi-algebraic sets. Namely, the frontier:
\be
\partial \cR= \{b\in \cR|\rho=1-|b_+|\}=\{(b_+,b_-,b_3)\in \R^3|b_+^2+b_-^2+b_3^2\leq 1 ~\&~ b_-^2+b_3^2=\frac{1}{4}(1+b_-^2+b_3^2-b_+^2)^2\}
\ee
decomposes into borderless manifolds $\partial_k \cR$ of dimensions
$k=0,1,2$:
\ben
\label{WhitneyR}
\partial\cR=\partial_0\cR\sqcup \partial_1\cR\sqcup \partial_2\cR~~,
\een
where:
\ben
\label{Rb}
\partial_0\cR\eqdef \partial I~,~\partial_1\cR\eqdef \partial D~,~\partial_2\cR\eqdef \partial \cR\setminus (\partial D\cup \partial I)~~.
\een
The set $\partial_1\cR$ coincides with the median circle and hence it
is connected. The set $\partial_0\cR$ is disconnected, being a
disjoint union of two singleton sets:
\be
\partial_0\cR=\partial_0^-\cR\sqcup \partial_0^+\cR~~,
\ee
where:
\be
\partial_0^-\cR\eqdef \{(-1,0,0)\}~~,~~\partial_0^+\cR\eqdef\{(1,0,0)\}~~
\ee
will be called the {\em left and right tips} of $\cR$. We have:
\be
\partial_0\cR\sqcup \partial_1\cR=\cR\cap S^2=\{(b_+,b_-,b_3)\in \R^3|b_+^2+b_-^2+b_3^2= 1 ~\&~ b_-^2+b_3^2=\frac{1}{4}(1+b_-^2+b_3^2-b_+^2)^2\}~~,
\ee
where $S^2$ denotes the unit sphere in the space $\R^3$. 

\begin{figure}[H]
\begin{center}
\includegraphics[scale=0.3]{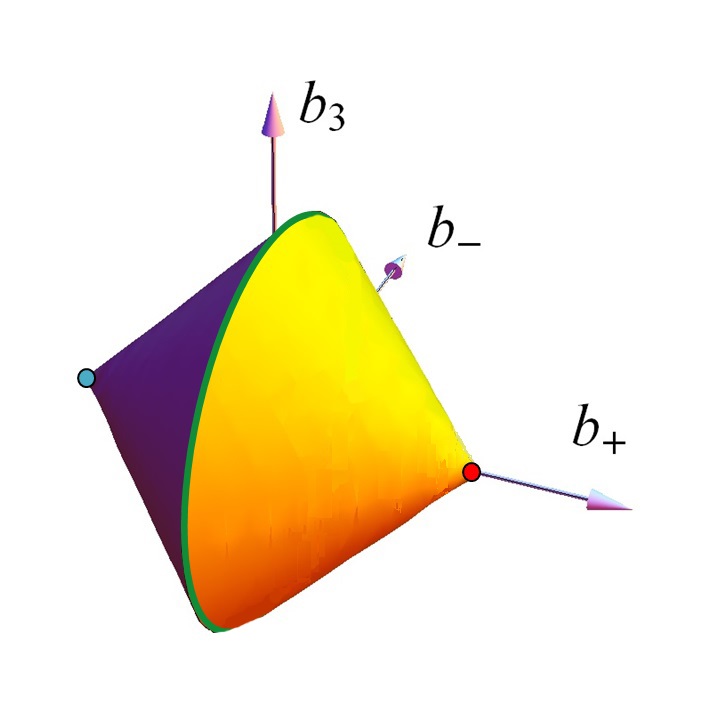}
\caption{The connected refinement of the canonical Whitney
  stratification of $\partial \cR$. We use green for the median circle
  $\partial_1\cR=\partial D$, purple for $\partial_2^-\cR$, yellow for
  $\partial_2^+ \cR$, blue for $\partial_0^-\cR$ and red for
  $\partial_0^+\cR$. Theorem 1 of Subsection \ref{sec:Theorem1} shows
  that the $b$-preimage of $\partial_1\cR$ equals $\cS_{11}$, while
  the $b$-preimages of $\partial_2^+\cR$ and $\partial_2^-\cR$ equal
  $\cS_{12}$ and $\cS_{21}$ respectively. The $b$-preimages of
  $\partial^+_0\cR$ and $\partial_0^-\cR$ are the sets $\cS_{02}$ and
  $\cS_{20}$.}
\label{fig:Rlr}
\end{center}
\end{figure}

The set $\partial_2\cR$ is relatively open in $\partial \cR$, being a
disjoint union of two connected components:
\be
\partial_2\cR=\partial_2^-\cR\sqcup \partial_2^+\cR~~,
\ee
where:
\be
\partial_2^-\cR\eqdef \{b\in \partial_2\cR|~b_+\in(-1,0)\}~~,~~\partial_2^+\cR\eqdef \{b\in \partial_2\cR|~b_+\in(0,1)\}
\ee
will be called the {\em left and right components} of $\partial_2\cR$.
The canonical Whitney stratification of $\partial \cR$ has strata
given by $\partial_0\cR$, $\partial_1\cR$ and $\partial_2\cR$ and
corresponds to the decomposition \eqref{WhitneyR}, while its connected
refinement (see Appendix \ref{app:stratif}) has strata given by
$\partial_0^\pm \cR$, $\partial_1\cR$ and $\partial_2^\pm \cR$ and
corresponds to the decomposition:
\ben
\label{WhitneyRconn}
\partial\cR=\partial_0^-\cR\sqcup\partial_0^+\cR\sqcup \partial_1\cR\sqcup \partial_2^-\cR\sqcup \partial_2^+\cR~~.
\een
The connected strata appearing in \eqref{WhitneyRconn} are depicted in
Figure \ref{fig:Rlr}, while the values of $b_+$ and $\rho$ on those
strata are summarized in Table \ref{table:partialR}. Together with
$\Int \cR$, the strata $\partial_k\cR$ give the canonical Whitney
stratification of $\cR$, whose connected refinement has strata
$\Int\cR, \partial_0^\pm\cR,\partial_1\cR$ and $\partial_2^\pm \cR$.
\begin{table}[H]
\centering {\footnotesize
\begin{tabular}{|c|c|c|c|c|c|c|}
\hline connected stratum& dimension & component of & topology & $b_+$ & $\rho$  \\ \hline
$\partial_0^\pm \cR$ & $0$ & $\partial_0\cR$ & point & $\pm 1$ & $0$ \\ \hline
$\partial_1\cR$ & $1$ & $\partial_1\cR$ & circle & $0$ & $1$\\ \hline
$\partial_2^\pm\cR$ & $2$ & $\partial_2\cR$ & open annulus & $\pm (1-\rho)$ & $(0,1)$\\ \hline
\end{tabular}
}
\caption{Connected strata of $\partial\cR$.}
\label{table:partialR}
\end{table}

\noindent For later reference, let:
\ben
\label{Rpm}
\cR^-\eqdef \{b\in \cR| b_+\leq 0\}~~,~~\cR^+\eqdef \{b\in \cR| b_+\geq 0\}
\een
denote the two closed halves of $\cR$ lying to the left and right of
the median disk. Notice that $\cR^\pm$ are three-dimensional compact
full cones. We have a disjoint union decomposition:
\be
\cR=\partial \cR\sqcup D\sqcup \Int (\cR^+)\sqcup \Int(\cR^-)~~.
\ee
We also define: 
\be
I^-\eqdef I\cap \cR^-=[-1,0]\times \{(0,0)\}~~,~~I^+\eqdef I\cap \cR^+=[0,1]\times \{(0,0)\}~~,
\ee
which give the decomposition: 
\be
I=\Int (I^+)\sqcup \Int (I^-) \sqcup\{(0,0,0)\}~~.
\ee

\subsection{The map $b$}
\label{sec:b}

Define the function $b\in \cC^\infty(M,\R^3)$ through:
\ben
\label{bdef}
b(p)\eqdef (b_+(p),b_-(p),b_3(p))~~.
\een

\paragraph{Proposition.} 
The image of $b$ is a subset of $\cR$.

\

\noindent{\bf Proof.}  
Let us separate $\xi_i$ into positive and negative chirality
parts: 
\be
\xi_i=\xi_i^++\xi_i^-~,~\mathrm{with}~~\xi_i^\pm\eqdef P_\pm\xi_i~~\mathrm{and}~~i=1,2~~.
\ee
The condition $\cB(\xi_i,\xi_j)=\delta_{ij}$ and the definitions of
$b_1,b_2$ and $b_3$ give the equations:
\beqa
&&||\xi_i^+||^2+||\xi_i^-||^2=1~~,~~||\xi_i^+||^2-||\xi_i^-||^2=b_i~~,\\
&& \cB(\xi_1^+,\xi_2^+)+\cB(\xi_1^-,\xi_2^-)=0~~,~~\cB(\xi_1^+,\xi_2^+)-\cB(\xi_1^-,\xi_2^-)=b_3~~,
\eeqa
which can be solved to give:
\ben
\label{xib}
||\xi_i^\pm||^2=\frac{1}{2}(1\pm b_i)~~,~~\cB(\xi_1^\pm,\xi_2^\pm)=\pm\frac{1}{2}b_3~~.
\een
The Gram matrix $\fG$ of the ordered system
$(\xi_1^+,\xi_2^+,\xi_1^-,\xi_2^-)$ takes the block diagonal
form\footnote{The Gram matrices considered here are defined at every
  point $p\in M$, hence they are matrix-valued functions defined on
  $M$.}:
\ben
\label{Gspinors}
\fG=\left[\begin{array}{cc}\fG_+ & 0\\ 0 & \fG_-\end{array}\right]
\een
where: 
\be
\fG_\pm\eqdef \left[\begin{array}{cc} ||\xi_1^\pm||^2 & \cB(\xi_1^\pm, \xi_2^\pm)\\  \cB(\xi_2^\pm, \xi_1^\pm) & ||\xi_2^\pm||^2\end{array}\right]= 
\left[\begin{array}{cc} \frac{1}{2}(1\pm b_1) & \pm \frac{1}{2}b_3\\ \pm \frac{1}{2}b_3 & \frac{1}{2}(1\pm b_2) \end{array}\right]
\ee
are the Gram matrices of the pairs $(\xi_1^\pm, \xi_2^\pm)$. A simple
computation gives:
\ben
\label{detGpm}
\det \fG_\pm=\frac{1}{4}(1+b_1b_2\pm b_1\pm b_2-b_3^2)=\frac{1}{4}\left[(1\pm b_+)^2-\rho^2\right]~~.
\een
The conclusion now follows from \eqref{detGpm} upon using the fact
that $\fG_\pm$ are semipositive, which by Sylvester's theorem amounts
to the conditions $(\fG_\pm)_{11}\geq 0$, $(\fG_\pm)_{22}\geq 0$ and
$\det \fG_\pm\geq 0$. $\blacksquare$

\paragraph{Remarks.} 
\begin{enumerate}
\itemsep 0.0em
\item We have $r_\pm=\rk \fG_\pm$ and $\rk(K_+\oplus K_-)=\rk \fG$.
\item The determinant $\det \fG=\det \fG_+\det \fG_-$ vanishes iff one
  of $\det \fG_\pm$ vanishes. The equality $\det \fG_\pm=0$ is
  attained on the locus where $r_\pm \leq 1$.
\item $\fG_\pm$ is symmetric under the exchange $\xi_1\leftrightarrow
  \xi_2$.
\end{enumerate}

\subsection{The map $b'$} 

Consider the determinant line bundle $\det K=\wedge^2 K$. The scalar
product $\cB|_K$ induces a norm on $\det K$ which we denote by
$||~||$. Since $(\xi_1,\xi_2)$ is an orthonormal frame of $(K,\cB)$,
we have $||\xi_1\wedge \xi_2||=1$ and hence $\xi_1\wedge \xi_2$ is an
orthonormal frame of $\det K$. The generalized bundles $K_\pm\subset
S_\pm$ inherit the Euclidean scalar product $\cB$ from $S$ and hence
$\wedge^2 K_\pm$ are normed generalized vector bundles of rank at most
one. The generalized bundle morphisms $P_\pm^K\eqdef P_\pm|_K^{K_\pm}:
K\rightarrow K_\pm$ induce generalized bundle morphisms $\wedge^2
P_\pm^K :\det K\rightarrow \wedge^2 K_\pm$.

\paragraph{Proposition.}  We have: 
\be
\det \fG_\pm=||\wedge^2 P_\pm^K||_\op~~,
\ee
where $||~||_\op$ denotes the fiberwise operator norm on the
generalized bundle $\Hom(\wedge^2 K, \wedge^2 K_\pm)$. In particular,
$\det \fG_\pm$ depend only on the subspace $\cK\subset \Gamma(M,S)$
and are independent of the choice of orthonormal basis for $\cK$.

\

\noindent{\bf Proof.}  By definition of $\wedge^2 P_\pm^K$, we have
$(\wedge^2 P_\pm^K)(\xi_1\wedge \xi_2)=P_\pm(\xi_1)\wedge
P_\pm(\xi_2)=\xi_1^\pm\wedge \xi_2^\pm$.  Using the Gram identity,
this gives $||(\wedge^2 P_\pm^K)(\xi_1\wedge
\xi_2)||^2=||\xi_1^\pm\wedge \xi_2^\pm||^2=\det \fG_\pm$, which
implies the conclusion. $\blacksquare$

\paragraph{Remark.} 
The proposition allows one to give a different proof of the fact that
the functions $b_+,\rho^2\in \cinf$ depend only on $\cK$. This follows
by taking the sum and difference of equations \eqref{detGpm}, which
allows one to express $\rho$ and $b_+$ in terms of $\det \Gamma_+$ and
$\det \Gamma_-$.

\

\noindent The map $b'\eqdef (b,\rho):M\rightarrow \R^2$ depends only
on $\cK$. Since the image of $b$ is contained inside $\cR$, we find:

\paragraph{Proposition.} The image of $b'$ is a subset of $\Delta$. 

\subsection{Relation to the rank stratifications of $\cD$ and $\cD_0$}

\paragraph{Lemma.} Let $p\in \cS$ be a $K$-special point. Then:
\begin{enumerate}
\itemsep 0.0em
\item When $p\in \cS_{11}\sqcup \cS_{12}$, we can rotate the
  orthonormal basis of $\cK$ such that either of the following holds,
  at our choice:
\begin{enumerate}[(a)]
\itemsep 0.0em
\item $\xi_1(p)\in S_p^+$, in which case $V_1(p)=V_3^+(p)=0$, 
  $V_3(p)=V_3^-(p)$ and $W(p)=-V_3(p)$
\item $\xi_2(p)\in S_p^+$, in which case $V_2(p)=V_3^-(p)=0$, 
  $V_3(p)=V_3^+(p)$ and $W(p)=V_3(p)$
\end{enumerate}
\item When $p\in \cS_{11}\sqcup \cS_{21}$, we can rotate the
  orthonormal basis of $\cK$ such that either of the following holds,
  at our choice:
\begin{enumerate}[(a)]
\itemsep 0.0em
\item $\xi_1(p)\in S_p^-$, in which case $V_1(p)=V_3^-(p)=0$,
  $V_3(p)=V_3^+(p)$ and $W(p)=V_3(p)$
\item $\xi_2(p)\in S_p^-$, in which case $V_2(p)=V_3^+(p)=0$,
  $V_3(p)=V_3^-(p)$ and $W(p)=-V_3(p)$
\end{enumerate}
\item When $p\in \cS_{11}$, we can rotate the orthonormal basis of
  $\cK$ such that either of the following holds, at our choice:
\begin{enumerate}[(a)]
\itemsep 0.0em
\item $\xi_1(p)\in S_p^+$ and $\xi_2(p)\in S_p^-$, in which case $V_1(p)=V_2(p)=V_3^+(p)=0$,
  $V_3(p)=V_3^-(p)$ and $W(p)=-V_3(p)$
\item $\xi_1\in S_p^-$ and $\xi_2(p)\in S_p^+$, in which case $V_1(p)=V_2(p)=V_3^-(p)=0$,
  $V_3(p)=V_3^+(p)$ and $W(p)=V_3(p)$.
\end{enumerate}
\end{enumerate}

\

\noindent{\bf Proof.}
\begin{enumerate}
\itemsep 0.0em
\item The condition $p\in \cS_{11}\sqcup \cS_{12}$ implies $r_-(p)=1$
  and hence $\det \Gamma_-(p)=0$.  Then $\xi_1^-(p)=\lambda_1 w$ and
  $\xi_2^-(p)=\lambda_2 w$ for some $w\in S_p^-\setminus\{0\}$, where
  $\lambda_1$ and $\lambda_2$ are real numbers, one of which may be
  zero. Under a rotation \eqref{basis_rot} of the basis of $\cK$, we
  have:
\be
(\xi'_1)^-(p)=\lambda_1'w~~,~~(\xi'_2)^-(p)=\lambda_2'w
\ee
with:
\beqa
\lambda_1'=\lambda_1\cos \left(\frac{u}{2}\right)+\lambda_2\sin \left(\frac{u}{2}\right) ~~,~~
\lambda_2'=-\lambda_1\sin \left(\frac{u}{2}\right)+\lambda_2\cos \left(\frac{u}{2}\right) ~~.
\eeqa
It is easy to see that we can choose $u$ such that either of the
combinations $\lambda_1\cos \left(\frac{u}{2}\right)+\lambda_2\sin
\left(\frac{u}{2}\right)$ or $-\lambda_1\sin
\left(\frac{u}{2}\right)+\lambda_2\cos \left(\frac{u}{2}\right)$
vanishes, at our choice. The statements about the 1-form spinor
bilinears follow immediately from the forms of $\xi_i$ after such a
rotation.
\item The case $p\in \cS_{11}\sqcup \cS_{21}$ proceeds similarly.
\item The condition $p\in \cS_{11}$ implies $\det
  \Gamma_+(p)=\det\Gamma_-(p)=0$. Using the result at point 2.,
  perform a rotation of the orthonormal basis of $\cK$ such that
  $\xi_2^+(p)=0$ for the new basis. Then $\xi_2^-(p)=\xi_2(p)$ and hence
  $||\xi_2^-(p)||=||\xi_2(p)||=1$ and
  $\cB_p(\xi_1^-(p),\xi_2^-(p))=-\cB_p(\xi_1^+(p),\xi_2^+(p))=0$,
  where the last relation follows from $\cB(\xi_1,\xi_2)=0$.  Thus:
  \ben
\label{dGm}
\det \fG_-(p)=||\xi_1^-(p)||^2~~.  
\een 
Since $\det \fG_-(p)$ is invariant under \eqref{basis_rot}, we also
have $\det \fG_-(p)=0$ after the rotation, which implies
$\xi_1^-(p)=0$ by \eqref{dGm}. Thus $\xi_1(p)\in S_p^+$ and
$\xi_2(p)\in S_p^-$ after the rotation. Had we rotated such that
$\xi_1^+(p)=0$, we would have similarly concluded that $\xi_1(p)\in
S_p^-$ and $\xi_2(p)\in S_p^+$.  The statements about the 1-form
spinor bilinears follow immediately.
\end{enumerate} $\blacksquare$

\paragraph{Remark.} 
For $p\in \cS_{02}\sqcup \cS_{20}$, we obviously have
$V_1(p)=V_2(p)=V_3(p)=W(p)=0$. The compactifications studied in
\cite{BeckerCY} correspond to the case $M=\cS_{02}$.

\paragraph{Proposition.} 
Let $p\in \cS$ be a $K$-special point. Then $\cD_0(p)=\cD(p)$ and we
can rotate the basis of $\cK$ such that either $V_3(p)=W(p)$ or
$V_3(p)=-W(p)$, at our choice. Moreover: 
\begin{itemize}
\itemsep 0.0em
\item For $p\in \cS_{20}\sqcup \cS_{02}$, we have $\cD(p)=T_pM$, hence $\rk \cD(p)=8$
\item For $p\in \cS_{11}$, we have $\rk \cD(p)=7$
\item For $p\in \cS_{12}\sqcup \cS_{21}$ we have $\rk \cD(p)=6$. 
\end{itemize}

\noindent{\bf Proof.}  Follows from the Lemma and from the remark
above upon using the fact that $\cD$ and $\cD_0$ are invariant under
rotations of the basis of $\cK$. The proposition also follows from Theorem
1 below and from the results of Subsection \ref{sec:pipreimage} and of
Appendix \ref{app:hatG}. $\blacksquare$

\paragraph{Remark.} 
It is shown in Appendix \ref{app:hatG} that, for $p\in \cG$, we have $\rk \cD(p)\in
\{5,6,7\}$ and $\rk \cD_0(p)\in \{4,6\}$, hence $\cD_0(p)$ and
$\cD(p)$ may differ; in fact, their ranks cannot be determined only
from the value of $b(p)$. Together with the Proposition, this gives:
\be
\cS_{02}\sqcup \cS_{20}=\cW_0=\cZ_0~~,~~\cS_{11}=\cZ_1~~,~~\cS_{12}\sqcup \cS_{21}\subset \cZ_2~~.
\ee
A precise description of the relation between the chirality
stratification and the rank stratifications of $\cD$ and $\cD_0$ can
be found in Section \ref{sec:degen}.

\subsection{Relation to the stabilizer group} 
\label{sec:spinorproof} 

\paragraph{Proposition.} 
Let $p$ be any point of $M$. Then the following statements hold:
\begin{enumerate}
\itemsep 0.0em
\item When $p\in \cS_{02}\sqcup \cS_{20}$ we have $H_p\simeq \SU(4)$
\item When $p\in \cS_{11}$, we have $H_p\simeq \G_2$
\item When $p\in \cS_{12}\sqcup \cS_{21}$, we have $H_p\simeq \SU(3)$
\item When $p\in \cG$, we have either $H_p\simeq \SU(2)$ or $H_p\simeq
  \SU(3)$, according to whether $\dim \cD_0(p)=4$ or $\dim \cD_0(p)=6$. 
\end{enumerate}

\noindent{\bf Proof.}
\begin{enumerate}
\itemsep 0.0em
\item In this case, $\xi_1$ and $\xi_2$ are chiral and of the same
  chirality at $p$, so their stabilizer inside $\Spin(8)$ equals
  $\SU(4)$.
\item After a rotation as in the Lemma given in the previous subsection, we have two non-vanishing
  spinors $\xi_1$ and $\xi_2$ of opposite chirality at $p$, whose stabilizer inside
  $\Spin(8)$ is isomorphic with $\G_2$.
\item Consider the case $p\in \cS_{12}$. The Lemma shows that (up to a
  rotation) we can assume $\xi_1(p)=\xi_1^+(p)$, $\xi_2^-(p)\neq 0$
  and that $\xi_1^+(p),\xi_2^+(p)$ are linearly independent. Since
  $S_+$ and $S_-$ are $\cB$-orthogonal sub-bundles of $S$, orthogonality of
  $\xi_1$ and $\xi_2$ implies that $\xi_1^+(p)$ and $\xi_2^+(p)$ are
  $\cB_p$-orthogonal. The stabilizer $H'_p$ of the pair
  $(\xi_2^+(p),\xi_2^-(p))$ inside $\Spin(8)$ is isomorphic with
  $\G_2$ and $S_p^\pm$ have the $\cB$-orthogonal decompositions: \be
  S_p^\pm=\Sigma_1^\pm(p)\oplus \Sigma_7^\pm(p)~~, \ee where $\Sigma_1^\pm(p)$
  are one-dimensional subspaces carrying trivial irreps while
  $\Sigma_7^\pm(p)$ are subspaces carrying the seven-dimensional irreps
  of $H'_p\simeq \G_2$. We have $\xi_2^\pm(p) \in \Sigma_1^\pm(p)$.
  Since $\xi_1^+(p)$ is $\cB_p$-orthogonal to $\xi_2^+(p)$, we have
  $\xi_1^+(p)\in \Sigma_7^+(p)$. $H_p$ is isomorphic with the stabilizer
  of the non-zero element $\xi_1^+(p)\in \Sigma_7^+(p)$ inside $H'_p$,
  which is known\footnote{The action of $\G_2$ on $S^6$ induced by
    this irrep. is transitive with stabilizer isomorphic with
    $\SU(3)$.} to be isomorphic with $\SU(3)$. This shows that
  $H_p\simeq \SU(3)$.  The case $p\in \cS_{21}$ proceeds similarly.
\item When $p\in \cG$, we have $H_p\simeq \Stab_{H''_p}(\xi_1^-(p))$,
  where $H''_p\eqdef
  \Stab_{\Spin(8)}(\xi_1^+(p),\xi_2^+(p),\xi_2^-(p))$. By point
  3. above, we have $H''_p\simeq \SU(3)$. The spaces $S_p^\pm$
  decompose as:
\be
S_p^\pm =\Sigma_1^\pm(p)\oplus \Sigma_1'^\pm(p)\oplus \Xi^\pm(p)~~,
\ee
where $\Sigma_1^\pm(p)$ and $\Sigma_1'^\pm(p)$ are trivial irreps
while $\Xi^\pm(p)\simeq \C^3$ are fundamental irreps of $\SU(3)$ such
that $\Sigma_7^\pm(p)=\Sigma_1'^\pm(p)\oplus \Xi^\pm(p)$ while
$\xi_1^+(p)\in \Sigma_1'^+(p)$ and $\xi_2^\pm(p)\in \Sigma_1^\pm(p)$.
Notice that $\xi_1^-(p)$ and $\xi_2^-(p)$ need not be
$\cB_p$-orthogonal. We have $H_p\simeq \Stab_{H_p''}(\zeta(p))$, where
$\zeta(p)$ denotes the $\cB_p$-orthogonal projection of $\xi_1^-(p)$
onto the subspace $\Xi^-(p)$. We distinguish the cases:
\begin{itemize}
\itemsep 0.0em
\item $\zeta(p)=0$. Then $H_p=H_p''\simeq \SU(3)$.
\item $\zeta(p)\neq 0$. Then $H_p\simeq \SU(2)$, since it is
  known\footnote{For any $n\geq 2$, the action of $\SU(n)$ on
    $S^{2n-1}$ induced from the fundamental representation of $\SU(n)$
    on $\C^n=\R^{2n}$ is transitive and has stabilizer isomorphic with
    $\SU(n-1)$.}  that $\SU(3)$ acts transitively on the sphere $S^5$,
  with stabilizer $\SU(2)$.
\end{itemize}
The results of Appendix \ref{app:def} show that the first case arises
iff $\rk\cD_0(p)=6$ while the second case arises iff $\rk
\cD_0(p)=4$. \end{enumerate} $\blacksquare$

\paragraph{Remark.} 
Appendix \ref{app:def} gives an explicit construction of a
one-parameter deformation of the pair $(\xi_1,\xi_2)$ which breaks the
stabilizer group from $\SU(3)$ to $\SU(2)$.

\paragraph{Corollary.} 
The stabilizer stratification coincides with the rank stratification
of $\cD_0$.

\subsection{Characterizing the chirality stratification}
\label{sec:Theorem1}

\paragraph{Theorem 1.} 
The $K$-special locus is given by:
\ben
\cS= b^{-1}(\partial \cR)=\{p\in M|b(p)\in \partial \cR\}~~.
\een
Furthermore, we have:
\begin{itemize}
\itemsep 0.0em 
\item $\cS_{11}=b^{-1}(\partial_1\cR)=b^{-1}(\partial D)$
 \item $\cS_{12}=b^{-1}(\partial_2^+\cR)$ and $\cS_{21}=b^{-1}(\partial_2^-\cR)$
\item $\cS_{02}=b^{-1}(\partial_0^+\cR)$ and $\cS_{20}=b^{-1}(\partial_0^-\cR)$.
\end{itemize}
Moreover, we have $\cG=b^{-1}(\Int \cR)$ and hence the chirality
stratification of $M$ coincides with the $b$-preimage of the connected
refinement of the canonical Whitney stratification of $\cR$.

\

\noindent{\bf Proof.} 
Relation \eqref{detGpm} implies: 
\ben
\label{detGpmv}
\det \fG_\pm(p)=0\Longleftrightarrow \rho(p)=1\pm b_+(p)~.
\een
Since $\det \fG_\pm(p) \geq 0$, we have $\rho(p)\leq 1\pm
b_+(p)$ and hence $\rho(p)\leq 1-|b_+(p)|$. Thus $\det
\fG_+(p)=0$ can be realized only for $b_+(p)\leq 0$ and $\det
\fG_-(p)=0$ can be realized only for $b_+(p)\geq 0$ and in
both cases we have $\rho(p)=1-|b_+(p)|$ i.e. $b(p)\in \partial
\cR$. The case $\det \fG_+(p)=\det \fG_-(p)=0$
occurs for $b_+(p)=0$ and $\rho(p)=1$, i.e. on the median circle
$\partial D$.  We also have:
\be
\fG_\pm(p)=0~\Longleftrightarrow ~b_3(p)=0~\&~b_1(p)=b_2(p)=\mp 1
~\Longleftrightarrow ~ \rho(p)=0~\&~~b_+(p)=\mp 1~~.
\ee
Hence $\fG_+(p)=0$ or $\fG_-(p)=0$ corresponds to $b(p)\in \partial
I$, namely $\fG_+(p)=0$ corresponds to the left tip
$(b_+(p),\rho(p))=(-1,0)$ of $\cR$ while $\fG_-(p)=0$ corresponds to
the right tip $(b_+(p),\rho(p))=(+1,0)$ of $\cR$. The remaining
statements follow since $b(M)\subset \cR$.  $\blacksquare$

\

\noindent The situation is summarized in Table \ref{table:S} and Figure \ref{fig:Rlr}.

\begin{table}[H]
\centering {\footnotesize
\begin{tabular}{|c|c|c|c|c|c|c|c|c|c|c|}
\hline stratum & $\cR$-description & $r_-(p)$ & $r_+(p)$ & $\rk \cD$ & $\rk \cD_0$ & $b_+$ & $\rho$ & $H_p$ & $\sigma_+(p)$ & $\sigma_-(p)$\\ \hline
\rowcolor{magenta}$\cS_{02}$ & $b^{-1}(\partial_0^+\cR)$ & $0$ & $2$ & $8$ & $8$  & $+1$ & $0$ &  $\SU(4)$ & $2$ & $0$\\ \hline
\rowcolor{magenta}$\cS_{20}$ & $b^{-1}(\partial_0^-\cR)$ & $2$ & $0$ & $8$ & $8$ & $- 1$  & $0$ & $\SU(4)$ & $0$ & $2$ \\ \hline
\rowcolor{yellow}$\cS_{11}$ & $b^{-1}(\partial_1\cR)$ & $1$ & $1$ & $7$ & $7$ & $0$ & $1$ &  $\G_2$ & $1$ & $1$ \\ \hline
\rowcolor{cyan}$\cS_{12}$ & $b^{-1}(\partial^+_2\cR)$ & $1$ & $2$ & $6$ & $6$ & $1-\rho$ & $(0,1)$  & $\SU(3)$ & $1$ & $0$\\ \hline 
\rowcolor{cyan}$\cS_{21}$ & $b^{-1}(\partial_2^-\cR)$& $2$ & $1$ & $6$ & $6$ & $-(1-\rho)$ & $(0,1)$ & $\SU(3)$ & $0$ & $1$ \\ \hline
\rowcolor{gray}$\cG$ & $b^{-1}(\Int \cR)$ & $2$ & $2$ & $5,6,7$  & $4,6$ & $(-1,1)$ & $<1-|b_+|$ &  $\SU(2)$ or $\SU(3)$ & $0$ & $0$ \\ \hline
\end{tabular}
}
\caption{Chirality stratification for $s=2$. The quantities $\sigma_\pm$ are defined through $\sigma_\pm(p)=\dim K^\pm(p)=2-r_\mp(p)$ (see 
\eqref{rhopm}).}
\label{table:S}
\end{table}

\paragraph{Remarks.}
\begin{enumerate}
\itemsep 0.0em
\item Theorem 1 implies a similar characterization of the
  stratification $\cS$ as the $b'$-preimage of the obvious
  stratification with connected strata of the set $\Delta\setminus
  (-1,1)\times \{0\}$; we leave the details of this to the reader.
\item For every $p\in M$, the dimensions $\sigma_\pm(p)\eqdef \dim
  K^\pm(p)=2-r_\mp(p)$ of the chiral slices of $K_p$ count the number
  of linearly independent spinors inside the space $K_p$ which have
  chirality $\pm 1$. In the case of compactifications down to
  $\AdS_3$, $\sigma_+(p)$ can be interpreted \cite{Palti} as the
  number of supersymmetries of the background which are preserved by a
  space-time filling M2-brane placed at $p$, while $\sigma_-(p)$
  counts the number of supersymmetries preserved by a space-time
  filling M2-antibrane placed at $p$; these numbers are indicated in
  the last column of the table.
\end{enumerate}

\section{Algebraic constraints}
\label{sec:alg}

The Fierz identities for $\xi_1^\pm, \xi_2^\pm$ imply that the
following relations hold (see Appendix \ref{app:alg}):
\ben
\label{VWsys}
\begin{split}
&||V_-||^2+b_-^2=||V_3||^2+b_3^2~~,~~||V_+||^2+b_+^2=1-(||V_3||^2+b_3^2)\\
& \langle V_+,V_-\rangle+b_+b_-=\langle V_+,V_3\rangle+b_+b_3=\langle V_-,V_3\rangle+b_-b_3=0~~\\
&||W||^2+||V_3||^2=1+b_-^2-b_+^2~\\
&\langle W, V_+\rangle =0~~,~~ \langle W, V_-\rangle =b_3 ~~,~~\langle W, V_3\rangle =-b_-~~.
\end{split}
\een
In particular, the first two rows of \eqref{VWsys} form the following
system for $V_r,b_r$:
\ben
\label{Vsys}
\begin{split}
&||V_-||^2+b_-^2=||V_3||^2+b_3^2\\
&||V_+||^2+b_+^2=1-(||V_3||^2+b_3^2)\\
& \langle V_+,V_-\rangle+b_+b_-=\langle V_+,V_3\rangle+b_+b_3=\langle V_-,V_3\rangle+b_-b_3=0~~.
\end{split}
\een
Relations \eqref{Vsys} constrain the norms $||V_r||^2$ and the angles
$\theta_{rs}=\theta_{sr}$ between $V_r$ and $V_s$ (a total of six
quantities) in terms of the three quantities $b_r$. Fixing the latter
generally fails to completely determine the former.

\paragraph{Remark.} 
For a general choice of $V_r$, one cannot find $b_r$ such that
\eqref{Vsys} is satisfied.  The conditions on $V_r$ under which it is
possible to solve for $b_r$ are given in Appendix \ref{app:proofs}.

\

\subsection{Reduction to a semipositivity problem}

\noindent Let us define:
\ben
\label{beta_def}
\beta\eqdef \sqrt{b_3^2+||V_3||^2}=\sqrt{b_-^2+||V_-||^2}
\een
as well as:
\ben
\label{rho_def}
\rho\eqdef \sqrt{b_-^2+b_3^2}
\een
and consider the smooth map $B\in \cC^\infty(M,\R^4)$ defined through:
\ben
\label{Bdef}
B(p)\eqdef (b(p),\beta(p))~,~~p\in M~~.
\een
The second line in \eqref{Vsys} gives: 
\ben
\label{normVplus}
||V_+||^2=1-b_+^2-\beta^2~~,
\een
which shows that $\beta$ contains the same information as the norm of
$V_+$, provided that $b_+$ is known.

When $\beta$ is fixed, the constraints \eqref{Vsys} amount to the
condition that the Gram matrix of $V_+,V_-,V_3$ be given by:
\ben
\label{G}
G(b,\beta)=\left[ \begin{array}{ccc}
1-\beta^2-b_+^2 &~-b_+b_-~& ~-b_+b_3~\\
-b_-b_+~ &~\beta^2-b_-^2 &~ -b_-b_3~\\ 
-b_3b_+~ &~ -b_3b_-~ & ~~\beta^2-b_3^2
\end{array}\right]~~.
\een
The system given by \eqref{beta_def} and the last two rows of
\eqref{Vsys} has solutions $V_r$ iff the symmetric matrix $G(b,\beta)$
is positive semidefinite; in this case, $V_r$ are determined by
$\beta$ and $b_r$ up to a common action of the group
$\Gamma(M,\O(TM,g))$.  Furthermore, $V_+,V_-$ and $V_3$ are linearly
independent at $p\in M$ iff $G(p)\eqdef G(b(p),\beta(p))$ is positive
definite.  Similarly, the system \eqref{VWsys} amounts to the
condition that the Gram matrix of $V_+,V_-,V_3,W$ be given by:
\ben
\label{Gext}
{\hat G}(b,\beta)=\left[ \begin{array}{cccc}
1-\beta^2-b_+^2 &~-b_+b_-~& ~-b_+b_3~ & 0\\
-b_-b_+~ &~\beta^2-b_-^2 &~ -b_-b_3~ & b_3 \\ 
-b_3b_+~ &~ -b_3b_-~ & ~~\beta^2-b_3^2 & -b_-\\
0 & b_3 & -b_- & 1-\beta^2-b_+^2+\rho^2
\end{array}\right]~~.
\een
Notice that $V_+\perp W$ and $||W||^2=||V_+||^2+\rho^2$. 

\paragraph{Remark.} 
Relation \eqref{normVplus} and the observations of Subsection
\ref{sec:rot} imply that $\beta$ is invariant under any proper or
improper rotation of the orthonormal basis of $\cK$. Hence $b_+$,
$\rho$ and $\beta$ depend only on $\cK$. Relations \eqref{G} and
\eqref{Gext} show that all scalar invariants under the transformations
\eqref{basis_rot} which can be constructed from $V_+,V_-, V_3$ and $W$
can be expressed as functions of $b_+,\rho$ and $\beta$.

\

\noindent The semipositivity conditions for $G(b,\beta)$ can be
analyzed using Sylvester's criterion, leading to a nonlinear
programming problem whose solution is given in Appendix
\ref{app:proofs}. To state the results concisely, we introduce a
compact four-dimensional semi-algebraic body $\fP$ which can be viewed
as a singular segment fibration over $\cR$.

\subsection{The four-dimensional body $\fP$}
\begin{figure}[H]
\centering
\begin{subfigure}{0.5\textwidth}
\centering \includegraphics[width=0.5\linewidth]{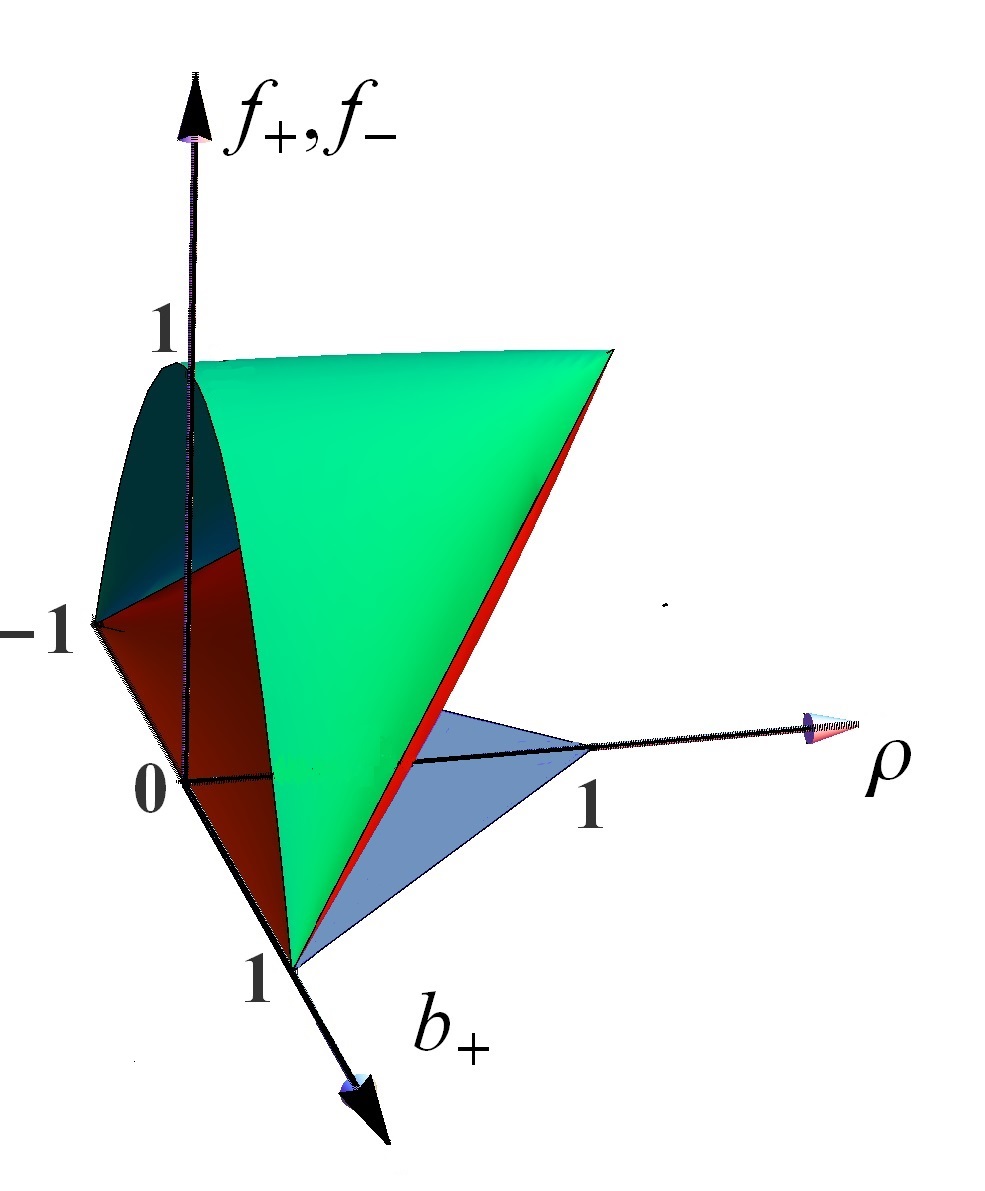}
\caption{Graphs of the functions $f_+(b_+,\rho)$ (green) \\ and
  $f_-(b_+,\rho)$ (red) for $(b_+,\rho)\in \Delta$ (blue). }
\end{subfigure}~~
\begin{subfigure}{0.49\textwidth}
\centering \includegraphics[width=0.7\linewidth]{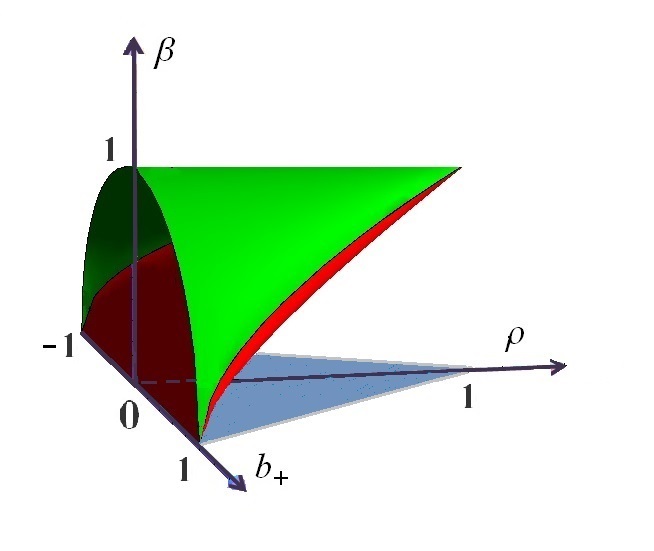}
\caption{Graphs of the functions $\sqrt{f_+(b_+,\rho)}$ (green) and
  $\sqrt{f_-(b_+,\rho)}$ (red) for $(b_+,\rho)\in \Delta$ (blue).}
\end{subfigure}
\caption{}
\label{fig:fDeltaM}
\end{figure}

\begin{figure}[H]
\begin{center}
\hskip 0.8in \includegraphics[scale=0.5]{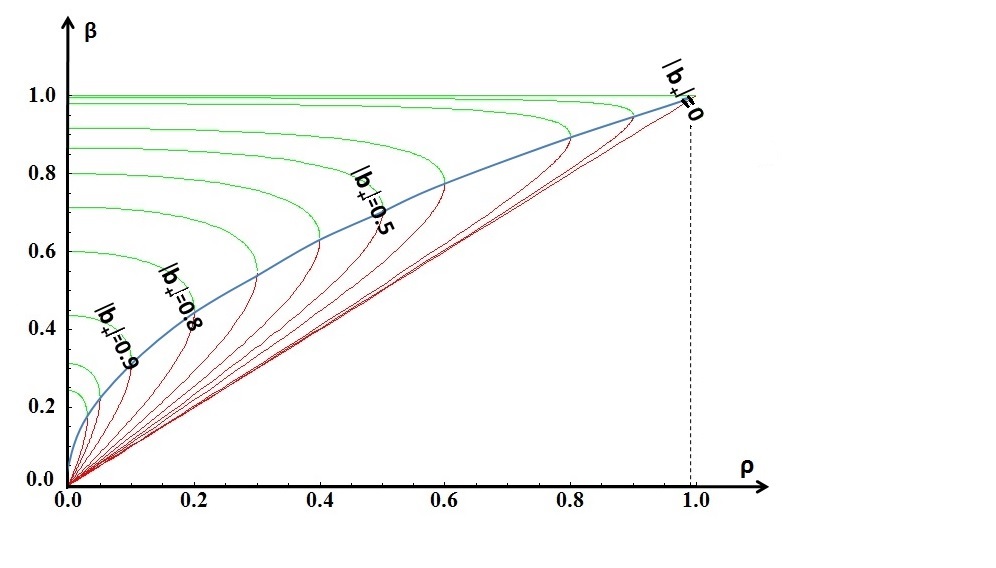}
\caption{Graphs of $\beta=\sqrt{f_+(b_+,\rho)}$ (green) and
  $\beta=\sqrt{f_-(b_+,\rho)}$ (red) for various fixed values of
  $|b_+|\in[0,1]$. Notice that the two graphs match each other
  smoothly at $|b_+|=1-\rho<1$ (corresponding to $\partial \cR$),
  where both $f_+$ and $f_-$ equal $\sqrt{\rho}$. The matching is
  non-smooth only when $\rho=\beta=1,~b_+=0$, which corresponds to the
  circle $\partial \fD$ defined below. }
\label{fig:betro}
\end{center}
\end{figure}

Recall that the image of $b$ is contained in $\cR$. The determinant of
the Gram matrix \eqref{G} takes the form:
\ben
\label{detG}
\det G=-\beta^2P(b,\beta)~~,
\een
where:
\ben
\label{P}
P(b,\beta)\eqdef \beta^4-\beta^2(1+b_3^2+b_-^2-b_+^2)+b_3^2+b_-^2=\beta^4-\beta^2(1+\rho^2-b_+^2)+\rho^2~~.
\een
Thus $\det G(b,\beta)$ vanishes for $\beta=0$ or
$\beta=\sqrt{f_\pm(b)}$, where the functions
$f_\pm:\cR\rightarrow \R$ (which give the roots of the second order
polynomial $x^2-(1+\rho^2-b_+^2)x+\rho^2$) are defined through:
\ben
\label{fpm}
f_\pm(b_+,b_-,b_3)=f_\pm(b_+,\rho)\eqdef\frac{1}{2}\left(1-b_+^2+\rho^2\pm \sqrt{h(b_+,\rho)}\right)~~.
\een 
The discriminant:
\ben
\label{hdef}
h(b)=h(b_+, \rho)\eqdef (1+b_++\rho)(1-b_++\rho)(1+b_+-\rho)(1-b_+-\rho)~~
\een
is non-negative on $\Delta$ and vanishes only for $\rho=1-|b_+|$,
i.e. on the left and right sides of $\Delta$. The functions $f_\pm$
satisfy:
\be
0\leq f_-(b)\leq f_+(b)\leq 1~~,~~\forall b\in \cR~~,
\ee
where: 
\begin{itemize}
\itemsep 0.0em
\item the equality $f_-(b)=f_+(b)$ is attained iff $b\in \partial
  \cR$, where we have $f_+|_{\partial \cR}=f_-|_{\partial \cR}=\rho$;
\item the equality $f_-(b)=0$ is attained iff $b\in I$;
\item the equality $f_+(b)=1$ is attained iff $b\in D$.
\end{itemize}
Notice that $f_\pm$ depend only on $b_+$ and $\rho$ and hence they can
be viewed as functions defined on $\Delta$ (see Figures
\ref{fig:fDeltaM} and \ref{fig:betro}). In fact, they are symmetric
under $b_+\rightarrow -b_+$, so they depend only on $|b_+|$ and
$\rho$.  Various special values of $f_\pm$ are summarized in Table
\ref{table:fpm}.

\begin{table}[H]
\centering
{\footnotesize
\begin{tabular}{|c|c|c|c|c|c|}
\hline &$\begin{array}{c}\!\!b\!\in\! \Int I
  \\ (b_+\!\in\!(-1,1),~\rho\!=\!0)\!\!  \end{array}$ &
$\begin{array}{c}\!\!b\!\in\!\partial I\!\! \\\!\!
  (b_+\!\!=\!\pm 1,~\rho\!\!=\!\!0)\!\!  \end{array}$ &
$\begin{array}{c}b\!\in\!\Int D\setminus\{(0,0,0)\}
  \!\!\\ (b_+\!\!=\!0,~\rho\!\in\!(0,1)) \end{array}$&
$\begin{array}{c}b\!\in\!\partial
  D\\(b_+\!=\!0~,~\rho\!=\!1) \end{array}$ &
$\begin{array}{c}b\!\in \!\partial \cR
  \!\!\\ (\rho\!=\!1-|b_+|) \end{array}$\\
\hline
$~f_+(b_+,\rho)~$ & $1-b_+^2$  & $0$& $1$ & $1$&$\rho$\\
\hline
$~f_-(b_+,\rho)~$ & $0$ & $0$ & $\rho^2$ & $1$&$\rho$\\
\hline
$ \beta $ & $[0,\sqrt{1-b_+^2}]$ & $0$ & $[\rho,1]$ & $1$ & $\sqrt{\rho}$\\
\hline
$\rk G$ & $\{1,2, 3\}$ & $0$ & $\{2,3\}$ & $1$ & $\{0,1,2\}$\\
\hline
\end{tabular}
}
\caption{Special values for $f_+$ and $f_-$. The values allowed for
  $\rk G$ on each region follow from Theorem 2 of Subsection
  \ref{sec:Theorems2and3}.}
\label{table:fpm}
\end{table}

\noindent For every $b\in \cR$, consider the closed interval:
\ben
\label{Jdef}
J(b)=J(b_+,\rho)\eqdef [\sqrt{f_-(b)},\sqrt{f_+(b)}]\subset [\sqrt{b_-^2+b_3^2}, \sqrt{1-b_+^2}]~~.  
\een
This interval degenerates to a single point for $b\in \partial \cR$,
namely $J|_{\partial \cR}=\{\sqrt{\rho}\}$. Finally, consider the
following four-dimensional compact body:
\ben
\label{Pdef}
\boxed{\fP\eqdef \{(b,\beta)\in \R^4|b\in \cR ~\&~ \beta\in J(b)\}}~~,
\een
which is fibered over $\cR$ via the projection
$(b,\beta)\stackrel{\pi}{\rightarrow} b$. The fiber over $b\in \cR$ is
the segment $J(b)$, which, as mentioned above, degenerates to a point
over $\partial \cR$.

\paragraph{The frontier of $\fP$.}

\begin{figure}[H]
\centering
\includegraphics[width=0.35\linewidth]{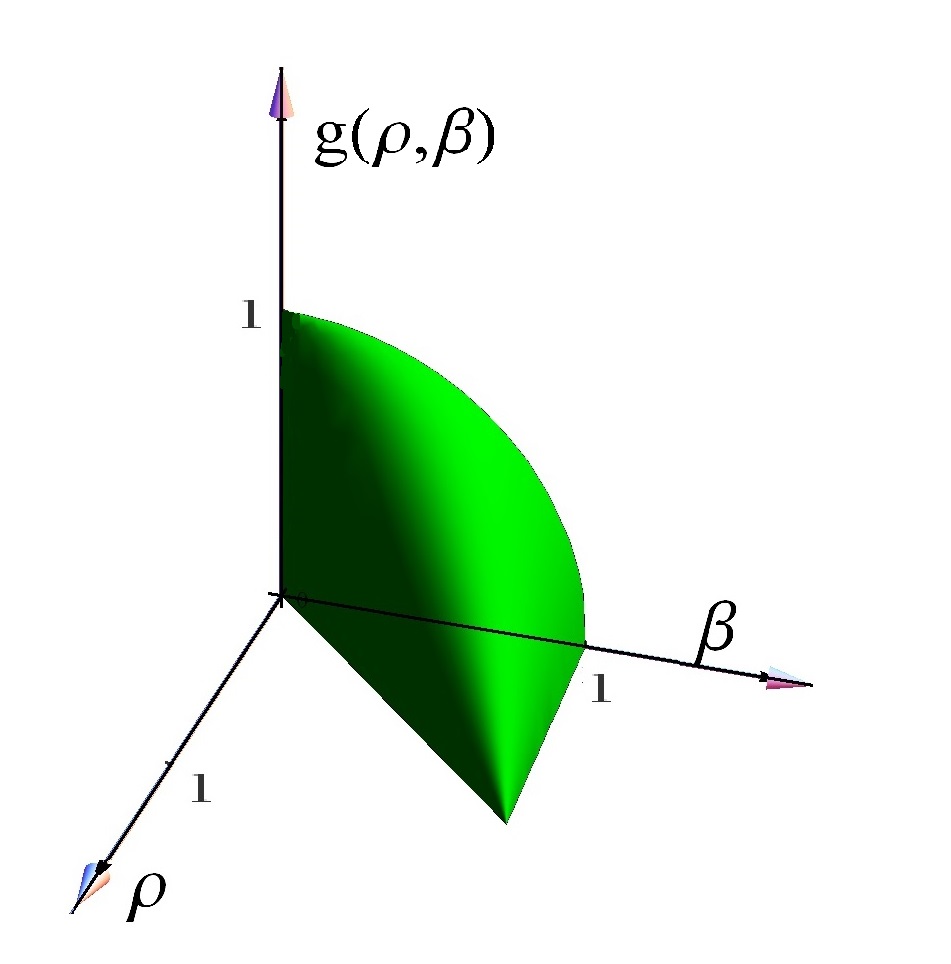}
\vskip 0.0in
\caption{Graph of the function $g(\rho,\beta)$ for $(\rho,\beta)$
  belonging to the triangular region defined by the inequalities
  $0<\rho\leq \beta\leq 1$. Notice that the directional limits of
  $g(\rho,\beta)$ at the point $\rho=\beta=0$ (taken from within this
  triangular region) take any value within the interval $[0,1]$.}
\label{fig:g}
\end{figure}

\begin{figure}[H]
 \centering
\begin{subfigure}{0.5\textwidth}
\centering
\includegraphics[width=0.9\linewidth]{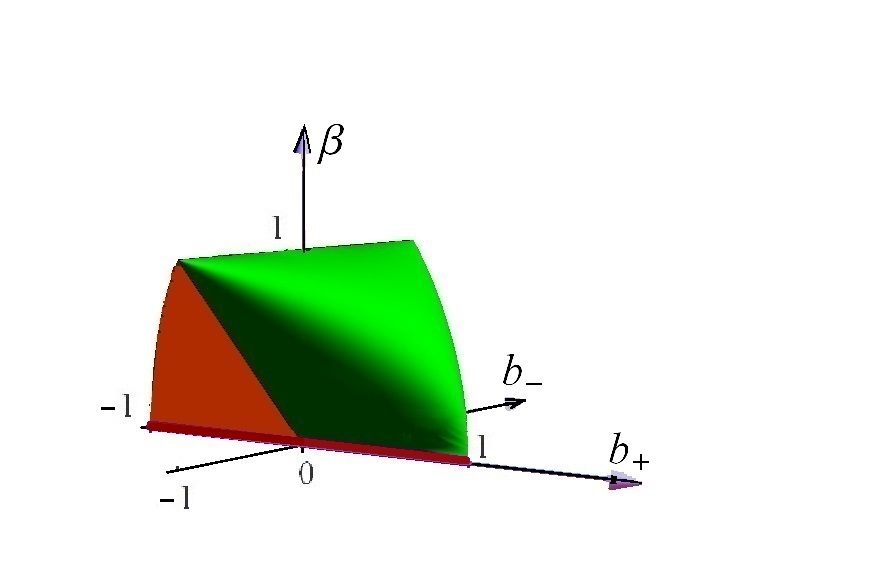}
\vskip 0.2in
\end{subfigure}~~~~~
\begin{subfigure}{0.5\textwidth}
\centering
\vskip 0.1in
\includegraphics[width=0.6\linewidth]{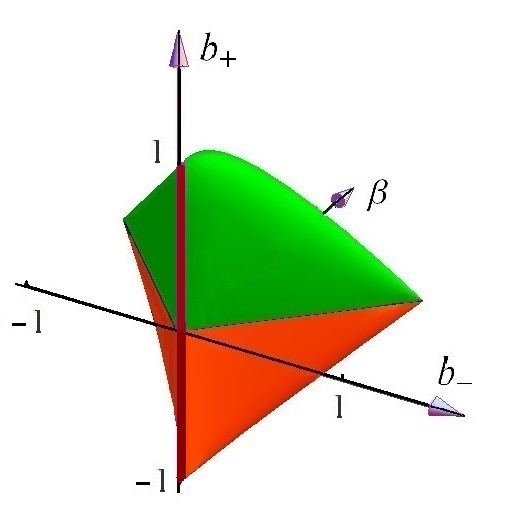}
\end{subfigure}
\vskip -0.1in
\caption{Section of $\partial \fP$ with the hyperplane $b_3=0$, where
  the corresponding sections of $\fC^-$, $\fC^+$ and $\fI$ are
  represented in orange, green and brown. }
\label{fig:Jpm}
\end{figure}

Let:
\ben
C\eqdef \{(b_-,b_3,\beta)\in \R^3|~0\leq \sqrt{b_-^2+b_3^2}\leq \beta\leq 1\}
\een
be the full compact cone in $\R^3$ with apex at the origin and base
given by the disk $D^2\times \{1\}$ and let:
\be
F\eqdef \partial C=\{(b_-,b_3,\sqrt{b_-^2+b_3^2})|(b_-,b_3)\in \Int(D^2)\}\sqcup \{(b_-,b_3,1)|(b_-,b_3)\in D^2\}~~
\ee
denote its frontier. Let $\dot{C}\eqdef C\setminus \{(0,0,0)\}$ and
$\dot{F}\eqdef F\setminus \{(0,0,0)\}$.  Notice that $C$ is homeomorphic
with the compact 3-dimensional ball, $F$ is homeomorphic with $S^2$
and $\dot{F}$ is homeomorphic with $\R^2$ (and hence with the interior
of the unit disk $D^2$). Consider the function $g:\dot{C}\rightarrow
\R$ given by (see Figure \ref{fig:g}):
\ben
\label{gdef}
g(b_-,b_3,\beta)=g(\rho,\beta)\eqdef \frac{1}{\beta}\sqrt{(1-\beta^2)(\beta^2-\rho^2)}~~.
\een
The quantity under the square root is non-negative for
$(b_-,b_3,\beta)\in C$ and we have $0\leq g(\rho,\beta)\leq
\frac{\sqrt{\beta^2-\rho^2}}{\beta}\leq 1$ for $(b_-,b_3,\beta)\in
\dot{C}$. Notice that $g$ vanishes on $\dot{F}$ and is
strictly positive in the interior of $C$.
\noindent Consider the following three-dimensional subsets of $\R^4$,
each of which is homeomorphic with $\dot{C}$:
\ben
\boxed{\fC^\pm \eqdef \{(\pm g(b_-,b_3,\beta),
 b_-,b_3,\beta)|~(b_-,b_3,\beta)\in \dot{C}\}}~~
\een
and the following compact interval sitting inside $\R^4$:
\ben
\boxed{\fI\eqdef [-1,1]\times \{(0,0,0)\}=I\times \{0\}}~~.
\een
The intersection of the sets $\fC^\pm$ is given by: 
\ben
\label{fF}
\boxed{\fF\eqdef \fC^+\cap \fC^-=\{0\}\times \dot{F}}
\een
and $\fC^\pm$ are disjoint from $\fI$ (since $\beta\neq 0$ on $\fC^\pm$
while $\beta=0$ on $\fI$). Notice that $\Int \fC^+$ and $\Int \fC^-$
are homeomorphic with $\Int \dot{C}=\Int C$ and hence with the
interior of the unit 3-ball while $\fF$ is homeomorphic with the
interior of the two-dimensional disk. Let:
\be
\fI^+\eqdef [0,1]\times \{0_{\R^3}\}=I^+\times \{0\}~~,~~\fI^-\eqdef [-1,0]\times \{0_{\R^3}\}=I^-\times \{0\}~~
\ee
be the compact right and left halves of $\fI$, which satisfy
$\fI^+\cap \fI^-=\{0_{\R^4}\}$. Figure \ref{fig:Jpm} shows the
sections of $\partial \fP$ with the hyperplane $b_3=0$.

\paragraph{Proposition.} 
The frontier of $\fP$ is given by:
\ben
\label{partialP}
\partial \fP=\fC^+\cup\fC^-\cup\fI=\Int\fC^+\sqcup\Int \fC^-\sqcup \fF\sqcup \fI~~,
\een
where the components can be identified as: 
\beqan
\label{partialPcomps}
&&\Int \fC^\pm=\{(b,\beta)\in \partial \fP|~\beta>0~\&~\pm b_+>0\}\nn\\
&&\fF=\{(b,\beta)\in \partial \fP|~\beta>0~\&~b_+=0\}\\
&&\fI=\{(b,\beta)\in \partial \fP|~\beta=0\}~~.\nn
\eeqan
Moreover, $\fI$ is closed (thus $\fr\fI=\emptyset$),
while\footnote{Since $\partial \fP$ is a closed subset of $\R^4$, the
  small frontier of a subset $A\subset \fP$ taken with respect to the
  topology induced on $\partial \fP$ from $\R^4$ coincides with the
  small frontier $\fr A$ of $A$ in $\R^4$.}:
\ben
\label{fr1}
\fr(\Int \fC^\pm)=\fF\sqcup \fI^\pm~~,~~\fr \fF=\{0_{\R^4}\}~~.
\een

\paragraph{Remark.} 
Topologically, $\fr(\Int \fC^\pm)=\fr (\fC^\pm)$ is obtained from the
compact disk upon picking two opposite points on the boundary circle
and identifying the resulting halves of the boundary to a segment
corresponding to $\fI^\pm$; the result is of course homeomorphic to a
sphere.

\

\noindent{\bf Proof.} 
We have: 
\be
P(b,\beta)=(1-\beta^2)(\rho^2-\beta^2)+b_+^2\beta^2~~.
\ee
The frontier of $\fP$ is the semi-algebraic set obtained by
intersecting $\cR$ with the hypersurface $P(b,\beta)=0$. This equation
can be written as:
\ben
\label{feq}
b_+^2\beta^2=(1-\beta^2)(\beta^2-\rho^2)
\een 
and requires that the right hand side be non-negative, which for $b\in
\cR$ is equivalent with the condition $\beta\in [\rho,1]$
i.e. $(b_-,b_3,\beta)\in C$. To study the solutions of \eqref{feq},
assume that this condition is satisfied and consider the cases:
\begin{itemize}
\item $\beta=0$. Then \eqref{feq} requires $\rho=0$ while $b_+$ is
  undetermined within the interval $[-1,1]$, which means that
  $(b,\beta)$ belongs to the interval $\fI$.
\item $\beta> 0$. Then \eqref{feq} requires $(b_-,b_3,\beta)\in
  \dot{C}$ as well as~ $b_+=\pm g(b_-,b_3,\beta)$~,~
where the function $g:\dot{C}\rightarrow \R$ was defined in
\eqref{gdef}. 
\end{itemize}
The above shows that $\partial \fP$ has the decomposition
\eqref{partialP} and that \eqref{partialPcomps} holds. The remaining
statements follow from \eqref{partialPcomps}. $\blacksquare$

\

\begin{figure}[H]
\begin{center}
\includegraphics[scale=0.44]{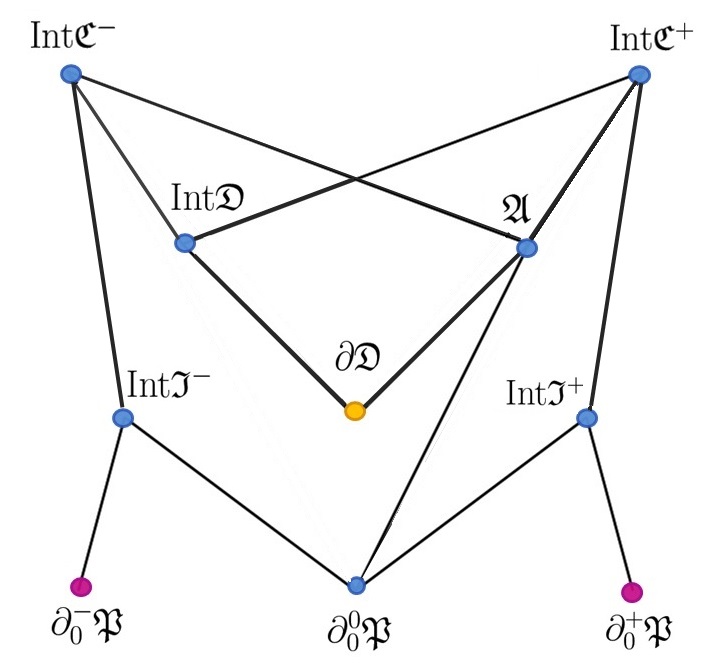}
\caption{The Hasse diagram of the incidence poset (see Appendix
  \ref{app:stratif}) of the connected refinement of the Whitney
  stratification of $\partial \fP$. The $B$-preimages of the connected
  components depicted as points colored in magenta, yellow and cyan
  are strata of $\SU(4)$, $G_2$ and $\SU(3)$ structure in $M$ (see
  Table \ref{table:HHprime} in Subsection \ref{sec:Gstratif}). The
  diagram depicts the covering relation of the incidence poset, namely
  an element of that poset covers another iff it sits above it in the
  diagram and there is an edge connecting the two elements. The small
  frontier of each connected Whitney stratum is the disjoint union of
  the strata covered by it in the diagram.}
\label{fig:PHasse}
\end{center}
\end{figure}

\noindent The body $\fP$ is a semi-algebraic set, hence it admits a
canonical Whitney stratification by smooth semi-algebraic subsets. To
describe this stratification, notice that the set defined in
\eqref{fF} decomposes as:
\ben
\label{fFdec}
\boxed{\fF=\Int \fD\sqcup \partial \fD\sqcup \fA}~~,
\een
where: 
\ben
\boxed{
\begin{split}
&\fD\eqdef \{(0,b_-,b_3,1)|(b_-,b_3)\in D^2\}=D\times \{1\}\\
&\fA\eqdef \{(0,b_-,b_3, \sqrt{b_-^2+b_3^2})|(b_-,b_3)\in \Int D^2\setminus \{0_{\R^2}\} \}~
\end{split}
}
\een
are homeomorphic with the compact disk and with an open annulus,
respectively. We have:
\be
\partial \fD =\{(0,b_-,b_3,1)|(b_-,b_3)\in \partial D^2\}=\partial D\times \{1\}~~.
\ee
The frontier $\partial \fP$ has the following decomposition into
borderless manifolds of dimensions $k=0,1,2,3$:
\ben
\boxed{\partial\fP=\partial_3\fP\sqcup \partial_2\fP\sqcup \partial_1\fP\sqcup\partial_0\fP}~~,
\een
where the $k$-dimensional pieces are the following unions of connected
components:
\ben
\label{Pfrontier}
\boxed{
\begin{split}
&\partial_3\fP=\Int \fC^+\sqcup \Int \fC^-\\
&\partial_2\fP=\Int\fD \sqcup \fA\\
&\partial_1\fP=\Int \fI^+\sqcup \Int \fI^- \sqcup \partial \fD\\
&\partial_0\fP=\partial_0^+\fP \sqcup \partial_0^0\fP\sqcup \partial_0^-\fP~~
\end{split}
}~~,
\een
with:
\ben
\partial_0^\pm \fP\eqdef \partial_0^\pm \cR\times \{0\}=\{(\pm 1, 0,0,0)\}~~,~~\partial_0^0\fP\eqdef \{0_{\R^4}\}~~.
\een
The ten connected components listed above give the connected refinement of
the canonical Whitney stratification of $\partial \fP$, whose
incidence poset is depicted in Figure \ref{fig:PHasse}. Using
relations \eqref{fr1} and \eqref{fFdec}, we find:
\beqan
\label{fr2}
&&\fr(\Int \fC^\pm)=\Int\fD\sqcup \partial \fD\sqcup \fA\sqcup \Int \fI^\pm\sqcup \partial_0^0\fP\sqcup \partial_0^\pm \fP\nn\\
&&\fr (\Int\fD)=\partial \fD~~,~~\fr\fA=\partial\fD\sqcup \partial_0^0\fP~~\\
&&\fr(\Int \fI^\pm)=\partial_0^0\fP\sqcup \partial_0^\pm \fP~~,~~\fr(\partial\fD)=\emptyset\nn~~,
\eeqan
which imply:
\beqan
\label{PWfrontiers}
&&\fr(\partial_3\fP)=\partial_2\fP\sqcup \partial_1\fP\sqcup \partial_0\fP=\fF\sqcup \fI\nn\\
&&\fr(\partial_2\fP)=\partial \fD\sqcup \partial_0^0\fP~~\\
&&\fr(\partial_1\fP)=\partial_0\fP=\partial\fI\sqcup \partial_0^0\fP\nn~~.
\eeqan
Notice that $\fr\fA=\partial \fD\sqcup \partial_0^0\fP$. 

\paragraph{Remark.} 
The canonical Whitney stratification of $\partial \fP$ has six strata
given by $\partial_3\fP$, $\partial_2\fP$, $\partial\fD$,
$\Int\fI^+\sqcup\Int \fI^-$, $\partial\fI=\partial_0^+\fP\sqcup
\partial_0^-\fP$ and $\partial_0^0\fP$. The canonical Whitney
stratification of $\fP$ is obtained from this by adding the stratum
$\Int\fP$ and similarly for its connected refinement.

\vskip-0.1in
\noindent The values of $b_+,\rho$ and $\beta$ on the connected strata of $\partial \fP$ are summarized in Table
\ref{table:partialP}.
\begin{table}[H]
\centering {\footnotesize
\begin{tabular}{|c|c|c|c|c|c|c|}
\hline connected stratum & dimension & component of & topology & $b_+$ & $\rho$ & $\beta$ \\ \hline
\rowcolor{magenta}$\partial_0^-\fP$ & $0$ & $\partial_0\fP$ & point & $-1$ & $0$ & $0$ \\ \hline
\rowcolor{magenta}$\partial_0^+\fP$ & $0$ & $\partial_0\fP$ & point & $+1$ & $0$ & $0$ \\ \hline
\rowcolor{cyan}$\partial_0^0\fP$ & $0$ & $\partial_0\fP$ & point & $0$ & $0$ & $0$ \\ \hline
\rowcolor{cyan}$\Int \fI^-$ & $1$ & $\partial_1\fP$ & open interval & $(-1,0)$ & $0$ & $0$\\ \hline
\rowcolor{cyan}$\Int \fI^+$ & $1$ & $\partial_1\fP$ & open interval & $(0,1)$ & $0$ & $0$\\ \hline
\rowcolor{yellow}$\partial \fD$ & $1$ & $\partial_1\fP$ & circle & $0$ & $1$ & $1$\\ \hline 
\rowcolor{cyan}$\Int \fD$ & $2$ & $\partial_2\fP$ & open disk & $0$ & $[0,1)$ & $1$\\ \hline
\rowcolor{cyan}$\fA$ & $2$ & $\partial_2\fP$ & open annulus & $0$ & $(0,1)$ & $\rho$ \\ \hline
\rowcolor{cyan}$\Int \fC^-$ & $3$ & $\partial_3\fP$ & open full cone & $-g(\rho,\beta)$ & $(0,1)$ & $(\rho,1)$\\ \hline
\rowcolor{cyan}$\Int \fC^+$ & $3$ & $\partial_3\fP$ & open full cone & $+g(\rho,\beta)$ & $(0,1)$ & $(\rho,1)$\\ \hline
\end{tabular}
}
\caption{Connected refinement of the Whitney stratification of $\partial\fP$.  The colors used
  in this table (magenta, yellow and cyan) correspond to loci of
  $\SU(4)$, $G_2$ and $\SU(3)$ structures on $M$.}
\label{table:partialP}
\end{table}

\noindent The following statement follows from the results of Appendix \ref{app:proofs}:

\paragraph{Proposition.} 
The locus $\beta=0$ on $\fP$ coincides with the compact segment $\fI$,
while the locus $\beta=1$ on $\fP$ coincides with the compact disk
$\fD=D\times\{1\}$.  The locus $\beta=\rho$ on $\fP$ coincides with
$\bar{\fA}=\fA\sqcup\partial \fD\sqcup \partial_0^0\fP$. In particular,
the only locus on $\cR$ where the value $\beta=0$ can be attained is
the interval $I$ while the only locus on $\cR$ where $\beta=1$ can be
attained is the median disk $D$.

\subsection{The preimage of $\partial\cR$ inside $\partial \fP$}
\label{sec:pipreimage}

Consider the surjection $\pi:\fP\rightarrow \cR$ given by
$\pi(b,\beta)=b$ (the projection on the first three coordinates).
Since $J|_{\partial \cR}=\{\sqrt{\rho}\}$, we have:
\be
\pi^{-1}(b)=\{(b,\sqrt{\rho})\}~~\mathrm{for}~~b\in \partial \cR~~.
\ee 
Hence the restriction of $\pi$ to the subset
$\pi^{-1}(\partial\cR)\subset \partial \fP$ is a bijection onto
$\partial \cR$. It is clear that $\partial \fI\cup \partial \fD$ is
contained in $\pi^{-1}(\partial \cR)$ while $\Int \fI$, $\fA$ and
$\Int \fD$ are disjoint from $\pi^{-1}(\partial \cR)$. Using
\eqref{Pfrontier}, this gives:
\ben
\label{pipreimage}
\pi^{-1}(\partial \cR)=\partial \fI\sqcup \partial \fD\sqcup \fS^+\sqcup \fS^-
\een
where: 
\be
\fS^\pm\eqdef \Int \fC^\pm \cap \pi^{-1}(\partial\cR)
\ee
We have:
\beqan
\label{piprojections}
&&\pi(\partial \fD)=\partial D~~\nn\\
&&\pi(\partial \fI)=\partial I~~\mathrm{namely}~~\pi(\partial_0^\pm \fP)=\partial_0^\pm \cR~~\nn\\
&& \pi(\fS^\pm)=\partial^\pm_2 \cR~~\mathrm{hence}~~\pi(\partial_3\fP)=\partial_2\cR. \\
&&\pi(\partial_2\fP)\subset \Int D~~,~~\pi(\Int \fI)\subset \Int I~~.\nn
\eeqan
In particular, $\pi(\partial_0^0\fP)$ and $\pi(\Int \fI^\pm)$ are contained in $\Int \cR$. 

\paragraph{Proposition.} 
$\fS^\pm$ are the following hypersurfaces contained in $\Int\fC^\pm$ 
(see Figure \ref{fig:Jpm}):
\ben
\label{fSpm}
\fS^\pm=\{(\pm (1-\rho),b_-,b_3,\sqrt{\rho})|\rho\eqdef \sqrt{b_-^2+b_3^3}\in (0,1)\}=
\{(b,(b_-^2+b_3^2)^{1/4})|b\in \partial_2^\pm \cR\}~~.
\een
\vskip -0.15in
\begin{figure}[H]
\centering 
\includegraphics[width=0.32\linewidth]{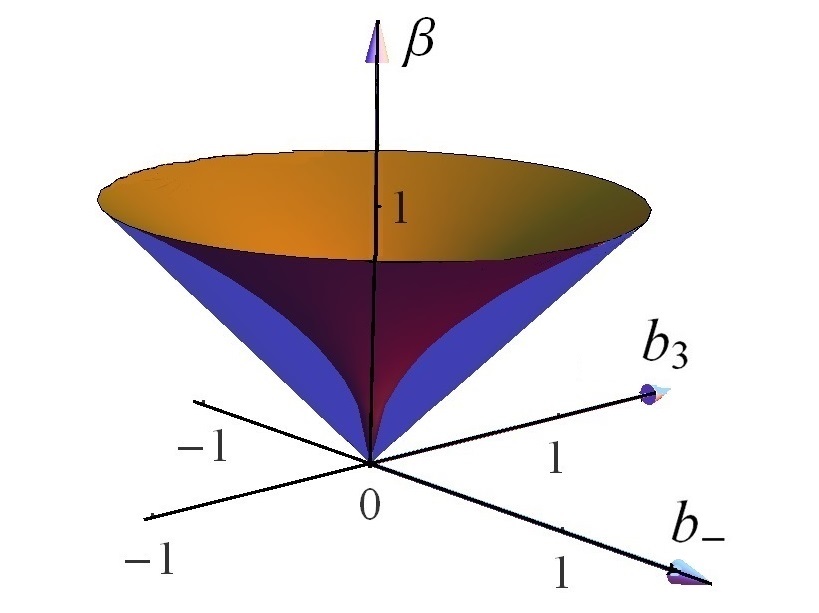}
\caption{The loci $\fS^\pm$ correspond to the hypersurface
  $\beta=\sqrt{\rho}$ (brown) inside $\Int C$ (blue). }
\label{fig:beta}
\end{figure}
\noindent{\bf Proof.} 
For $(b,\beta)\in \Int \fC^\pm$, we have $b_+=\pm g(\rho,\beta)$,
where $g(\rho,\beta)$ was defined in \eqref{gdef}. The condition
$(b,\beta)\in \fS^\pm$ further requires $b\in \partial \cR$,
i.e. $b_+=\pm (1-\rho)$.  This gives $1-\rho=g(\rho,\beta)$, which
(upon squaring both sides) is easily seen to be equivalent with
$\beta=\sqrt{\rho}$. The condition $(b,\beta)\in \Int\fC^\pm$ excludes
the values $\beta=\rho=0$ and $\beta=\rho=1$, hence we must have $\rho\in
(0,1)$. $\blacksquare$

\paragraph{Remark.}
Since $\rho\leq 1$, we have $\beta|_{\partial \cR}=\sqrt{\rho}\geq
\rho$, with equality iff $b\in \partial D$, which corresponds to
$(b,\beta)\in \partial\fD$.

\paragraph{Sections of $\fP$ with the hyperplanes $b_+=$ const.}

\begin{figure}[H]
\centering
\begin{subfigure}{0.6\textwidth}
\centering 
\includegraphics[width=0.55\linewidth]{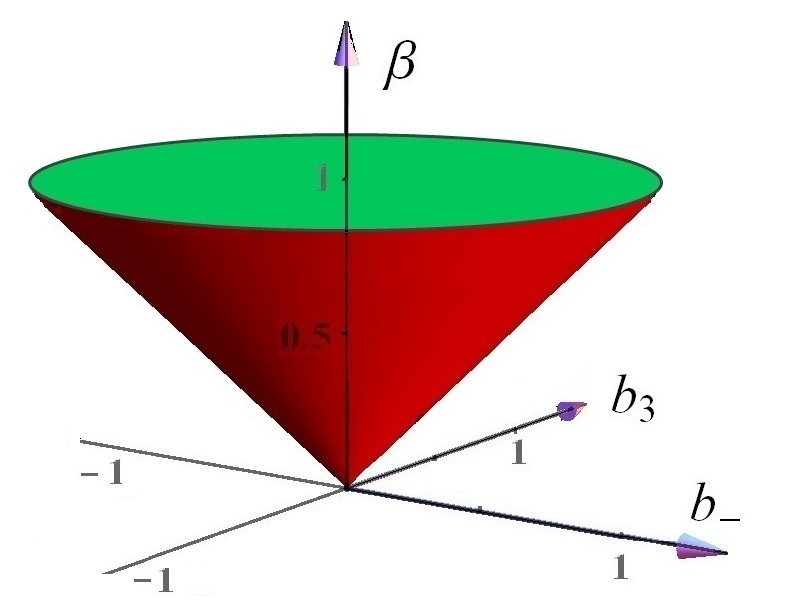}
\vskip 0.2in
\caption{Plot of $\sqrt{f_+(0,b_-,b_3)}$ (green) and
  $\sqrt{f_-(0,b_-,b_3)}$ (red) for $(b_-,b_3)$ belonging to the unit
  disk. The section of $\fP$ with the hyperplane $b_+=0$ is the
  compact full cone $\fK=\{0\}\times C$ contained between these two
graphs, whose basis is the disk $\fD$ (green).  This disk coincides
with the locus on $\fP$ where $\beta=1$.  The apex of the cone is the
midpoint of the interval $\fI$.}
\end{subfigure}~~~~
\begin{subfigure}{0.37\textwidth}
\centering
\vskip 0.2in
\includegraphics[width=0.71\linewidth]{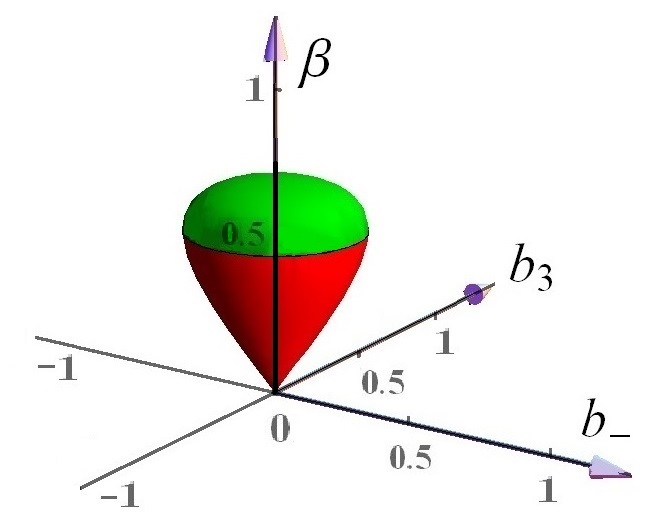}
\vskip 0.2in
\caption{Plot of $\sqrt{f_+(b)}$ and $\sqrt{f_-(b)}$ for $b\in \cR$
  with $b_+=0.5$ (thus $\rho\leq 1-0.5$). The section of
  $\fP$ with the hyperplane $b_+=0.5$ is the body of revolution
  contained between these two graphs. The boundary of this body is the
  union of a cone with a ``cap'' (a curved disk).}
\end{subfigure}
\caption{Presentation of $\fP$ as a singular fibration over the
  interval $[-1,1]$. The sections of $\fP$ with planes $b_+=$
  const. $\neq \pm 1$ are 3-dimensional bodies of revolution around
  the $\beta$-axis, obtained by rotating the graphs of Figure
  \ref{fig:betro}.  The points of $\Int \fI$ are conical singularities
  for these bodies. The bodies degenerate to points for $b_+=\pm 1$.}
\label{fig:Psections}
\end{figure}

The sections of $\fP$ with such hyperplanes are depicted in Figure
\ref{fig:Psections}; they allow one to present $\fP$ as a fibration
over the interval $[-1,1]$.  In particular, the section with the
hyperplane $b_+=0$ is the compact full 3-dimensional cone
$\fK=\{0\}\times C$, whose frontier equals $\fF$.
\vskip -0.12in

\section{Description of the rank stratifications of $\cD$ and $\cD_0$}
\label{sec:degen}

\subsection{Description of the rank stratification of $\cD$}
\label{sec:Theorems2and3}

The following result shows that the map $B$ has image contained in
$\fP$ and that the rank stratification of $\cD$ is a certain
coarsening of the $B$-preimage of the connected refinement of the
Whitney stratification of $\fP$.

\paragraph{Theorem 2.} 
The image of the map $B$ defined in \eqref{Bdef} is contained in $\fP$:
\be
\im B\subset \fP
\ee
Furthermore, the following hold for $p\in M$:
\begin{itemize}
\itemsep 0.0em
\item $\rk \cD(p)=5$ iff $B(p)\in \Int \fP$
\item $\rk \cD(p)=6$ iff $B(p)\in \partial_2 \fP\cup \partial_3 \fP=\Int \fD\sqcup \fA\sqcup \Int \fC^+\sqcup \Int \fC^-$
\item $\rk \cD(p)=7$ iff $B(p)\in \partial_0^0\fP\sqcup \partial_1\fP=\partial\fD\sqcup \Int \fI$ 
\item $\rk \cD(p)=8$ iff $B(p)\in \partial_0^+\fP\sqcup \partial_0^-\fP=\partial \fI$. 
\end{itemize}
In particular, the rank stratification of $\cD$ is given by: 
\be
\cU=B^{-1}(\Int \fP)~~,~~\cW_2=B^{-1}(\partial_2 \fP\cup \partial_3
\fP)~~,~~\cW_1=B^{-1}(\partial\fD\sqcup \Int
\fI)~~,~~\cW_0=B^{-1}(\partial \fI)~~
\ee
and we have $\cW= B^{-1}(\partial \fP)$.

\

\noindent{\bf Proof.} See Appendix \ref{app:proofs}.

\

\paragraph{Remark.} 
The map $b$ of \eqref{bdef} is related to the map $B$ of
\eqref{Bdef} by:
\be
b=\pi\circ B~~.
\ee
Using relations \eqref{pipreimage} and \eqref{piprojections}, this implies:
\be
b^{-1}(\partial \cR)=B^{-1}(\pi^{-1}(\partial \cR))~~,
\ee
namely: 
\ben
\label{bpreimages}
b^{-1}(\partial_0^\pm\cR)=B^{-1}(\partial_0^\pm \fP)~,~b^{-1}(\partial D)=B^{-1}(\partial \fD)~,~b^{-1}(\partial_2^\pm \cR)=B^{-1}(\fS^\pm)~~.
\een

\

\noindent The behavior of the one-forms $V_r$ on the locus $\cW$ is
given by the following result, whose proof can be found in Appendix
\ref{app:proofs}:

\paragraph{Theorem 3.} Let $p\in \cW$ and write: 
\be
 b_-(p)=\rho(p)\cos\psi~,~b_3(p)=\rho(p)\sin\psi~
\ee
with $\psi\in [0,2\pi)$. Then $V_r$ and $b_r$ behave as follows:
\begin{enumerate}[\leftmargin=0em 1.]
\itemsep 0.0em
\item When $p\in \cW_2$, we have:
\begin{enumerate}[\leftmargin=0em (a)]
\itemsep 0.0em

\item For $p \in b^{-1}(\Int \fD)$ we have:
\beqa
&&\beta(p)=1~~,~~b_+(p)=0~~,~~\rho(p)\in [0,1)\nn\\
&&V_+(p)=0~~,~~||V_-(p)||=\sqrt{1-\rho(p)^2\cos^2\psi}~~\\~
&& ||V_3(p)||=\sqrt{1-\rho(p)^2\sin^2\psi}~~,~~\cos\theta_{-3} =-\frac{\rho^2(p)\sin\psi\cos\psi}{||V_-(p)||~||V_3(p)||}~~.\nn
\eeqa

\item When $p\in B^{-1}(\fA)$, we have:
\beqa
&& \beta(p)=\rho(p)~~,~~b_+(p)=0~~,~~\rho(p)\in (0,1)\nn\\
&& ||V_+(p)||=\sqrt{1-\rho(p)^2}~~,~~V_-(p)=(\rho(p) \sin\psi) v~~,~~V_3(p)=-(\rho(p) \cos \psi) v~~
\eeqa
with $v\in T_p^\ast M$ an arbitrary 1-form of unit norm such that
$V_+(p)\perp v$.

\item When $p\in B^{-1}(\Int \fC^\pm)$, we have:
\beqa
&&b_+(p)=\pm g(\rho(p),\beta(p))~~,~~0<\rho(p)<\beta(p)<1\nn\\
&&V_+(p)=-\frac{||V_+||^2}{b_+\rho(p)}(\cos\psi V_-(p)+\sin \psi V_3(p))~\nn\\
&&||V_-(p)||=\sqrt{\beta^2\!\!-\!\!\rho(p)^2\cos^2\!\psi}~~,~~||V_3(p)||=\sqrt{\beta^2\!\!-\!\!\rho(p)^2\sin^2\!\psi}\\
&& \cos \theta_{-3}(p)=-\frac{\rho(p)^2\sin2\psi}{2||V_-(p)||~||V_3(p)||}~~.
\eeqa
\end{enumerate}
\item When $p\in \cW_1$, we have:
\begin{enumerate}[\leftmargin=0em (a)]
\itemsep 0.0em
\item For $p\in B^{-1}(\partial \fD)$ we have:
\beqa
&&\beta(p)=1~~,~~b_+(p)=0~~,~~\rho(p)=1\nn\\
&&V_+(p)=0~~,~~V_-(p)= (\sin \psi)v~~, ~~V_3(p)=-(\cos \psi) v~~,
\eeqa
where $v\in T_p^\ast M$ is an arbitrary 1-form of unit norm. 
\item For $p\in B^{-1}(\Int \fI)$ we have:
\beqa
&& \beta(p)=0~~,~~b_+(p)\in (-1,1)~~,~~\rho(p)=0~~\\
&& ||V_+(p)||=\sqrt{1-b_+(p)^2}~~,~~V_-(p)=V_3(p)=0~~.
\eeqa
\end{enumerate}
\item When $p\in \cW_0$ we have:
\beqa
&& \beta(p)=0~~,~~b_+(p)=\pm 1~~,~~\rho(p)=0~~\\
&& V_+(p)=V_-(p)=V_3(p)=0~~.
\eeqa
\end{enumerate}

\subsection{Description of the rank stratification of $\cD_0$ and of the stabilizer stratification}
\label{sec:Gstratif}

The following result shows that the rank stratification of $\cD_0$
(which coincides with the stabilizer stratification) is given by
another coarsening of the $B$-preimage of the connected refinement of
the canonical Whitney stratification of $\fP$.

\paragraph{Theorem 4.} For $p\in M$, we have:
\begin{itemize}
\itemsep 0.0em
\item $\rk \cD_0(p)=4$ iff $B(p)\in \Int \fP$ i.e. iff $p\in \cU$
\item $\rk \cD_0(p)=6$ iff $B(p)\in \Int \fI \sqcup \Int \fD\sqcup
  \fA\sqcup \Int \fC^+\sqcup \Int \fC^-=\Int \fI \sqcup \partial_2\fP\sqcup \partial_3\fP$
\item $\rk \cD_0(p)=7$ iff $B(p)\in \partial \fD$
\item $\rk \cD_0(p)=8$ (i.e. $\cD(p)=T_pM$) iff $B(p)\in \partial
  \fI$.
\end{itemize}
Hence the rank stratification of $\cD_0$ is given by: 
\be
\cU_0=\cU~,~\cZ_3=\emptyset~,~\cZ_2=B^{-1}(\Int \fI \sqcup \partial_2\fP\sqcup
  \partial_3\fP)~,~\cZ_1=B^{-1}(\partial \fD)~,~\cZ_0=B^{-1}(\partial \fI)=\cW_0~~~
\ee
and the stabilizer group $H_p$ is given by:
\begin{itemize}
\itemsep 0.0em
\item $H_p\simeq \SU(2)$ if $p\in \cU_0=\cU$
\item $H_p\simeq \SU(3)$ if $p\in \cZ_2$
\item $H_p\simeq \G_2$ if $p\in \cZ_1$
\item $H_p\simeq \SU(4)$ if $p\in \cZ_0$ . 
\end{itemize}

\

\noindent{\bf Proof.}  Follows immediately from Theorem 1 of Section
\ref{sec:spinors} together with the Lemma of Appendix
\ref{app:hatG}. $\blacksquare$

\

\noindent The situation is summarized in Table \ref{table:HHprime}. 
\begin{table}[H]
\centering
{\footnotesize
\begin{tabular}{|c|c|c|c|c|c|}
\hline $\fP$-description & $\cD$-stratum & $\cD_0$-stratum &  $\rk \cD$ &  $\rk\cD_0$  & $H_p$\\
\hline
\rowcolor{magenta}$B^{-1}(\partial \fI)$  & $\cW_0$ & $\cZ_0$ & $8$  &  $8$   & $\SU(4)$ \\
\hline
\rowcolor{yellow}$B^{-1}(\partial \fD)$  & $\cW_1^1$ & $\cZ_1$ & $7$  & $7$   & $\G_2$ \\
\hline
\rowcolor{cyan}$B^{-1}(\Int \fI)$  & $\cW_1^0$  & $\subset \cZ_2$ & $7$  & $6$   & $\SU(3)$ \\
\hline
\rowcolor{cyan}$B^{-1}(\partial_2\fP\sqcup \partial_3\fP)$  & $\cW_2$  & $\subset \cZ_2$ & $6$  & $6$ & $\SU(3)$  \\
\hline
$\Int\fP$ & $\cU$ & $\cU_0$ & $5$  & $4$  & $\SU(2)$\\\hline
\end{tabular}
}
\caption{The ranks of $\cD$ and $\cD_0$ on various loci and the isomorphism type of $H_p$.}
\label{table:HHprime}
\end{table}

\begin{figure}[H]
\centering \includegraphics[width=0.34\linewidth]{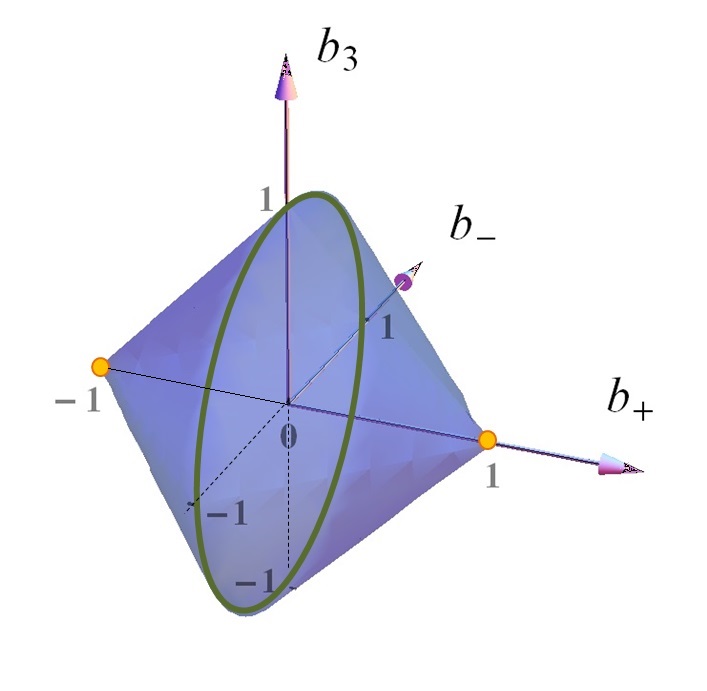}
\caption{The $b$-image of the $\SU(4)$ locus is contained in $\partial
  I$ (orange). The $b$-image of the $\G_2$ locus is contained in $\partial
  D$ (green). The $b$-image of the $\SU(3)$ locus is contained in $\cR\setminus (\partial I\cup \partial D)$ (blue),
  while the $b$-image of the $\SU(2)$ locus is contained in $\Int
  \cR$ (blue).}
\label{fig:degen}
\end{figure}

\vskip 0.1in
\noindent The $b$-image of the G-structure stratification is depicted in Figure
\ref{fig:degen}.

\subsection{Comparing the rank stratifications of $\cD$ and $\cD_0$}
Using relations \eqref{Pfrontier}, Theorem 2 shows that $\cW_k$ decompose as follows:
\beqa
&&\cW_0=\cW_0^+\sqcup \cW_0^-~~\mathrm{where}~~\cW_0^\pm=B^{-1}(\partial_0^\pm\fP)\\
&&\cW_1=\cW_1^0\sqcup\cW_1^1~~~\mathrm{where}~~\cW_1^0\eqdef B^{-1}(\Int \fI)~,~\cW_1^1\eqdef B^{-1}(\partial\fD)\\
&&\cW_2=\cW_2^2\sqcup\cW_2^3~~~\mathrm{where}~~\cW_2^2\eqdef B^{-1}(\partial_2\fP)~,~\cW_2^3\eqdef B^{-1}(\partial_3 \fP)~~,
\eeqa
where $\cW_2^2$ and $\cW_2^3$ decompose further as: 
\beqa
&&\cW_2^2=\cW_2^{2+}\sqcup \cW_2^{2-}~~\mathrm{with}~~\cW_2^{2+}\eqdef B^{-1}(\Int \fD)~~,~~\cW_2^{2-}\eqdef B^{-1}(\fA)~~\\
&&\cW_2^3=\cW_2^{3+}\sqcup \cW_2^{3-}~~\mathrm{with}~~\cW_2^{3\pm}\eqdef B^{-1}(\Int \fC^\pm)~~,
\eeqa
so that: 
\be
\cW_2=\cW_2^{2+}\sqcup \cW_2^{2-}\sqcup\cW_2^{3+}\sqcup\cW_2^{3-}~~.
\ee
Finally, $\cW_1^0$ decomposes as: 
\be
\cW_1^0=\cW_1^{0+}\sqcup \cW_1^{0-}\sqcup \cW_1^{00}~~\mathrm{with}~~\cW_1^{0\pm}=B^{-1}(\Int\fI^\pm)~~,~~\cW_1^{00}=B^{-1}(\partial_0^0\fP)~~.
\ee
The components listed above give the $B$-preimage of the connected
refinement of the canonical Whitney stratification of $\partial \fP$,
to which we can add $B^{-1}(\Int \fP)$ to obtain the $V$-preimage of
the connected refinement of the Whitney stratification of $\fP$ (see
Table \ref{table:W}). Theorems 2 and 4 give:
\be
\cU_0=\cU~~,~~\cZ_3=\emptyset~~,~~\cZ_2=\cW_1^0\sqcup \cW_2~~,~~\cZ_1=\cW_1^1~~,~~\cZ_0=\cW_0~~.
\ee
In view of the last equality, we define $\cZ_0^\pm\eqdef \cW_0^\pm$.

\begin{table}[H]
\centering {\footnotesize
\begin{tabular}{|c|c|c|c|c|c|c|c|c|}
\hline & $\fP$-description & $b$-image & $\cD$-stratum & $\cD_0$-stratum & $b_+$ & $\rho$ & $\beta$ & $H_p$\\ \hline
\rowcolor{magenta}$\cW_0^+$ & $B^{-1}(\partial_0^+ \fP)$ & $\partial_0^+\cR$ & $\cW_0$ & $\cZ_0$ & $+ 1$ & $0$ & $0$ & $\SU(4)$\\ \hline
\rowcolor{magenta}$\cW_0^-$ & $B^{-1}(\partial_0^-\fP)$ & $\partial_0^-\cR$ & $\cW_0$ & $\cZ_0$ & $- 1$ & $0$ & $0$ & $\SU(4)$\\ \hline
\rowcolor{yellow}$\cW_1^1$ & $B^{-1}(\partial \fD)$ & $\partial_1\cR=\partial D$ & $\cW_1$ & $\cZ_1$ & $0$ & $1$ & $1$ & $\G_2$\\ \hline 
\rowcolor{cyan}$\cW_1^{0+}$ & $B^{-1}(\Int \fI^+)$ & $\Int(I^+)$ & $\cW_1$ & $\cZ_2$ & $(0,+1)$ & $0$ & $0$ & $\SU(3)$\\ \hline
\rowcolor{cyan}$\cW_1^{0-}$ & $B^{-1}(\Int \fI^-)$ & $\Int(I^-)$ & $\cW_1$ & $\cZ_2$ & $(-1,0)$ & $0$ & $0$ & $\SU(3)$\\ \hline
\rowcolor{cyan}$\cW_1^{00}$ & $B^{-1}(\partial_0^0\fP)$ & $\{0_{\R^3}\}$ & $\cW_1$ & $\cZ_2$ & $0$ & $0$ & $0$ & $\SU(3)$\\ \hline
\rowcolor{cyan}$\cW_2^{2+}$ & $B^{-1}(\Int \fD)$ & $\Int D$ & $\cW_2$ & $\cZ_2$ & $0$ & $[0,1)$ & $1$ & $\SU(3)$\\ \hline
\rowcolor{cyan}$\cW_2^{2-}$ & $B^{-1}(\fA)$ & $\Int D\setminus \{0\}$ & $\cW_2$ & $\cZ_2$ & $0$ & $(0,1)$ & $\rho$ & $\SU(3)$\\ \hline
\rowcolor{cyan}$\cW_2^{3+}$ & $B^{-1}(\Int \fC^+)$ & $\Int(\cR^+)$ & $\cW_2$ & $\cZ_2$ & $+g(\rho,\beta)$ & $[0,1)$ & $(\rho,1)$ & $\SU(3)$\\ \hline
\rowcolor{cyan}$\cW_2^{3-}$ & $B^{-1}(\Int \fC^-)$ & $\Int(\cR^-)$ & $\cW_2$ & $\cZ_2$ & $-g(\rho,\beta)$ & $[0,1)$ & $(\rho,1)$ & $\SU(3)$\\ \hline
\rowcolor{white}$\cU$ & $B^{-1}(\Int \fP)$ & $\Int \cR$ & $\cU$ & $\cU_0$ & $(-1,1)$ & $[0,1)$ & $J(b_+,\rho)$ & $\SU(2)$\\ \hline
\end{tabular}
}
\caption{Preimage of the connected refinement of the canonical Whitney
  stratification of $\fP$.}
\label{table:W}
\end{table}

\subsection{Description of the chirality stratification}
\label{sec:cstratif}

We saw in Section \ref{sec:spinors} that $\cS=b^{-1}(\partial
\cR)$. Since $b=B\circ \pi$, this gives
$\cS=B^{-1}(\pi^{-1}(\partial\cR))$.  The set
$\pi^{-1}(\partial\cR)\subset \partial \fP$ which was discussed in
Section \ref{sec:pipreimage}.  Together with Theorem 1, decomposition
\eqref{pipreimage} and relations \eqref{piprojections} imply:
\beqa
&&\cS_{02}=B^{-1}(\partial_0^+\fP)=\cW_0^+~~,~~\cS_{20}=B^{-1}(\partial_0^-\fP)=\cW_0^-~~\\
&&\cS_{12}=B^{-1}(\fS^+)\subset \cW_2^{3+}~~,~~\cS_{21}=B^{-1}(\fS^-)\subset \cW_2^{3-}~~\\
&& \cS_{11}=B^{-1}(\partial \fD)=\cW_1^1=\cZ_1~.
\eeqa
In particular, we have $\cS\subset \cW_0\sqcup \cW_1^1\sqcup \cW_2^3\subset \cW$ and 
\be
\cG=\cU\sqcup B^{-1}(\Int \fI)\sqcup B^{-1}(\partial_2\fP)\sqcup B^{-1}(\partial_3\fP\setminus \fS)~~,
\ee
where $\fS\eqdef \fS^+\sqcup \fS^-$. The situation is summarized in Table \ref{table:S2}, where we remind the 
reader that the restrictions of $\cD$ and $\cD_0$ to the special locus $\cS$ coincide (see Section \ref{sec:spinors}).

\begin{table}[H]
\centering
{\footnotesize
\begin{tabular}{|c|c|c|c|c|c|c|}
\hline $\fP$-description & $\cS$-stratum& $\cD$-stratum & $\cD_0$-stratum &  $\rk \cD$ &  $\rk\cD_0$  & $H_p$\\
\hline
\rowcolor{magenta} $B^{-1}(\partial_0^+\fP)$ & $\cS_{02}$ & $\cW_0^+$ & $\cZ_0^+$ & $8$  &  $8$   & $\SU(4)$ \\
\hline
\rowcolor{magenta} $B^{-1}(\partial_0^-\fP)$  & $\cS_{20}$ & $\cW_0^-$ & $\cZ_0^-$ & $8$  & $8$   &  $\SU(4)$ \\
\hline
\rowcolor{yellow} $B^{-1}(\partial \fD)$ & $\cS_{11}$ & $\cW_1^1$ & $\cZ_1$ & $7$  & $7$   & $\G_2$\\
\hline
\rowcolor{cyan} $B^{-1}(\fS^+)$ & $\cS_{12}$ & $\subset \cW_2^{3+}$  & $\subset \cZ_2$ & $6$  & $6$   & $\SU(3)$ \\
\hline
\rowcolor{cyan}$B^{-1}(\fS^-)$  & $\cS_{21}$ & $\subset \cW_2^{3-}$  & $\subset \cZ_2$ & $6$  & $6$  & $\SU(3)$  \\
\hline
\end{tabular}
}
\caption{Description of the special strata of the chirality stratification. The table does not show the non-special locus $\cG$.}
\label{table:S2}
\end{table}

\subsection{Relation to previous work}
\label{sec:Palti}

Some aspects of $\cN=2$ compactifications of eleven-dimensional
supergravity down to $\AdS_3$ were approached in \cite{Palti} using a
nine-dimensional formalism based on the auxiliary 9-manifold ${\hat
  M}\eqdef M\times S^1$, but without carefully exploring the
consequences of that formalism for the geometry of $M$. Sections 3-5
of \cite{Palti} also discuss some consequences of the supersymmetry
equations (which were also derived in \cite{ga2}) using the
nine-dimensional formalism. Reference \cite{Palti} makes intensive use
of an assumption (equation (3.9) of loc. cit.) which, as we show in
Appendix \ref{app:P}, can only hold when the $\SU(2)$ locus $\cU$ of
$M$ is empty. Since most results of \cite{Palti} (including the count
of the number of supersymmetries preserved by membranes transverse to
$M$ as well as the discussion of Sections 3-6 of that reference) rely
on that assumption, those results can apply only to the highly
non-generic case when $\cU=\emptyset$. As we explain in detail in
forthcoming work, failure of \cite[eq. (3.9)]{Palti} is related to the
transversal vs. non-transversal character of the intersection of a
certain distribution ${\hat \cD}$ defined on ${\hat M}$ with the
pullback to ${\hat M}$ of the tangent bundle of $M$.

\section{Conclusions}
\label{sec:concl}

We studied the conditions for ``off-shell'' extended supersymmetry in
compactifications of eleven-dimensional supergravity on
eight-manifolds $M$. We gave an explicit description of the stabilizer
stratification induced by two globally-defined Majorana spinors as a
certain coarsening of the preimage of the connected refinement of the
Whitney stratification of a four-dimensional compact semi-algebraic
set $\fP$ through a smooth map $B:M\rightarrow \R^4$ whose image is
contained in $\fP$. We also described the chirality stratification as
a coarsening of the preimage of the connected refinement of the
Whitney stratification of a 3-dimensional compact semi-algebraic set
$\cR$ through a smooth map $b:M\rightarrow \R^3$ whose image is
contained in $\cR$. Unlike the case of $\cN=1$ compactifications, the
stabilizer and chirality stratifications do not coincide. We found a
rich landscape of reductions of structure group along the various
strata, which we classified explicitly. The open strata of the
chirality and stabilizer stratifications coincide and correspond to an
open subset $\cU\subset M$ which carries an $\SU(2)$ structure. This
locus is present in generic $\cN=2$ flux compactifications of
eleven-dimensional supergravity on eight manifolds, for example in
generic $\cN=2$ compactifications down to $\AdS_3$ spaces.

We also discussed two natural cosmooth generalized distributions $\cD$
and $\cD_0$ which exist on $M$ when considering such
backgrounds. These are defined by the four one-form spinor bilinears
$V_1,V_2,V_3$ and $W$ which are induced by two independent globally-defined
Majorana spinors given on $M$, namely $\cD$ is the intersection of the
kernel distributions of $V_1, V_2$ and $V_3$ while $\cD_0$ is the
intersection of $\cD$ with the kernel distribution of $W$. We showed
that the rank stratification of $\cD_0$ coincides with the stabilizer
stratification, while the rank stratification of $\cD$ is another
coarsening of the $B$-preimage of the connected refinement of the
Whitney stratification of $\fP$. The restriction of $\cD$ to the open
stratum $\cU$ is a rank five regular Frobenius distribution which
carries an $\SU(2)$ structure in the sense of \cite{ContiSalamon},
while the restriction of $\cD_0$ to $\cU$ is a rank four Frobenius
distribution (the almost contact distribution of
\cite{Bedulli}). Since the $\SO(8)$ image $G_p=\q(H_p)$ of the
pointwise stabilizer group $H_p$ of two independent Majorana spinors
fixes the forms $V_1(p),V_2(p),V_3(p)$ and $W(p)$, the distribution
$\cD_0|_\cU$ carries the $\SU(2)$ structure of $\cD|_{\cU}$ in the
sense that $G_p$ is contained in the group $\SO(\cD_0(p),g_p)\simeq
\SO(4)$ for any point $p\in \cU$. In this paper, we focused on the
classification of spinor positions and stabilizer groups, which we
treated in detail given its complexity. We mention that considerably
more can be said about the chirality and stabilizer stratifications
provided that one makes appropriate Thom-Boardman type genericity
assumptions which allow one to apply results from the singularity
theory of differentiable maps \cite{GG, AVZ, Mather, MatherSM}.

Since the manifolds $M$ considered in this paper are
eight-dimensional, it is not entirely clear how a description of such
backgrounds may be given within the framework of exceptional
generalized geometry \cite{Hull, Waldram1, Waldram2,
  Waldram3,Waldram4, Waldram5, Baraglia}, similar to the one given in
\cite{Waldram3, Waldram4, Waldram5} for 7-dimensional backgrounds of
eleven-dimensional supergravity and in \cite{MinasianGG1, MinasianGG2, GLSM,GranaOrsi} for
six-dimensional type II backgrounds.  This stems from
difficulties\footnote{The precise problem (see \cite{Waldram3}) is
  that one wants to consider generalized connections which are
  compatible with the generalized metric as well as torsion-free in an
  appropriate sense, however one does not have a natural definition of
  the torsion of a generalized connection when $\dim M>7$.} in
building an appropriate generalized connection in eight dimensions,
which in turn relates to the presence of Kaluza-Klein monopoles in the
U-duality algebra and hence to the problem of including ``dual
gravitons'' at the nonlinear level in $E_{8(8)}$-covariant
formulations of eleven-dimensional supergravity \cite{CurtrightDG,
  HullDG1, WestDG, HullDG2} (which is obstructed by the no-go results
of \cite{DGnogo1, DGnogo2}). A solution to this problem was recently
proposed in \cite{Samtleben} within the framework of exceptional field
theory but, as pointed out in \cite{Rosabal}, that solution may be
incomplete. It would be interesting to understand what light may be
shed on our results by exceptional generalized geometry.

The results of this paper show that the rich landscape of G-structures
arising in $\cN=2$ flux compactifications of eleven-dimensional
supergravity on eight-manifolds admits a natural description using
stratification theory and standard constructions of real
semi-algebraic geometry \cite{BCR, AK, BPR}, thus giving clues about
the mathematical tools required for general treatments of flux
backgrounds. We note that the approach via cosmooth generalized
distributions, stratified G-structures and semi-algebraic sets appears
to be quite general and thus could be applied to flux backgrounds of
any supergravity theory. In general, the complexity of the
stratifications involved grows rather fast with the number of spinors
(as implied by the results of \cite{Rannou}), but such stratifications
can be computed algorithmically. We mention that powerful 
algorithms exist \cite{BPR} for the study of semi-algebraic sets.

\acknowledgments{The work of E.M.B. was partly supported by the
  strategic grant POSDRU/159/1.5/S/133255, Project ID 133255 (2014),
  co-financed by the European Social Fund within the Sectorial
  Operational Program Human Resources Development 2007 -- 2013 and
  partly by the CNCS-UEFISCDI grant PN-II-ID-PCE 121/2011 and by 
 PN 09370102/2009.  The work of C.I.L was supported by the research
  grants IBS-R003-G1 and IBS-R003-S1. C.I.L. acknowledges hospitality
  of the University of Nis and of the SEENET-MTP Office, where part of
  this work was completed. His visit to Nis, Sebia, was supported by the ICTP
  - SEENET-MTP network project PRJ-09 titled ``Cosmology and
  Strings''. E.M.B. acknowledges the invitation, financial support 
and hospitality of IPhT-CEA, Paris-Saclay, for her visit at 
the institute during the preparation of this article.}

\appendix

\section{Notations and conventions}
\label{app:notations}

Throughout this paper, $(M,g)$ denotes a connected and compact smooth
Riemannian eight-manifold, which we assume to be oriented and spin.
The unital commutative $\R$-algebra of smooth real-valued functions on
$M$ is denoted by $\cinf$.  The fact that $M$ is orientable and spin
means that its first two Stiefel-Whitney classes vanish,
i.e. $w_1(M)=w_2(M)=0$. All fiber bundles we consider are
smooth\footnote{The generalized bundles \cite{Drager, BulloLewis}
  considered in this paper are {\em not} fiber bundles
  and they will be either smooth or cosmooth.}. We use freely the
results and notations of \cite{ga1,ga2,gf}, with the same
conventions as there. 

Recall that the set of isomorphism classes of spin structures of $M$ is a torsor
for the finite group $H^1(M,\Z_2)$. Let $(T^\ast M,\diamond)$ denote
the \KA bundle of $(M,g)$, which is a bundle of unital associative
$\R$-algebras.  Consider the set $\cA$ consisting of all pairs
$(S,\gamma)$, where $S$ is a vector bundle of rank $16$ over $M$ and
$\gamma:(T^\ast
M,\diamond)\stackrel{\sim}{\rightarrow}(\End(S),\circ)$ is a unital
isomorphism of bundles of $\R$-algebras. Two pairs
$(S,\gamma),(S',\gamma')$ are called equivalent (and we write
$(S,\gamma)\sim (S',\gamma')$) if there exists an isomorphism of
$\Z_2$-graded vector bundles $f:S\stackrel{\sim}{\rightarrow} S'$ such
that $\gamma'={\tilde f}\circ \gamma$, where ${\tilde
  f}:\End(S)\rightarrow \End(S')$ is the unital isomorphism of bundles
of algebras corresponding to ${\tilde f}(Q)\eqdef f\circ \cA\circ
f^{-1}$ for all $Q\in \Gamma(M,\End(S))$. Given a spin structure on
$M$, let $S^\pm$ be the corresponding bundles of spinors of positive
and negative chirality and $S\eqdef S^+\oplus S^-$ denote the
corresponding bundle of real pinors (a.k.a. Majorana spinors). Then
$S$ is a bundle of modules over the \KA bundle $(T^\ast M,\diamond)$
whose structure morphism is an isomorphism of bundles of algebras
$\gamma:(T^\ast M,\diamond)\stackrel{\sim}{\rightarrow}(\End(S),\circ)$ and hence the
pair $(S,\gamma)$ is an element of $\cA$. This gives a map which
associates an element of $\cA$ to every spin structure of $M$.  It is
easy to see that two spin structures are equivalent iff the
corresponding pairs $(S,\gamma)$ and $(S',\gamma')$ are equivalent in
the sense described above, hence we have a bijection between
$H^1(M,\Z_2)$ and the set $\cA/_\sim$. Throughout the paper, we assume
that a spin structure has been chosen for $M$ and we work with the
corresponding pair $(S,\gamma)\in \cA$. 

Up to rescalings by smooth nowhere-vanishing real-valued functions
defined on $M$, the bundle $S$ of Majorana spinors has two admissible
pairings $\cB_\pm$ (see \cite{gf, AC1, AC2}), both of which are
symmetric. These pairings are distinguished by their types
$\epsilon_{\cB_\pm}=\pm 1$. Throughout the paper, we work with
$\cB\eqdef \cB_+$, which we can take to be a scalar product on $S$,
denoting the induced norm on $S$ by $||~||$.

Our convention for the Clifford algebra $\Cl(h)$ of a bilinear form
$h$ is that common in Physics, i.e. the generators satisfy
$e_ke_l+e_le_k=2h_{kl}$; the convention common in Mathematics has a
minus on the right hand side. One recovers the Mathematics convention by
multiplying all $e_k$ with the imaginary unit $i$; accordingly, the
Killing constant of a Killing spinor is multiplied by $i$.  Unlike in
some of the literature on flux compactifications, we reserve the name
``Killing spinor'' for the mathematically consecrated notion, i.e. for
a spinor $\xi$ which satisfies $\nabla_k\xi=\lambda e_k\xi$, where
$\lambda$ is the Killing constant and the right hand side involves
Clifford multiplication; spinors which satisfy generalizations of this
equation in which the right hand side contains a polynomial in $e_i$
are called {\em generalized} Killing spinors, as usual in the
Mathematics literature. 

The generalized distributions \cite{Drager, BulloLewis} $\cD$
and $\cD_0$ considered in this paper are cosmooth in the sense of
\cite{Drager} rather than smooth. As explained in Appendix D of
\cite{g2s}, their integrability theory (see \cite{Freeman}) is in some
sense ``orthogonal'' to that of smooth generalized distributions
\cite{Stefan, Sussmann, Michor, Ratiu}. When integrable, a cosmooth generalized
distribution integrates to a Haefliger structure (a.k.a. a singular
foliation in the sense of Haefliger) while a smooth generalized
distribution integrates to a singular foliation in the sense of
\cite{Androulidakis1, Androulidakis2}.

We use the ``mostly plus'' convention for pseudo-Riemannian metrics of
Minkowski signature. Given a subset $A$ of $M$, we let ${\bar A}$
denote the closure of $A$ in $M$ (taken with respect to the manifold
topology of $M$). The {\em frontier} (also called {\em topological
  boundary}) of $A$ is defined as $\partial A \eqdef {\bar A}\setminus
\Int A$, where $\Int A$ denotes the interior of $A$. The {\em small
  topological frontier} is $\fr A \eqdef {\bar A}\setminus A$. When
considering the canonical Whitney stratification of a semi-algebraic
set, we always work with its connected refinement (see Appendix
\ref{app:alg}). In some references (such as \cite{Rannou}) it is this
connected refinement which is called the canonical Whitney
stratification of that semi-algebraic set.

\section{Algebraic constraints for $V_r,W$ and $b$}
\label{app:alg}

Relations \eqref{VWsys} can be obtained through direct computation
using Fierz identities. Here, we give a proof which relies on reducing
\eqref{Vsys} to a Fierz identity satisfied by a single spinor. Consider the Majorana 
spinor:
\be
\xi(x)\eqdef x_{1+}\xi_1^++x_{1-}\xi_1^-+x_{2+}\xi_2^++x_{2-}\xi_2^-\in \Gamma(M,S)~~
\ee
and the corresponding one-form:
\be
V(x)\eqdef_U  \cB(\xi(x), \gamma_a\xi(x))e^a~~,
\ee
where $U\subset M$ and $x_{i\pm}$ are arbitrary real numbers. This satisfies the
relation \cite{ga1}: 
\ben
\label{Vbc}
||V(x)||^2=||\xi(x)||^4-b(x)^2~~,
\een
where: 
\be
b(x)\eqdef_U \cB(\xi(x),\gamma(\nu)\xi(x))~~.
\ee
The relations $\gamma_a^t=\gamma_a$ and $\gamma(\nu)^t=\gamma(\nu)$ give: 
\be
\cB(\xi_i^\alpha,\gamma_a\xi_j^\beta)=\cB(\xi_j^\beta,\gamma_a\xi_i^\alpha)~~,~~\cB(\xi_i^\alpha,\gamma(\nu)\xi_j^\beta)=\cB(\xi_j^\beta,\gamma(\nu)\xi_i^\alpha)
\ee
for all $i,j=1,2$ and all $\alpha,\beta\in \{-,+\}$. Using these as
well as $\cB(\xi_i^\pm,\xi_j^\mp)=0$, we find:
\be
\begin{split}
& V(x)=x_{1+}x_{1-}V_1+x_{2+}x_{2-}V_2+2x_{1-}x_{2+}V_3^++2x_{1+}x_{2-}V_3^-\\
& ||\xi(x)||^2=x_{1+}^2||\xi_1^+||^2+x_{2+}^2||\xi_2^+||^2+x_{1-}^2||\xi_1^-||^2+x_{2-}^2||\xi_2^-||^2+2 x_{1+}x_{2+}\cB(\xi_1^+,\xi_2^+)+2x_{1-}x_{2-}\cB(\xi_1^-,\xi_2^-)\\
& b(x)=x_{1+}^2||\xi_1^+||^2+x_{2+}^2||\xi_2^+||^2-x_{1-}^2||\xi_1^-||^2-x_{2-}^2||\xi_2^-||^2+2 x_{1+}x_{2+}\cB(\xi_1^+,\xi_2^+)-2x_{1-}x_{2-}\cB(\xi_1^-,\xi_2^-)~~.
\end{split}
\ee
Using \eqref{xib}, these relations become:
\be
\begin{split}
& V(x)=x_{1+}x_{1-}V_1+x_{2+}x_{2-}V_2+2x_{1-}x_{2+}V_3^++2x_{1+}x_{2-}V_3^-\\
& ||\xi(x)||^2=\frac{1}{2}\left[x_{1+}^2(1+b_1)+x_{2+}^2(1+b_2)+x_{1-}^2(1-b_1)+x_{2-}^2(1-b_2)\right]+x_{1+}x_{2+}b_3-x_{1-}x_{2-} b_3\\
& b(x)=\frac{1}{2}\left[x_{1+}^2(1+b_1)+x_{2+}^2(1+b_2)-x_{1-}^2(1-b_1)-x_{2-}^2(1-b_2)\right]+x_{1+}x_{2+}b_3+x_{1-}x_{2-}b_3~~.
\end{split}
\ee
Substituting these expressions into \eqref{Vbc} gives an algebraic equation
which must hold for all $x_{i\alpha}$, i.e. a certain polynomial in
the variables $x_{i\alpha}$ must vanish identically. This means that
the coefficients of all monomials in $x_{i\alpha}$ in that polynomial
must vanish, giving the relations:
\beqan
\label{VWsys2}
&&||V_1||^2=1-b_1^2~~,~~||V_2||^2=1-b_2^2~~,\nn\\
&&||V_3^+||^2=\frac{1}{4}(1-b_1+b_2-b_1b_2)~~,~~||V_3^-||^2=\frac{1}{4}(1+b_1-b_2-b_1b_2)~~,\nn\\
&&\langle V_1,V_2\rangle +4\langle V_3^-,V_3^+\rangle =2b_3^2~~,\\
&&\langle V_1,V_3^+\rangle =1/2(1-b_1)b_3~~,~~\langle V_1,V_3^-\rangle =-1/2(1+b_1)b_3~~,\nn\\
&&\langle V_2,V_3^+\rangle =-1/2(1+b_2)b_3~~,~~\langle V_2,V_3^-\rangle =1/2(1-b_2)b_3~~.\nn
\eeqan
Using $V_3=V_{3}^++V_3^-$, $W=V_{3}^+-V_3^-$ and
$V_\pm=\frac{1}{2}(V_1\pm V_2)$, we can write \eqref{VWsys2} in the
form \eqref{VWsys}. The system \eqref{VWsys} can also be written as:
\beqan
\label{Fierz2f}
&&||V_1||^2=1-b_1^2~~,~~||V_2||^2=1-b_2^2~~,~~||V_3||^2+||W||^2=1-b_1 b_2~~,\nn\\
&&\langle V_1,V_2\rangle +2||V_3||^2=1-b_1b_2-2b_3^2~~,~~\langle V_1,V_3\rangle =-b_1b_3~~,
~~\langle V_2,V_3\rangle =-b_2b_3~~,\\
&&\langle V_1,W\rangle =b_3~~,~~\langle V_2,W\rangle =-b_3~~,~~\langle V_3,W\rangle =\frac{1}{2}(b_2-b_1)~~.\nn
\eeqan
Using \eqref{xib}, we find:
\beqan
\label{brels}
&&1-b_i^2=(1-b_i)(1+b_i)=4||\xi_i^+||^2||\xi_i^-||^2~~(i=1,2)\nn\\
&& 1\mp b_1\pm b_2-b_1b_2=(1\mp b_1)(1\pm b_2)=4||\xi_1^\mp||^2||\xi_2^\pm||^2~~.
\eeqan
This allows us to write the norms of $V_i, V_3^\pm$ given in \eqref{VWsys} in the form: 
\ben
\label{Vnorms}
||V_i||=2||\xi_i^+||~||\xi_i^-||~~,~~||V_3^\pm||=||\xi_1^\mp||~||\xi_2^\pm||~~.
\een

\paragraph{Proposition.} 
Assume that $\xi_j^\pm$ does not vanish anywhere on the open subset
$U\subset M$ which supports the local orthonormal coframe
$(e^a)_{a=1\ldots 8}$ of $(M,g)$. Then $(\gamma^a\xi_j^\pm)_{a=1\ldots 8}$
is an orthogonal frame of $S^\mp$ defined above $U$ which satisfies
$||\gamma^a\xi_j^\pm||^2=||\xi_j^\pm||^2$ and we have:
\be
\xi_i^\mp=_U\frac{1}{||\xi_j^\pm||^2}\sum_{a=1}^8 \cB(\xi_i^\mp,\gamma_a\xi_j^\pm)\gamma^a\xi_j^\pm~~.
\ee
In particular, if $\xi_1^+,\xi_1^-,\xi_2^+$ and $\xi_2^-$ are all non-vanishing 
on $U$ then: 
\beqan
\label{xiexp}
&&\xi_1^+=_U\frac{1}{2||\xi_1^-||^2}\gamma(V_1)\xi_1^-=\frac{1}{||\xi_2^-||^2}\gamma(V_3^-)\xi_2^-~~,~~
\xi_2^+=_U\frac{1}{2||\xi_2^-||^2}\gamma(V_2)\xi_2^-=\frac{1}{||\xi_1^-||^2}\gamma(V_3^+)\xi_1^-\nn\\ 
&& \xi_1^-=_U\frac{1}{2||\xi_1^+||^2}\gamma(V_1)\xi_1^+=\frac{1}{||\xi_2^+||^2}\gamma(V_3^+)\xi_2^+~~,~~
\xi_2^-=_U\frac{1}{2||\xi_2^+||^2}\gamma(V_2)\xi_2^+=\frac{1}{||\xi_1^+||^2}\gamma(V_3^-)\xi_1^+~~.~~~~~~~
\eeqan

\

\noindent{\bf Proof.}  
Follows immediately by applying a result proved in \cite[Section
  2.6]{g2s} (the Corollary on page 14 of loc. cit.). $\blacksquare$

\paragraph{Remark.} 
Under the assumption of the second part of the proposition, relations
\eqref{Vnorms} show that $V_1$, $V_2$ and $V_3^\pm$ are
nowhere-vanishing on $U$ and that the following rescaled 1-forms have
unit norm everywhere on $U$, where $i=1,2$:
\be
\hV_i\eqdef \frac{1}{2||\xi_i^+||~||\xi_i^-||}V_i~~,~~\hV_3^\pm \eqdef \frac{1}{||\xi_1^\mp||~||\xi_2^\pm||}V_3^\pm~~.
\ee
Using these normalized 1-forms, relations \eqref{xiexp} can be written as:
\beqan
\label{xiexpnorm}
&&\frac{\xi_1^+}{||\xi_1^+||}=_U\gamma(\hV_1)\frac{\xi_1^-}{||\xi_1^-||}=\gamma(\hV_3^-)\frac{\xi_2^-}{||\xi_2^-||}~~,~~
\frac{\xi_2^+}{||\xi_2^+||}=_U\gamma(\hV_2)\frac{\xi_2^-}{||\xi_2^-||}=\gamma(\hV_3^+)\frac{\xi_1^-}{||\xi_1^-||}~~\nn\\
&&\frac{\xi_1^-}{||\xi_1^-||}=_U\gamma(\hV_1)\frac{\xi_1^+}{||\xi_1^+||}=\gamma(\hV_3^+)\frac{\xi_2^+}{||\xi_2^+||}~~,~~
\frac{\xi_2^-}{||\xi_2^-||}=_U\gamma(\hV_2)\frac{\xi_2^+}{||\xi_2^+||}=\gamma(\hV_3^-)\frac{\xi_1^+}{||\xi_1^+||}~~.
\eeqan
Notice $\hV_1, \hV_2,\hV_3^\pm$ square to one in the \KA algebra of
$(U,g)$ and hence the endomorphisms $\gamma(\hV_i)$, 
$\gamma(\hV_3^\pm)$ square to the identity automorphism of the bundle
$S|_U$. Also note that the second part of the proposition applies to
any open subset of the non-special locus $\cG\subset M$ which supports
a local orthonormal coframe of $(M,g)$.

\paragraph{Remark.} 
The sub-system \eqref{Vsys} can be obtained more directly as follows. 
An arbitrary norm one element $\xi$ of $\cK$ has the form:
\ben
\label{xiu}
\xi(u)=\cos\left (\frac{u}{2}\right) \xi_1+\sin\left(\frac{u}{2}\right) \xi_2~
\een
where $u \in \R$ is constant on $M$. This induces a function $b(u)\in
\cinf$ and a one-form $V(u)\in \Omega^1(M)$ given by:
\ben
\label{forms8alt}
b(u)=_U\cB(\xi(u), \gamma(\nu)\xi(u))~~,~~V(u)=_U\cB(\xi(u),\gamma_a\xi(u))e^a~~.
\een
Relation \eqref{xiu} gives:
\ben
\label{Fexp}
b(u)~=b_++b_-\cos u+b_3\sin u~~,~~V(u)=V_++V_-\cos u+V_3\sin u~~.
\een
Since $||\xi(u)||=1$, relation \eqref{Vbc} implies that the following
equality must hold for all $u$ (cf. \cite{MartelliSparks, g2,g2s}):
\ben
\label{Fierzu}
||V(u)||^2=1-b(u)^2~~.
\een
Substituting \eqref{Fexp} into \eqref{Fierzu}, we can separate the
Fourier components in $u$, using the fact that
$\{1,\cos(u),\sin(u)|n\in \N^\ast\}$ form an orthogonal basis of the
Hilbert space $\mathrm{L}^2(S^1)$ of (complex-valued) square
integrable functions on the circle. This leads to a system of
algebraic constraints for $b_r$ and $V_r$ which is equivalent with
\eqref{Fierzu}. Expanding in Fourier components, one finds after some
computation that \eqref{Fierzu} is equivalent with \eqref{Vsys}.

\section{Stratified spaces} 
\label{app:stratif}

We recall some basic notions from stratification theory in order to fix 
terminology. In this paper, a finite stratification of a topological space $X$ is
understood in the most general sense, i.e. as a finite partition of
$X$ into non-empty subsets called {\em strata}. We say that the stratification
is {\em connected} if all strata are connected. We let $\Sigma \subset
\cP(X)$ (where $\cP(X)$ is the power set of $X$) denote the set of all
strata, thus:
\be
X=\sqcup_{S\in \Sigma}S~~.
\ee

\subsection{Incidence poset of a stratification}

Consider the partial order relation $\leq$ defined on
$\Sigma$ through:
\be
S'\leq S ~~\mathrm{iff}~~S'\subseteq \overline{S}~~.
\ee
Then $(\Sigma,\leq)$ is a finite poset called the {\em incidence
  poset} of the stratification.  We let $<$ denote the transitive
binary relation defined on $\Sigma$ through:
\be
S'<S ~~\mathrm{iff}~~S'\leq S~\mathrm{and}~S'\neq S
\ee
i.e.:
\be
S'<S~~\mathrm{iff}~~S'\subseteq \fr(S)~~,
\ee
where $\fr(S)$ denotes the {\em small} frontier of $S$ (see Appendix
\ref{app:notations}).  For any $S\in \Sigma$, let $\cC(S)$ denote the
strict lower set of $S$:
\be
\cC(S)\eqdef \{S'\in \Sigma|S'<S\}=\{S'\in \Sigma|S'\subseteq \fr S\}~.
\ee
For all $S\in \Sigma$, we have the obvious inclusion: 
\ben
\label{frineq}
\sqcup_{S'\in \cC(S)} S'\subseteq \fr(S)~~.
\een 

\subsection{The adjointness relation}

We say that a stratum $S'$ adjoins a stratum $S$ (and write
$S'\trianglelefteq S$) if the intersection $S'\cap {\bar S}$ is
non-empty. This defines a reflexive (but generally non-transitive)
binary relation on $\Sigma$.  We say that $S'$ {\em strictly adjoins}
$S$ (and write $S'\triangleleft S$) if $S'\trianglelefteq S$ and
$S'\neq S$ i.e. if $S'$ intersects $\fr S$. We have:
\be
\fr S\subseteq \sqcup_{S'\triangleleft S}{S'}~~,~~\forall S\in \Sigma~~
\ee
and: 
\ben
\label{Phiineq}
\cC(S)\subseteq \{S'\in \Sigma|S'\triangleleft S\}~~.
\een

\subsection{The frontier condition}
We say that the stratification satisfies the {\em frontier condition}
if the {\em small} frontier of each stratum is a union of strata. This
amounts to the requirement that equality is always realized in
\eqref{frineq}:
\be 
\fr(S)=\sqcup_{S'\in  \cC(S)}S'~~,~~\forall S\in \Sigma~~
\ee
and with the condition that equality is realized in \eqref{Phiineq}.
This happens iff the binary relations $<$ and $\triangleleft$
coincide, in which case $\leq$ and $\trianglelefteq$ also coincide i.e. iff 
$S'\cap {\bar S}\neq \emptyset$ implies $S'\subseteq {\bar S}$. 
When the frontier condition is satisfied, the small frontier of any
stratum can be determined immediately by looking at the Hasse diagram
of the incidence poset of the stratification.
\subsection{Refinements and coarsenings}

We say that a stratification $\Sigma'$ is a {\em refinement} of
$\Sigma$ if any stratum of $\Sigma$ is a union of strata of
$\Sigma'$. In this case, we also say that $\Sigma$ is a {\em
  coarsening} of $\Sigma'$. The {\em connected refinement} of
$\Sigma$ is the refinement whose strata are the connected components
of the strata of $\Sigma$; it is the coarsest connected stratification 
which is a refinement of $\Sigma$. We say that two stratifications $\Sigma$ and 
$\Sigma'$ {\em agree} if one of them is a refinement of the other. 

\section{The semipositivity conditions for $G$}
\label{app:proofs}

Consider the Gram matrix \eqref{G}. We use the notation $G_{[ij|ij]}$
for the $2$ by $2$ submatrix of $G$ obtained by keeping only the
$i$-th and $j$-th rows and columns of $G$, where $1\leq i<j\leq 3$. By
Sylvester's criterion:
\begin{itemize}
\itemsep 0.0em
\item $G$ is positive semidefinite, iff  each of its principal (unsigned)
  minors:
\be
\det G~,~\det G_{[12|12]}~,~\det G_{[23|23]}~,~\det G_{[13|13]}~,~G_{11}~,~G_{22}~,~G_{33}
\ee
is non-negative. 
\item $G$ is positive definite iff  each of its {\em leading}
  principal minors $\det G$, $\det G_{[12|12]}$ and $G_{11}$ is
  positive; in this case, the non-leading principal minors are
  automatically positive.
\end{itemize}

\paragraph{Remark.} 
When $G$ is positive semidefinite, Kosteljanski's inequality \cite{HornJohnson} gives:
\be
\det G[I\cup J]\det G[I\cap J]\leq \det G[I]\det G[J]~~,
\ee
where $G[I]$ denotes the unsigned principal minor defined by keeping
only those rows and columns of $G$ indexed by elements of the subset
$I$ of the set $\{1,2,3\}$.  For $I\cap J=\emptyset$, this reduces to
Fisher's inequality:
\be
\det G[I\cup J] \leq \det G[I]\det G[J]~~\mathrm{when} ~~I\cap J=\emptyset~~,
\ee
which gives:
\beqan
\label{Fischer}
&&\det G\leq \min (G_{11}\det G_{[23|23]}, G_{22}\det G_{[13|13]}, G_{33}\det G_{[12|12]})\\
&&\det G_{[12|12]}\leq G_{11}G_{22}~,~\det G_{[13|13]}\leq G_{11}G_{33}~,~\det G_{[23|23]}\leq G_{22}G_{33}\nn~~.
\eeqan
To study Sylvester's conditions, we start by computing the
determinants of the various submatrices of $G$.  Consider the
polynomial \eqref{P}, which we reproduce here for convenience:
\ben
P(b,\beta)=\beta^4-\beta^2(1+\rho^2-b_+^2)+\rho^2~~.
\een
Notice that: 
\ben
\label{Prho}
P(b,\rho)=b_+^2\rho^2~~.
\een
Direct computation gives: 
\beqan
\label{minors}
&&G_{11}= 1-\beta^2-b_+^2~~,~G_{22}=~\beta^2-b_-^2~~,~~G_{33}=\beta^2-b_3^2\nn\\
&&\det G_{[12|12]}=-P(b_+,b_-,0,\beta)~~,~~\det G_{[13|13]}=-P(b_+,0,b_3,\beta)\\
&&\det G_{[23|23]}=\beta^2\left[\beta^2-(b_-^2+b_3^2)\right]~,~\det G=-\beta^2 P(b_+,b_-,b_3,\beta)~~.\nn
\eeqan
When viewing $P$ as a quadratic polynomial in $\beta^2$, its
discriminant equals the function $h(b_+,\rho)$ defined in
\eqref{hdef}.

\paragraph{Proposition.} 
We have $h(b)\geq 0$ for $b\in \cR$, with equality iff $b\in \partial
\cR$.

\

\noindent{\bf Proof.} The statement follows by noticing that:
\be
h(b_+,\rho)=[(1+b_+)^2-\rho^2][(1-b_+)^2-\rho^2]=[(1+|b_+|)^2-\rho^2][(1-|b_+|)^2-\rho^2]~~
\ee
and using the fact that $b\in \cR$ implies $(b_+,\rho)\in \Delta$,
which in turn means that $|b_+|\leq 1$ and $\rho\leq
1-|b_+|$. $\blacksquare$

\

\noindent It follows that for any $b\in \cR$ we can factorize $P(b,\beta)$ as:
\ben
\label{factorization}
P(b,\beta)=(\beta^2-f_+(b))(\beta^2-f_-(b))~~,
\een
where $f_\pm(b)$ are given in \eqref{fpm}. This allows us to write:
\beqan
\label{minorsfact}
 -\det G_{[12|12]} &=& (\beta^2-f_+(b_+,b_-,0))(\beta^2-f_-(b_+,b_-,0))\nn\\
 \det G_{[23|23]} &=& \beta^2(\beta^2-\rho^2)\\
 -\det G_{[13|13]} &=& (\beta^2-f_+(b_+,0,b_3))(\beta^2-f_-(b_+,0,b_3))~~.\nn
\eeqan

\subsection{Proof of Theorem 2}

Theorem 2 is an immediate consequence of Lemmas A, B and C proved below. 

\paragraph{Proposition.} 
The following inequality holds for $b\in \cR$:
\ben
\label{ineq0}
\sqrt{h(b_+,\rho)}\leq 1-b_+^2-\rho^2~~,
\een
with equality iff $b_+\rho=0$. 

\

\noindent{\bf Proof.} 
For $b\in \cR$, we have $(b_+,\rho)\in \Delta$ and hence $\rho\leq
1-|b_+|$, which implies $\rho^2\leq (1-|b_+|)^2\leq
(1-|b_+|)(1+|b_+|)=1-b_+^2$. Hence the right hand side of
\eqref{ineq0} is non-negative for $b\in \cR$. It follows that
\eqref{ineq0} is equivalent with the inequality obtained by squaring
both of its sides, which can be seen by direct computation to be
equivalent with $4b_+^2\rho^2\geq 0$. $\blacksquare$

\paragraph{Proposition.} For $b\in \cR$, we have:
\ben
\label{ineq1}
\rho^2\leq f_-(b_+,\rho)\leq f_+(b_+,\rho)\leq 1-b_+^2~~.
\een
The first and third inequalities in \eqref{ineq1} are both strict
unless $b_+\rho=0$, in which case both of them become equalities.  In
particular, we have $J(b)\subset [\rho, \sqrt{1-b_+^2}]$, where the
interval $J(b)$ was defined in \eqref{Jdef}.

\

\noindent{\bf Proof.} 
The middle inequality is obvious, while the first and third
inequalities are both equivalent with \eqref{ineq0}. The other
statements follow immediately. $\blacksquare$

\paragraph{Remark.} For $(b,\beta)\in \fP$, we have:
\begin{enumerate}
\itemsep 0.0em
\item $G_{11}=0$ (i.e. $||V_+||=0$) iff one of the following holds:
\begin{itemize}
\itemsep 0.0em
\item $\beta=1$ or
\item $b_+=b_-=0$ and $\beta=\sqrt{1-b_+^2}$, which requires
  $||V_-||=||V_3||=\sqrt{1-b_+^2}$ and $\langle V_-,V_3\rangle=0$
\end{itemize}
\item $\det G_{[23|23]}=0$ (i.e. $V_-$ and $V_3$ are linearly dependent) iff one of the following holds:
\begin{itemize}
\itemsep 0.0em
\item $\beta=\rho=0$ and hence $V_-=V_3=0$ and $||V_+||=\sqrt{1-b_+^2}$ or
\item $b_+=0$ and $\beta=\rho=\sqrt{b_-^2+b_3^2}$, which requires
  $||V_+||^2=1-b_-^2-b_3^2$, $||V_-||=|b_3|$, $||V_3||=|b_-|$,
  $V_+\perp ( V_-,V_3) $ and $\langle
  V_-,V_3\rangle=-b_-b_3$~.
\end{itemize}
\end{enumerate}

\paragraph{Lemma A.} 
Let $b\in \cR$. Then the condition $\det G(b,\beta)\geq 0$ is
equivalent with the condition that $B=(b,\beta)$ belong to the body $\fP$.
Furthermore, this condition implies that $G_{11}$, $G_{22}$, $G_{33}$
and $\det G_{[23|23]}$ are non-negative.

\

\noindent{\bf Proof.} 
Equation \eqref{factorization} shows that condition $\det G\geq 0$ is
equivalent with\footnote{The case $\beta=0$ requires $\rho=0$, which
  gives $f_+(b_+,0)=1-b_+^2$ and $f_-(b_+,0)=0$, in which case
  \eqref{betacond} is satisfied.}:
\ben
\label{betacond}
f_-(b)\leq \beta^2\leq f_+(b)~~\mathrm{i.e.}~~\beta\in J(b)~~,
\een
which is equivalent with $(b_+,\beta)\in \fP$ (see \eqref{Pdef}).  By
\eqref{ineq1}, this implies $\rho^2\leq \beta^2\leq 1-b_+^2$, which
upon using \eqref{minorsfact} implies that $G_{ii}$ and $\det
G_{[23|23]}$ are non-negative. $\blacksquare$

\paragraph{Proposition.} 
For each fixed value of $b_+\in [-1,1]$, $f_-(b_+,\rho)$ is
monotonically increasing while $f_+(b_+,\rho)$ is monotonically
decreasing as a function of $\rho\in [0,1-|b_+|]$. Moreover:
\begin{itemize}
\itemsep 0.0em
\item $f_-(b_+,\rho)$ is strictly increasing as a function of $\rho\in
  (0,1-|b_+|)$ for any $b_+\in [-1,1]$
\item $f_+(b_+,\rho)$ is strictly decreasing as a function of $\rho\in
  (0,1-|b_+|)$ for any $b_+\in [-1,0)\cup (0,1]$ while $f_+(0,\rho)=1$
  for any $\rho\in [0,1]$.
\end{itemize}

\

\noindent{\bf Proof.} 
We have\footnote{These relations should be interpreted in a limiting
  sense for $\rho=1-|b_+|$.}:
\beqa
&&\frac{\partial f_-(b_+,\rho)}{\partial\rho}=\rho\frac{1+b_+^2-\rho^2+\sqrt{(1+b_+^2-\rho^2)^2-4b_+^2}}{\sqrt{(1+b_+^2-\rho^2)^2-4b_+^2}}\geq 0  \\
&&\frac{\partial f_+(b_+,\rho)}{\partial\rho}=-\rho\frac{1+b_+^2-\rho^2-\sqrt{(1+b_+^2-\rho^2)^2-4b_+^2}}{\sqrt{(1+b_+^2-\rho^2)^2-4b_+^2}}\leq 0~~,
\eeqa
where the inequalities follow using $\rho^2\leq 1$. The first
inequality is strict unless $\rho=0$ or $(b_+,\rho)=(0,1)$.  The
second inequality is strict unless $\rho=0$ or $b_+=0$. Notice that
$f_+(0,\rho)=1$ for $\rho\in [0,1]$. $\blacksquare$

\paragraph{Proposition.} 
For any $b\in \cR$, we have:
\beqan
\label{ineq2}
&& f_-(b_+,b_-,0)\leq f_-(b_+,b_-,b_3)\leq f_+(b_+,b_-,b_3)\leq f_+(b_+,b_-,0)~~\nn\\
&& f_-(b_+,0,b_3)\leq f_-(b_+,b_-,b_3)\leq f_+(b_+,b_-,b_3)\leq f_+(b_+,0,b_3)~~.
\eeqan
In particular, the condition $\beta^2\in
[f_-(b_+,\rho),f_+(b_+,\rho)]$ implies $\det G_{[12|12]}\geq 0$ and
$\det G_{[13|13]}\geq 0$. Furthermore, we have:
\begin{itemize}
\itemsep 0.0em
\item $\det G_{[12|12]}=0$ iff  $\beta=1$ or ($b_3=0$ and $\beta^2 \in \{f_-(b_+,b_-,0),f_+(b_+,b_-,0)\}$)
\item $\det G_{[13|13]}=0$ iff  $\beta=1$ or ($b_-=0$ and $\beta^2 \in \{f_-(b_+,0,b_3),f_+(b_+,0,b_3)\}$).
\end{itemize}

\noindent{\bf Proof.} 
Inequalities \eqref{ineq2} follow immediately from the Lemma. When
$\beta^2\in [f_-(b_+,\rho),f_+(b_+,\rho)]$, these inequalities imply
that $\beta$ lies between the two roots of $P( b_+,b_-,0;\beta))$ and
$P(b_+,0,b_3;\beta)$ (viewed as polynomials in $\beta^2$), which shows
that $\det G_{[12|12]}\geq 0$ and $\det G_{[13|13]}\geq 0$ (see
\eqref{minors}). The other statements follow from the strict
monotonicity properties listed in the lemma, recalling that $\beta=1$
requires $b_+=0$ . $\blacksquare$

\paragraph{Lemma B.} 
The determinants $\det G_{[12|12]}$ and $\det G_{[23|23]}$ are
non-negative for any $(b,\beta)\in \fP$.

\

\noindent{\bf Proof.} 
Follows immediately from the previous proposition upon recalling that
the body $\fP$ is a fibration over $\cR$ with fiber given by the
interval $J(b)$ defined in \eqref{Jdef}. $\blacksquare$

\paragraph{Remark.} 
Lemma B implies that we have $f_-(b_+,\rho)\geq 0$, with equality iff
$\rho=0$ and $|b_+|=1$.

\paragraph{Proposition.} 
For $(b,\beta)\in \fP$, the equality $\beta=\rho$ can be attained only
for $(b,\beta)\in \fI\cup {\bar \fA}$.

\paragraph{Proof.} 
For $(b,\beta)\in \fP$, we have $\beta\in J(b)$ and hence $\rho^2\leq
f_-(b)\leq \beta^2$ by \eqref{ineq1}.  Thus $\beta=\rho$ means that
equality is realized in the first inequality of \eqref{ineq1}, which
requires $b_+\rho=0$, i.e. $b_+=0$ or $\rho=0$. In the first case we
have $(b,\beta)\in {\bar \fA}$ while in the second case we have
$(b,\beta)\in \fI$. $\blacksquare$.

\paragraph{Lemma C.} 
Let $B=(b,\beta)\in \fP$. Then $\rk G(B)\leq 1$ iff $B \in \fI \sqcup
\partial \fD$. Furthermore, we have $\rk G=0$ iff $B\in \partial \fI$.

\

\noindent{\bf Proof.} 

\

\noindent (Necessity) The condition $\rk G(B)\leq 1$ requires that all
two by two minors of $G$ vanish. Relations \eqref{minorsfact} show
that $\det G_{[23|23]}=0$ implies $\beta=0$ or $\beta=\rho$.  In the
first case, the first row of \eqref{minors} and the conditions
$G_{22}\geq 0$ and $G_{33}\geq 0$ imply $\rho=0$, hence the first case
is contained in the second.  Thus we must have $\beta=\rho$ and the
Proposition gives $(b_+,\rho)\in \fI\cup {\bar \fA}$.  Consider the
case $(b_+,\rho)\in {\bar \fA}$, i.e. $b_+=0$. Substituting
$\beta=\rho$ and $b_+=0$ in \eqref{minors}, we find:
\ben
\label{minorsdisk}
\det G_{[12|12]}=b_3^2(1-\rho^2)~~,~~\det G_{[13|13]}=b_-^2(1-\rho^2)~~.
\een
Hence these two by two minors of $G(B)$ vanish simultaneously iff
$\rho=1$ or $\rho=0$, i.e.  iff $(b_+,\beta)$  belongs to
$\partial_0^0\fP\sqcup \partial \fD$. Since $\partial_0^0\fP$ is the
midpoint of $\fI$, we conclude that $\rk G(B)\leq 1$ requires
$(b_+,\rho)\in \fI\sqcup \partial \fD$.

\

\noindent (Sufficiency) For $B\in \fI$ (i.e. for $\rho=\beta=0$), we
have:
\be
G(B)=\left[\begin{array}{ccc} 1-b_+^2 & 0 & 0\\ 0 & 0 &0 \\ 0 & 0 & 0\end{array}\right]
\ee
and hence $\rk G\leq 1$. Notice that $\rk G=0$ iff $b_+=\pm 1$
i.e. iff $B\in \partial\fI$.  For $B\in \partial D$ (i.e. for $b_+=0$
and $\beta=\rho=1$), we have:
\be
G(B)=\left[\begin{array}{ccc} 0 & 0 & 0 \\ 0 & 1-b_-^2 & -b_-b_3 \\ 0 & -b_-b_3 & 1-b_3^2 \end{array}\right]=
\left[\begin{array}{ccc} 0 & 0 & 0 \\ 0 & b_3^2 & -b_-b_3 \\ 0 & -b_-b_3 & b_-^2 \end{array}\right]~~,
\ee
where in the second row we used the relation
$b_-^2+b_3^2=\rho^2=1$. Thus $\rk G\leq 1$, since the two by two minor
in the lower right corner has vanishing determinant. In this case, we
cannot have $\rk G=0$ (i.e. $G=0$) since $b_-^2+b_3^2=1$ and hence
$b_-$ and $b_3$ cannot vanish simultaneously. $\blacksquare$

\paragraph{Proof of Theorem 2.}

The following result follows by combining Lemmas A, B and C:

\paragraph{Theorem 2'.} 
Let $b\in \cR$. Then the matrix $G(b,\beta)$ is semipositive iff
$\beta\in J(b)$, i.e. iff $B\eqdef (b,\beta)\in \fP$.  It is strictly
positive iff $B\in \Int \fP$.  In particular, we have $\rk G(B)< 3$ at
a point $p\in M$ iff $B(p)\in \partial \fP$. When $B\in \partial\fP$,
we have:
\begin{itemize}
\itemsep 0.0em
\item $\rk G(B)=0$ iff $B\in \partial_0^+\fP\sqcup \partial_0^-\fP=\partial \fI$  
\item $\rk G(B)=1$ iff $B\in \partial_0^0\fP\sqcup \partial_1\fP=\partial\fD\sqcup \Int \fI$
\item $\rk G(B)=2$ iff $B\in \partial_2 \fP\cup \partial_3 \fP=\Int \fD\sqcup \fA\sqcup \Int \fC^+\sqcup \Int \fC^-$. 
\end{itemize}

\

\noindent We know from Subsection \ref{sec:b} that $\im b\subset
\cR$. Combining this with Theorem 2', we find that the image of $B$ is
contained in $\fP$.  Theorem 2 now follows immediately.

\subsection{Proof of Theorem 3} 
\label{app:deg}

Theorem 3 is an immediate consequence of Lemma D proved below. 

\paragraph{Lemma D.} Let $p\in M$. Then:
\begin{enumerate}
\itemsep 0.0em
\item The value $\beta(p)=0$ is attained iff $B(p)\in \fI$. At such points,
  we have $b_-(p)=b_3(p)=0$, $V_-(p)=V_3(p)=0$ and $||V_+(p)||=\sqrt{1-b_+(p)^2}$, thus
  $\cD(p)$ has dimension seven or eight, according to whether $|b_+|<1$ or $|b_+|=1$.
\item The value $\beta(p)=1$ is attained iff $B(p)\in \fD$. At such points,
  we have $V_+(p)=0$, $\det G_{[12|12]}(p)=\det G_{[13|13]}(p)=0$ and:
\be
||V_-(p)||=\sqrt{1-b_-(p)^2}~~,~~||V_3(p)||=\sqrt{1-b_3(p)^2}~~,~~\langle V_-(p),V_3(p)\rangle =-b_-(p)b_3(p)~~.
\ee
The space $\cD(p)$ has dimension six when $B(p)\in \Int \fD$ and dimension seven when
$B(p)\in \partial \fD$.
\item When $B(p)\in \partial \fD$ (i.e. when $\beta(p)=\rho(p)=1$), we have
  $b_+(p)=0$~,~$V_+(p)=0$,
\be
V_-(p)=(\sin \psi) v~,~V_3(p)=-(\cos\psi) v~,~b_-(p)=\cos\psi~,~b_3(p)=\sin\psi
\ee
and $\langle V_+(p),V_-(p),V_3(p)\rangle=\langle v\rangle$, where $\psi\in
[0,2\pi)$ and $v\in T_p^\ast M$ is an arbitrary 1-form of norm one. 
\item When $B(p)\in {\bar \fA}$ (i.e. when $\beta(p)=\rho(p)$), we have $\det
  G_{[23|23]}(p)=0$ and $||V_+(p)||=\sqrt{1-\rho(p)^2}$, $V_-(p)=(\rho(p)
  \sin\psi) v$, $V_3(p)=-(\rho(p) \cos \psi) v$ with $\psi\in [0,2\pi)$ and
    $v\in T_p^\ast M$ an arbitrary 1-form of unit norm such that $V_+(p)\perp v$. The
    space $\cD(p)$ has dimension six when $B(p)\in \fA$ and dimension seven when
    $B(p)\in \fr\fA=\partial_0^0\fP\sqcup \partial \fD$.
\end{enumerate}

\noindent{\bf Proof.}  Inequalities \eqref{ineq1} imply that $\beta=0$
can be attained only at $\rho=0$, i.e. only for $B(p)\in \fI$. They also
imply that $\beta=1$ can be attained only at $b_+=0$, i.e. only for
$B(p)\in {\bar \fA}$. The remaining statements follow immediately using 
the system \eqref{Vsys}. $\blacksquare$

\subsection{Solving for $b_r$ in terms of $V_r$}

Notice that $G_{12}G_{23}G_{13}=-(b_+b_-b_3)^2$, so the condition
$b_+b_-b_3\neq 0$ amounts to the requirement that no two of the
vectors $V_r$ are orthogonal. In this case, we have:
\ben
\label{bsol}
b_+=\epsilon \frac{\sqrt{-G_{12}G_{23}G_{13}}}{G_{23}}~~,~~b_-=\epsilon \frac{\sqrt{-G_{12}G_{23}G_{13}}}{G_{13}}~~,~~b_3=\epsilon \frac{\sqrt{-G_{12}G_{23}G_{13}}}{G_{12}}
\een
where $\epsilon\in \{-1,1\}$ and hence \eqref{Vsys} can be solved for
$b_r$ iff the following conditions are satisfied:
\ben
\label{Vcond}
0\leq 1-G_{11}+\frac{G_{12}G_{13}}{G_{23}}=G_{22}-\frac{G_{12}G_{23}}{G_{13}}=G_{33}-\frac{G_{13}G_{23}}{G_{12}}(=\beta^2)~~.
\een
Conditions \eqref{Vcond} show that the triples of vectors allowed by
\eqref{Vsys} are constrained.

\section{The rank of ${\hat G}$}
\label{app:hatG}

Direct computation using \eqref{Gext} gives: 
\ben
\label{dethG}
\det\hG(b,\beta)=P(b,\beta)^2~~,
\een
The determinants of the 3 by 3 principal minors of $\hG$ are given by:
\beqan
\label{3minorshG}
&& \det{{\hat G}_{[123|123]}}=\det G=-\beta^2P(b,\beta)~~,\nn\\
&& \det{{\hat G}_{[124|124]}}=-(1-b_+^2-\beta^2+b_-^2)P(b,\beta)~~,\\
&& \det{{\hat G}_{[134|134]}}=-(1-b_+^2-\beta^2+b_3^2)P(b,\beta)~~,\nn\\
&& \det{{\hat G}_{[234|234]}}=(\rho^2-\beta^2)P(b,\beta)~~,\nn
\eeqan
where $P(b,\beta)$ was defined in \eqref{P}:
\be
P(b,\beta)=(1-\beta^2)(\rho^2-\beta^2)+\beta^2b_+^2=-\beta^2(1-b_+^2-\beta^2+\rho^2)+\rho^2~~,
\ee
while the determinants of the 2 by 2 principal minors are:
\beqan
\label{2minorshG}
&& \det {\hat G}_{[12|12]}=\det G_{[12|12]}=\beta^2(1-b_+^2-\beta^2+b_-^2)-b_-^2=-P(b,\beta)+b_3^2(1-\beta^2)~~,\nn\\
&& \det \hat G_{[13|13]}=\det G_{[13|13]}=\beta^2(1-b_+^2-\beta^2+b_3^2)-b_3^2=-P(b,\beta)+b_-^2(1-\beta^2)~~,\nn\\
&& \det \hat G_{[14|14]}=(1-b_+^2-\beta^2)(1-b_+^2-\beta^2+\rho^2)~~,\nn\\
&& \det \hat G_{[23|23]}=\det G_{[23|23]}=\beta^2(\beta^2-\rho^2)~~,\\
&& \det \hat G_{[24|24]}=(1-b_+^2-\beta^2+\rho^2)(\beta^2-b_-^2) -b_3^2~~,\nn\\
&& \det \hat G_{[34|34]}=(1-b_+^2-\beta^2+\rho^2)(\beta^2-b_3^2) -b_-^2~~.\nn
\eeqan

\paragraph{Lemma.} 
The rank of $\hG(B)$ is given as follows:
\begin{enumerate}
\itemsep 0.0em
\item For $B\in \Int \fP$, we have $\rk\hG(B)=4$.
\item For $B\in \Int \fI\sqcup \Int \fD\sqcup \fA\sqcup \Int
  \fC^+\sqcup \Int \fC^-$, we have $\rk\hG(B)=2$.
\item For $B\in \partial \fD$, we have $\rk \hG(B)=1$.
\item For $B\in \partial \fI$, we have $\rk\hG(B)=0$.
\end{enumerate}

\noindent{\bf Proof.} 
Since $P(B)$ vanishes iff $B\in \fP$, relation \eqref{dethG} implies
that $\hG$ is non-degenerate on $\Int \fP$ and degenerate on $\partial
\fP$. In particular, we have $\rk\hG(B)=4$ for $B\in \Int \fP$.  For
$B\in \partial \fP$, we have $P(B)=0$ and hence $\det
\hG=0$. Furthermore, all 3 by 3 minors of $\hG$ vanish by relations
\eqref{3minorshG}. We distinguish the cases:
\begin{itemize}
\itemsep 0.0em
\item $B\in \fI$. Then $\beta=\rho=0$ and $\rk G(B)\leq 1$, thus $\det
  \hat G_{[12|12]}=\det \hat G_{[13|13]}=\det \hat
  G_{[23|23]}=0$. Relations \eqref{2minorshG} give:
\be
 \det \hat G_{[14|14]}=(1-b_+^2)^2~~,~~\det \hat G_{[24|24]}=\det \hat G_{[34|34]}=0~~~,
\ee
which show that $\rk \hG(B)=2$ for $B\in \Int \fI$. The case $B\in
\partial \fI=\partial_0^+\fP\sqcup \partial_0^-\fP$ corresponds to
$\rho=\beta=b_-=b_3=0$ with $b_+^2=1$.  For these values, \eqref{Gext}
gives $\hG(B)=0$ and hence $\rk \hG(B)=0$.
\item $B\in \partial \fD$, i.e. $b_+=0$ and $\beta=\rho=1$. Then
  \eqref{2minorshG} shows that all 2 by 2 minors of $\hG$ vanish while
  \eqref{Gext} shows that $\hG\neq 0$, which means that we must have
  $\rk \hG(B)=1$.
\item $B\in \Int \fD$, i.e. $b_+=0$, $\beta=1$ and $\rho\in
  [0,1)$. Then \eqref{2minorshG} gives $\det \hG_{[23|23]}=1-\rho^2>0$
    and hence $\rk \hG(B)=2$.
\item $B\in \fA$, i.e. $b_+=0$ and $\beta=\rho\in (0,1)$. Then
  \eqref{2minorshG} gives $\det \hG_{[14|14]}=1-\rho^2>0$ and hence
  $\rk \hG(B)=2$.
\item $B\in \Int \fC^+\sqcup \Int \fC^-$, i.e. $P(b,\beta)=0$ with
  $b_+=\pm g(\rho,\beta)$ and $0\leq \rho <\beta <1$, where $g$ is the
  function defined in \eqref{gdef}. Then $\det \hG_{[23|23]}=\det
  G_{[23|23]}>0$ and hence $\rk \hG(B)=2$.
\end{itemize}
The Lemma follows by combining these results. $\blacksquare$

\paragraph{Proposition.} 
For $p\in \cG$, we have $\dim\cD_0(p)\in \{4,6\}$.

\

\noindent{\bf Proof.} 
Follows immediately from the Lemma upon noticing that $b(\cG)\subset
\Int \cR$ while $\pi(\partial \fD),\pi(\partial \fI)\\
\subset \partial
\cR$.  $\blacksquare$

\section{On certain deformations of  $(\xi_1,\xi_2)$}
\label{app:def}

\subsection{A family of special deformations}

Consider a locally non-degenerate and $\cB$-compatible two-dimensional
subspace $\cK\subset \Gamma(M,S)$ and let $(\xi_1,\xi_2)$ be an
orthonormal basis of $\cK$. Thus $\xi_1(p)$ and $\xi_2(p)$ form an
orthonormal system of Majorana spinors for any $p\in M$. Let $\cG$
denote the non-special locus of $\cK$, i.e. the set consisting of those
points $p\in M$ such that the positive chirality components
$\xi_1^+(p)$ and $\xi_2^+(p)$ are linearly independent and such that
the same holds for the negative chirality components $\xi_1^-(p)$
and $\xi_2^-(p)$.

Consider the special class of deformations of the pair $(\xi_1,\xi_2)$
to another pair of Majorana spinors $(\txi_1, \txi_2)$ such that only
$\xi_1^-$ changes:
\ben
\label{tconds}
\txi_2=\xi_2~~\mathrm{and}~~\txi_1^+=\xi_1^+~~.
\een
Recall that $\xi_1^\pm$ and $\xi_2^\pm$ generate the chiral
projections $K_\pm$ of the spinor sub-bundle $K$ associated to
$S$. Under a special deformation obeying \eqref{tconds}, the positive
chirality projection is invariant while the negative chirality
projection may change:
\be
{\tilde K}_+=K_+~~,~~K_-\rightarrow {\tilde K}_-~~.
\ee
As a result, the bundle $K$ changes to ${\tilde K}$ and the space
$\cK$ changes to the space ${\tilde \cK}=\R\txi_1+\R\txi_2\subset
\Gamma(M,S)$.  We require that the system $(\txi_1,\txi_2)$ is
everywhere orthonormal, so that ${\tilde \cK}$ is again a
two-dimensional and $\cB$-compatible locally-nondegenerate subspace of
$\Gamma(M,S)$.

For the remainder of this appendix, consider two Majorana spinors
$\txi_1,\txi_2\in \Gamma(M,S)$ which satisfy \eqref{tconds} and are
everywhere orthonormal.  Let $\tb_1,\tb_2,\tb_3$ and
$\tV_1,\tV_2,\tV_3, \tW$ denote the zero- and one-forms defined by the
spinors $\txi_1,\txi_2$ according to relations \eqref{forms0} and
\eqref{forms1} and $\tb_\pm,\tV_3^\pm$ denote the associated
quantities defined as in Section \ref{sec:cD}. Let $\tbeta\in
\cC^\infty(M,\R^+)$ denote the function defined according to
\eqref{beta_def}. Notice that $\txi_1^-$ has the form:
\ben
\label{txim}
\txi_1^- =\alpha_1\xi_1^-+\alpha_2\xi_2^- +\zeta\in \Gamma(M,S^-)~~,
\een
where $\alpha_1,\alpha_2\in \cinf$ and $\zeta\in 
\Gamma(M,S^-)$ is the projection of $\txi_1$ onto the
$\cB$-orthocomplement of $K^-$ inside $S^-$. Hence $\zeta$ is a
section of $S^-$ which is everywhere orthogonal to $K_-$ and whose
norm we shall denote by:
\ben
\label{lambda_def}
\lambda\eqdef ||\zeta||~~.
\een 
Recall that $b(p)\in \Int\cR$ for any $p\in \cG$. 

\paragraph{Lemma.} 
The following inequalities hold for any point $p\in \cG$: 
\ben
\label{pos}
|b_1(p)|<1~~,~~|b_2(p)|<1~~,~~\rho(p)<1-|b_+(p)|
\een

\

\noindent{\bf Proof.} 
For any point $p\in \cG$, we have $b(p)\in \Int \cR$ and hence
$\rho(p) <1-|b_+(p)|\leq 1-b_+(p)$, which shows that $\det A(p)>0$.
On the other hand, the planes $b_1=\pm 1\leftrightarrow b_++b_-=1$ and
$b_2=\pm 1\leftrightarrow b_+-b_-=\pm 1$ in the space $\R^3$ with
coordinates $b_+,b_-,b_3$ intersect the body $\cR$ along two segments
which lie within $\partial \cR$ and hence we have $|b_1(p)|<1$ and
$|b_2(p)|<1$. $\blacksquare$

\paragraph{Proposition} 
We have $\tb_i=b_i$ for all $i=1,2,3$ and hence $\tb_\pm=b_\pm$. On the
locus $\cG$, we have:
\ben
\label{alpha1ineq}
|\alpha_1|\leq_\cG 1~~
\een
and:
\ben
\label{alpha2}
\alpha_2=_\cG\frac{b_3}{1-b_2}(\alpha_1-1)
\een
Furthermore, the norm of $\zeta$ has the following form on the locus $\cG$:
\ben
\label{lambda}
\lambda\eqdef ||\zeta||=_\cG\lambda_M\sqrt{1-\alpha_1^2}~~.
\een
where:
\ben
\label{lambdaM}
\lambda_M\eqdef \sqrt{ \frac{(1-b_+)^2-\rho^2}{2(1+b_- - b_+)}}=\sqrt{\frac{(1-b_1)(1-b_2)-b_3^2}{2(1-b_2)}}\in \cC^\infty(\cG,\R)~~.
\een


\

\noindent{\bf Proof.} 
Consider the scalars \eqref{forms0} defined by the orthonormal
Majorana spinors $\txi_1(p)$ and $\xi_2(p)$, namely $\tb_1\eqdef
\cB(\txi_1,\gamma(\nu)\txi_1)~~,~~\tb_3\eqdef
\cB(\txi_1,\gamma(\nu)\xi_2)$ and $\tb_2\eqdef
\cB(\xi_2,\gamma(\nu)\xi_2)=b_2$. Since \eqref{xib} hold for
$\txi_1,\xi_2$ and $\tb_r$ and since the positive chirality components
of $\txi_1$ and $\xi_1$ coincide, we find
$\tb_3=2\cB(\xi_1^+,\xi_2^+)=b_3$ and
$\tb_1=2||\xi_1^+||^2-1=b_1$. Thus $\tb_i=b_i$ for all $i=1,2,3$.

It is clear that $\txi_1^-|_\cG$ has the form \eqref{txim}, 
where $\zeta=(\id_{S^-}-P_-)\txi_1^-$ is the projection of
$\txi_1^-|_\cG$ onto the orthocomplement of $K^-|_\cG$ inside
$S^-|_\cG$. Since $\zeta$ is $\cB_p$-orthonormal on
$\xi_1^-(p)$ and $\xi_2^-(p)$, we have:
\beqa
&&||\txi_1^-||^2=||\alpha_1\xi_1^-+\alpha_2\xi_2^-||^2+ \lambda^2~~,\nn\\
&&\cB(\txi_1^-,\xi_2^-)=\alpha_1\cB(\xi_1^-,\xi_2^-)+\alpha_2||\xi_2^-||^2~~,
\eeqa
where we set $\lambda\eqdef ||\zeta||$. 
Since $(\xi_1,\xi_2)$ is $\cB_p$-orthonormal and since $\txi_1^+=\xi_1^+$,
the condition that $(\txi_1,\xi_2)$ be orthonormal amounts to the constraints:
\beqa
&&||\txi_1^-||^2=||\xi_1^-||^2~(=1-||\xi_1^+||^2)~~,\nn\\
&&\cB_p(\txi_1^-,\xi_2^-)=\cB_p(\xi_1^-,\xi_2^-)~ (=-\cB_p(\xi_1^+,\xi_2^+))~~,
\eeqa
which upon using \eqref{xib} gives the system:
\beqan
\label{alphasys}
&& (1-b_1)\alpha_1^2+(1-b_2)\alpha_2^2-2b_3\alpha_1\alpha_2=1-b_1-2\lambda^2\nn\\
&& b_3 (1-\alpha_1)+(1-b_2)\alpha_2=0~~.
\eeqan
The left hand side of the first equation defines the quadratic form
$\alpha^TA(p)\alpha$, where $A$ is the symmetric matrix-valued function:
\be
A\eqdef \left[\begin{array}{cc} 1-b_1 & -b_3\\ -b_3 & 1-b_2\end{array}\right]~~,
\ee
whose determinant equals:
\be
\det A=(1-b_1)(1-b_2)-b_3^2=(1-b_+)^2-\rho^2~~.
\ee
The inequalities $|b_1|\leq 1$, $|b_2|\leq 1$ and $\rho\leq 1-|b_+|$
imply that $A(p)$ is a semi-positive matrix for any $p\in M$ while
\eqref{pos} imply that $A(p)$ is strictly positive for $p\in \cG$.
The eigenvalues $a_-,a_+$ of $A$ are given by: 
\be
a_\pm=1-b_+\pm \rho~~.
\ee
Since $A$ is semipositive on $M$, we have $\alpha^TA\alpha\geq 0$,
which shows that the first equation in \eqref{alphasys} has solutions
iff the right hand side is non-negative, i.e. only for $\lambda\leq
\lambda_0$, where $\lambda_0\eqdef \sqrt{\frac{1-b_1}{2}}\in \cinf$.
For any $\lambda(p)\leq \lambda_0(p)$ in this interval, the first
equation of \eqref{alphasys} considered at the point $p\in M$ defines
an ellipse $E_\lambda(p)$ in the $\alpha(p)$-plane, whose half-axes
have length $\frac{1}{\sqrt{a_\pm(p)}}$. This ellipse degenerates to a
single point (namely the origin $\alpha_1(p)=\alpha_2(p)=0$) for
$\lambda(p)=\lambda_0(p)$. For $b_2(p)\neq 1$, the second equation in
\eqref{alphasys} (considered at $p$) defines a line in the
$\alpha(p)$-plane which passes through the points $(1,0)$ and
$(0,-\frac{b_3(p)}{1-b_2(p)})$. This equation implies
$\alpha_2(1-b_2)=b_3(\alpha_1-1)$, which combines with the first
relation of \eqref{alphasys} to give:
\be
2\lambda^2(1-b_2)=(1-\alpha_1^2)[(1-b_+)^2-\rho^2]~~.
\ee
Since the left hand side is non-negative and since $\rho^2<(1-b_+)^2$
on the locus $\cG$, this implies \eqref{alpha1ineq}. Provided that
\eqref{alpha1ineq} is satisfied, we can solve \eqref{alphasys} in
terms of $\alpha_1$. This gives \eqref{alpha2} and \eqref{lambda},
with $\lambda_M$ is as in \eqref{lambdaM}. The second equation in
\eqref{alphasys} shows that solutions of \eqref{alphasys} exist only
for $\lambda\leq \lambda_M$. $\blacksquare$

\paragraph{Proposition.} 
The 1-forms defined by $\txi_1$ and $\txi_2=\xi_2$ are given by:
\beqan
\label{tVtW}
&& \tV_1=\alpha_1V_1+\alpha_2V_3-\alpha_2 W+2 U_1\nn\\
&&\tV_2=V_2~\\
&& \tV_3=\tV_3^++V_3^-=\frac{1}{2}\alpha_2V_2+\frac{1}{2}(1+\alpha_1)V_3-\frac{1}{2}(1-\alpha_1)W+ U_2~~\nn\\
&& \tW=\tV_3^+-V_3^-=\frac{1}{2}\alpha_2V_2-\frac{1}{2}(1-\alpha_1)V_3+\frac{1}{2}(1+\alpha_1)W+U_2~~,\nn
\eeqan
where:
\ben
\label{Udef}
U_i\eqdef \cB(\zeta,\gamma_a\xi_i^+)e^a\in \Omega^1(M)~~(i=1,2)
\een
and $\tV_3^-=V_3^-$. 

\

\noindent{\bf Proof.}
Using \eqref{VWspinors}, we find that the following relations hold on
$M$:
\beqan
\label{tV}
&& \tV_1\eqdef 2\cB(\txi_1^-,\gamma_a\xi_1^+)e^a=\alpha_1V_1+2\alpha_2V_3^-+2U_1\nn\\
&& \tV_3^+\eqdef \cB(\txi_1^-,\gamma_a\xi_2^+)e^a=\alpha_1 V_3^++\frac{1}{2}\alpha_2 V_2+U_2\\
&& \tV_2=V_2~~,~~\tV_3^-=V_3^-~~.\nn
\eeqan
Recall that $\tV_\pm\eqdef \frac{1}{2}(\tV_1\pm \tV_2)$ and $\tW\eqdef
\tV_3^+-\tV_3^-=\cB(\txi_1,\gamma_a\gamma(\nu)\xi_2)$. Equations
\eqref{tV} give \eqref{tVtW}, where we used
\eqref{forms1pm3}. $\blacksquare$

\

\noindent Consider the following open subset of $\cG$:
\be
\cG_0\eqdef \{p\in \cG|\zeta(p)\neq 0\}=\{p\in \cG|\lambda(p)\neq 0\}=\{p\in \cG|\alpha_1(p)\neq \pm 1\}~~.
\ee

\

\paragraph{Proposition.} 
We have $U_1(p)\neq 0$ and $U_2(p)\neq 0$ at any point $p\in \cG_0$.

\

\noindent{\bf Proof.}  Since $p\in \cG$, the spinors $\xi_1^+(p)$ and
$\xi_2^-(p)$ are linearly independent and in particular
non-vanishing. It was shown in \cite[Section 2.6]{g2s} that, for any
non-vanishing spinor $\eta\in S_p^+\setminus \{0\}$, the spinors
$(\gamma_a\eta)_{a=1\ldots 8}$ form a basis of $S_p^-$. Thus
$(\gamma_a\xi_1^+(p))_{a=1\ldots 8}$ is a basis of $S_p^-$ and the
same is true for $(\gamma_a\xi_2^+(p))_{a=1\ldots 8}$. Since
$\zeta(p)$ is non-zero, this gives the conclusion. $\blacksquare$

\paragraph{Proposition.} 
The one-forms $U_1$ and $U_2$ satisfy the following relations on the
locus $\cG$:
\beqan
\label{Urels}
&& \langle U_1,V_1\rangle = \langle U_2,V_2\rangle =0~~,~~\langle U_1,U_2\rangle =\frac{b_3}{2}\lambda^2~~\nn\\
&& ||U_1||^2=\frac{1+b_1}{2}\lambda^2~~,~~||U_2||^2=\frac{1+b_2}{2}\lambda^2~~,\\
&& \langle U_1,V_3 \rangle=\langle U_1,W \rangle =-\frac{1}{2}\langle U_2,V_1 \rangle~~,\nn\\
&&\langle U_2,V_3 \rangle=-\langle U_2,W \rangle =-\frac{1}{2}\langle U_1,V_2 \rangle~~.\nn
\eeqan
while $\tbeta$ is given by the following expression on the same locus:
\ben
\label{tbeta}
\tbeta^2=\alpha_1 \beta^2+\frac{1-\alpha_1}{2}(1+\rho^2-b_+^2)-\langle U_1,V_2\rangle~~.
\een

\

\noindent{\bf Proof.} 
Since the system $(\txi_1,\txi_2)$ is everywhere-orthonormal, the
1-forms $\tV_+,\tV_-,\tV_3, \tW$ satisfy \eqref{VWsys} and hence
their Gram matrix ${\hat {\tilde G}}$ must have the form (see
\eqref{Gext}):
\ben
\label{tGext}
{\hat {\tilde G}}={\hat G}(b,\tbeta)=\left[ \begin{array}{cccc}
1-\tbeta^2-b_+^2 &~-b_+b_-~& ~-b_+b_3~ & 0\\
-b_-b_+~ &~\tbeta^2-b_-^2 &~ -b_-b_3~ & b_3 \\ 
-b_3b_+~ &~ -b_3b_-~ & ~~\tbeta^2-b_3^2 & -b_-\\
0 & b_3 & -b_- & 1-\tbeta^2-b_+^2+\rho^2
\end{array}\right]~~,
\een
where we used the fact that $\tb_i=b_i$ and thus $\tbeta^2=b_3+||V_3||^2$. 
The Gram determinant is given by \eqref{dethG}:
\be
\det {\hat G}(b,\tbeta)=P(b,\tbeta)^2=\big[( b_3^2+b_-^2)(\tbeta^2-1)-\tbeta^2(\tbeta^2 -1 + b_+^2) \big]^2
=\big[ (\tbeta^2-1)(\rho^2-\tbeta^2) -\tbeta^2 b_+^2 \big]^2~~,
\ee
where $P$ is the polynomial given in \eqref{P}. Using \eqref{VWsys},
\eqref{tVtW} and \eqref{alpha2}, we find that $\tbeta$ can be
expressed as follows as a function of $\alpha_1$ on the locus $\cG$:
\beqan
\label{tbeta0}
&&\tbeta^2=\frac{1}{4}\left(  1+\rho^2-b_+^2+2b_-+\frac{2 b_3^2}{1+b_--b_+} \right)+
\langle U_2,V_3 \rangle-\langle U_2, W\rangle+||U_2||^2 -\frac{b_3 \langle U_2,V_2 \rangle}{1+b_- -b_+} \nn\\ 
&&~~~~~~ +\alpha_1\left[\frac{1}{2}\left( -1+2\beta^2-\rho^2+b_+^2 \right) +
\langle U_2,V_3\rangle+ \langle U_2,W\rangle+\frac{b_3}{1+b_--b_+}||U_2||^2\right]\nn\\ 
&&~~~~~~+\frac{\alpha_1^2(-1+b_--b_+)[\rho^2-(1-b_+)^2]}{1+b_--b_+}~~.
\eeqan
On the locus $\cG_0$, we have
$\zeta=\lambda\hzeta$, where $\hzeta\eqdef \frac{\zeta}{\lambda}$ is a
unit norm spinor of negative chirality defined on $\cG_0$ and which is
orthonormal to $\xi_1^-$ and $\xi_2^-$ at every point of $\cG_0$. On
this locus, we can write $U_i=\lambda \hU_i$, with:
\ben
\label{hatU}
\hU_i\eqdef \cB(\hzeta,\gamma_a\xi_i^+)e^a\in \Omega^1(\cG_0)~~(i=1,2)~~.
\een
Substituting this into \eqref{tVtW}, we find an expression for the
Gram matrix ${\hat {\tilde G}}$ as a function of $\alpha_1,\alpha_2$
and $\lambda$, where $\alpha_2$ and $\lambda$ can be expressed as
functions of $\alpha_1$ using the previous proposition. Thus ${\hat
  {\tilde G}}(\alpha_1)$ must equal the matrix ${\hat
  G}(b,\tbeta(\alpha_1))$ of \eqref{tGext} (where $\tbeta(\alpha_1)$
is given by \eqref{tbeta0}) for any $\alpha_1\in [-1,1]$. Expanding
both of these matrices to order two in $\alpha_1$, we find three
linear systems in the quantities $\langle U_i, V_+\rangle, \langle
U_i, V_-\rangle, \langle U_i, V_3\rangle$ and $\langle U_i, W\rangle$,
which can be shown to be equivalent\footnote{At this step we used
  ${\tt Mathematica}^{\textregistered}$, which we acknowledge here.}
with \eqref{Urels}. Using \eqref{Urels}, relation \eqref{tbeta0}
simplifies to \eqref{tbeta}.  Substituting \eqref{Urels} into ${\hat
  {\tilde G}}$, we find that ${\hat {\tilde G}}$ equals the
matrix ${\hat G}(b,\tbeta)$ of \eqref{tGext}, where $\tbeta$ is given
by \eqref{tbeta}. It follows that there are no further constrains on
$U_1$ and $U_2$ and hence that equality of ${\hat {\tilde
    G}}(\alpha_1)$ and \eqref{tGext} is equivalent with relations
\eqref{Urels} and \eqref{tbeta} on the locus $\cG_0$. These relation
also hold on $\cG\setminus \cG_0$ since $U_1,U_2$ and $\lambda$ vanish
on that locus.  $\blacksquare$

\

\noindent  Since $U_1$ depends continuously on $\alpha_1$,
relation \eqref{tbeta} shows that:
\be
\tbeta^2=t(B,\alpha_1)
\ee
where $B=(b,\beta)\in \fP$ and $t:\fP\times [-1,1]\rightarrow \R$ is a
continuous function.  Since $\tbeta$ is the function associated by
relation \eqref{beta_def} to the system of everywhere orthonormal
spinors $(\txi_1,\txi_2)$, we know that $\tbeta(p)$ must belong to the
interval $J(p)=J(b(p))$ for any value of $\alpha_1(p)$, where $J(b)$
was defined in \eqref{Jdef}. Hence the image of the function
$t_B:[-1,1]\rightarrow \R$ defined through:
\be
t_B(\alpha_1)\eqdef t(B,\alpha_1)~~(B\in \fP)
\ee
is contained in the interval $[f_-(b),f_+(b)]$. On the sub-locus of
$\cG\setminus\cG_0$ where $\alpha_1=\pm 1$, we have $\zeta=0$ and
$U_1=0$, hence \eqref{tbeta} gives:
\ben
\label{tbeta1}
t(B,+1)=\beta^2~~,~~t(B,-1)=-\beta^2+1+\rho^2-b_+^2~~
\een
while on the locus $\cG_0$, relation \eqref{tbeta} gives:
\ben
\label{tbeta2}
t(B,\alpha_1)=\alpha_1\beta^2+\frac{1-\alpha_1}{2}(1+\rho^2-b_+^2)-\lambda_M\sqrt{1-\alpha_1^2} \langle \hU_1,V_2\rangle~~(\alpha_1\in (-1,1))~~,
\een
which shows that $t$ is differentiable on $\fP\times (-1,1)$.

\paragraph{Proposition.} 
Let $B\in \partial \fP$. Then the image of $t_B$ equals the interval
$[f_-(b),f_+(b)]$ and hence the image of the function $\sqrt{t_B}$
equals the interval $J(b)$ defined in \eqref{Jdef}.

\

\noindent{\bf Proof.}
The condition $B\in \partial \fP$ means that $\beta=\sqrt{f_\pm(b)}$,
where the functions $f_\pm(b)=f_\pm(b_+,\rho)$ were defined in
\eqref{fpm}.  Then $t_B(+1)=\beta^2=f_\pm (b)$ while
$t_B(-1)=1+\rho^2-b_+^2-f_\pm(b)=f_\mp(b)$, where we used
\eqref{fpm}. Thus:
\ben
\label{tbetalimits}
t_B(+1)=f_\pm(p)~~\mathrm{and}~~t_B(-1)=f_\mp(p)~~.
\een
Since $t_B$ is continuous, its image (which is contained in
$[f_-(b),f_+(b)]$) is an interval which must contain the two values
\eqref{tbetalimits} and hence must equal
$[f_-(b),f_+(b)]$. $\blacksquare$

\subsection{Explicit spinor deformations which break the stabilizer from $\SU(3)$ to $\SU(2)$}

Let ${\tilde B}=(b,\tbeta):M\rightarrow \fP$ be the function
\eqref{Bdef} defined by the system of spinors $(\txi_1,\txi_2)$ and
let ${\tilde \cD}_0\eqdef \ker \tV_1\cap \ker \tV_2\cap \ker \tV_3\cap
\ker \tW$.

\paragraph{Proposition.}
Let $p\in \cG$ be such that $B(p)=(b(p),\beta(p))$ belongs to
$\partial \fP$ and let $\beta_0$ be any point in the interior of the
interval $J(b)$. Then we can find a deformation
$(\xi_1,\xi_2)\rightarrow (\txi_1,\txi_2)$ such that $(\txi_1,\txi_2)$
is a system of everywhere-orthonormal Majorana spinors on $M$ and such
that ${\tilde B}=(b,\tbeta)$ with $\tbeta(p)=\beta_0$.

\

\noindent{\bf Proof.}  Follows immediately from the results of the
previous subsection. $\blacksquare$

\paragraph{Remark.} 
Together with the results of Subsection \ref{sec:spinorproof}, the
proposition implies that, for every value $B_0\in \fP$ and every point
$p\in M$, there exists a pair of everywhere-orthonormal Majorana
spinors $(\xi_1,\xi_2)$ on $M$ whose function $B$ satisfies $B(p)=B_0$. In
particular, all points of $\fP$ can be realized by some
two-dimensional and $\cB$-compatible locally-nondegenerate subspace
$\cK\subset \Gamma(M,S)$.

\paragraph{Corollary.} 
Let $p\in \cG$ be such that $H_p\simeq \SU(3)$. Then $\dim \cD_0(p)=6$
and $B(p)\in \Int \fI\sqcup \Int \fD\sqcup \fA\sqcup \Int \fC^+\sqcup
\Int \fC^-\subset \fP$. Moreover, we can find a deformation
$(\xi_1,\xi_2)\rightarrow (\txi_1,\txi_2)$ (given explicitly in the
previous subsection) such that $(\txi_1,\txi_2)$ is a system of
everywhere-orthonormal Majorana spinors on $M$ and such that:
\begin{itemize}
\itemsep 0.0em
\item $\dim {\tilde \cD}_0(p)=4$
\item The stabilizer ${\tilde H}_p$ of $(\txi_1(p),\txi_2(p))$ inside
  $\Spin(T_pM,g_p)\simeq \Spin(8)$ is isomorphic with $\SU(2)$.
\end{itemize}

\

\noindent{\bf Proof.}  For $p\in \cG$ such that $B(p)\in \partial
\fP$, we have $\rk {\hat G}(B(p))=2$ and \eqref{Ginclusion} implies
that the 1-forms $V_1(p),V_2(p),V_3(p)$ and $W(p)$ are stabilized by a
subgroup containing $\SU(3)$. Since $\dim \cD_0(p)\in \{4,6\}$ for
$p\in \cG$ (see Appendix \ref{app:hatG}) and since $\SU(3)$ does not
embed into $\SO(4)$, we must have $\dim \cD_0(p)=6$ and the common
stabilizer of the one-forms must equal $\SO(6)$. In particular, the
space spanned by $V_1(p),V_2(p),V_3(p)$ and $W(p)$ inside $T_p^\ast M$
has dimension two. Since $\dim\cD_0(p)=6$, the results of Appendix
\ref{app:hatG}, imply that the point $B(p)$ belongs to the subset
$\Int \fI\sqcup \Int \fD\sqcup \fA\sqcup \Int \fC^+\sqcup \Int \fC^-$
of the frontier $\partial \fP$. Let $(\txi_1,\txi_2)$ be chosen as in
the previous proposition. Then we have ${\tilde B}(p)\in \Int \fP$ and
hence $\rk {\tilde {\hat G}}(p)=\rk {\hat G}(b(p),\tbeta(p))=4$ by the
results of Appendix \ref{app:hatG}. Thus the 1-forms ${\tilde
  V}_1(p),{\tilde V}_2(p),{\tilde V}_3(p)$ and ${\tilde W}(p)$ are
linearly independent at $p$ and we have $\dim {\tilde
  \cD}_0(p)=4$. Moreover, the spinor $\zeta(p)$ of the previous
subsection is non-zero and hence ${\tilde H}_p$ is isomorphic with
$\SU(2)$ (see Subsection \ref{sec:spinorproof}). $\blacksquare$

\paragraph{Remark.} 
The orthogonal complement of $K_-(p)$ inside $S^-_p$ equals the space
$\Xi^-(p)$ considered in the proof of point 4 of the Proposition of
Subsection \ref{sec:spinorproof}, a space which carries the
fundamental representation of the group $H_p''\eqdef
\Stab_{\Spin(T_pM,g_p)}(\xi_1^+(p),\xi_2^+(p),\xi_2^-(p))\simeq
\SU(3)$. The fact that the deformed spinor $\txi_1^-(p)$ has non-zero
projection $\zeta(p)$ on the space $\Xi^-(p)$ is responsible for
breaking the stabilizer group at $p$ from $\SU(3)$ to $\SU(2)$.

\section{The non-generic assumption made in \cite{Palti}}
\label{app:P}

Let $\pi_1:{\hat M}\rightarrow M$ and $\pi_2:{\hat M}\rightarrow S^1$
denote the projections on the first and second factor of the direct
product ${\hat M}=M\times S^1$ (which, as in \cite{Palti}, we endow
with the direct product metric). Let $\theta\in \Omega^1({\hat M})$ be
the $\pi_2$-pullback of the canonical normalized 1-form of $S^1$
(notice that $\theta$ is the normalized Killing form on ${\hat M}$
corresponding to the symmetry given by rotations along the circle). 
Loc. cit.  uses three one-forms\footnote{The one-forms used by
  \cite{Palti} on ${\hat M}$ are denoted there by $V_+, V_-$ and
  $V_3$.  The relation with our notation is ${\hat V}_\pm^{\rm here}
  =\frac{1}{2}V_\pm^{\rm there}$ and ${\hat V}_3^{\rm here}=V_3^{\rm
    there}$, cf. \cite[eq. (2.26)]{Palti}. } ${\hat V}_+,{\hat
  V}_-,{\hat V}_3\in \Omega^1({\hat M})$ defined on the 9-manifold
${\hat M}$ which are invariant under $S^1$-rotations and hence are given by:
\ben
\label{hV}
{\hat V}_r=\pi_1^\ast(V_r)+(b_r\circ \pi_1)\theta~~,~~\forall r\in \{+,-,3\}
\een
where $V_r\in \Omega^1(M)$ and $b_r\in \cinf$. The quantities
$V_r,b_r$ turn out to coincide with the 0-forms and 1-forms given in
\eqref{forms0pm3} and \eqref{forms1pm3}. Indeed, it is easy to see
that the algebraic constraints for \eqref{hV} given in equations
\cite[eq. (2.15)]{Palti} are equivalent with the system \eqref{Vsys}
for $V_r$ if one takes into account relation \cite[eq. (2.26)]{Palti}.
Since $\theta$ and $\pi_1^\ast(V_r)$ are orthogonal at every point of
${\hat M}$, relations \eqref{hV} give:
\ben
\label{sprod}
\langle \theta, {\hat V}_r\rangle=b_r\circ \pi_1~~.
\een
Loc cit. makes intensive use of the assumption
(cf. \cite[eq. (3.9)]{Palti}) that the following relation holds on
${\hat M}$: 
\ben
\label{theta}
\theta=\frac{2}{1+ {\hat \alpha}}\langle \theta, {\hat V}_+\rangle
            {\hat V}_++\frac{2}{1-{\hat \alpha}}\langle \theta,
            {\hat V}_-\rangle {\hat V}_-+ \frac{2}{1-{\hat
                \alpha}}\langle \theta,{\hat V}_3\rangle {\hat V}_3~~,
\een
where\footnote{The function ${\hat \alpha}$ is denoted by $\alpha$ in
  \cite{Palti}.} ${\hat \alpha}\in \cC^\infty({\hat M},\R)$ is a function
independent of the $S^1$ coordinate, hence ${\hat \alpha}=\alpha\circ
\pi_1$ for any  $\alpha\in \cinf$. To arrive at \eqref{theta}, we used
the fact that ${\hat V}_\pm^{\rm here} =\frac{1}{2}V_\pm^{\rm there}$
and ${\hat V}_3^{\rm here}=V_3^{\rm there}$. Comparing with
\eqref{Vsys}, it is not hard to check that $\alpha=1-2\beta^2$,
where $\beta$ was defined in \eqref{beta_def}. Equations \eqref{sprod}
give:
\be
\frac{2}{1\pm {\hat \alpha}}\langle \theta, {\hat V}_\pm\rangle=a_\pm \circ \pi_1~~,~~
\frac{2}{1-{\hat \alpha}}\langle \theta,{\hat V}_3\rangle =a_3\circ \pi_1~~,
\ee
where $a_{\pm, 3}\in \cinf$ are given by: 
\be
a_\pm \eqdef \frac{2b_\pm}{1\pm \alpha}~~,~~a_3\eqdef\frac{2b_3}{1-\alpha}~~.
\ee
Hence \eqref{theta} takes the form:
\ben
\label{theta0}
\theta=(a_+\circ \pi_1){\hat V_+}+(a_-\circ \pi_1){\hat V}_-
+(a_3\circ \pi_1){\hat V}_3~~.
\een
Since $\theta$ and $\pi_1^\ast(V_r)$ are orthogonal at every point of
${\hat M}$, substituting \eqref{hV} into \eqref{theta0} and projecting
onto $\pi_1^\ast(T^\ast M)$ gives:
\ben
\label{lc}
a_+ V_++a_- V_-+a_3V_3=0~~.
\een
Hence equation \cite[eq. (3.9)]{Palti} requires that $V_+,V_-$ and
$V_3$ be linearly dependent at every point of $M$, a requirement which
cannot be satisfied in the generic case. In the non-generic case when
\eqref{lc} holds, we have $\rk \cD\geq 6$ on $M$ and hence the
$\SU(2)$ locus $\cU$ of $M$ must be empty (see Table
\ref{table:HHprime}). 

The fact that the $\SU(2)$ locus $\cU$ need not be empty follows from
the results of Subsection \ref{sec:spinorproof} (which gives a proof
of this fact directly in terms of spinors), from the results of
Appendix \ref{app:hatG} (which shows that the 1-forms
$V_1(p),V_2(p),V_3(p)$ and $W(p)$ are linearly independent in the
generic case) and also from the results of Appendix \ref{app:def},
which gives an explicit construction of a family of spinor
deformations which can be used to break the stabilizer group $H_p$
from $\SU(3)$ to $\SU(2)$.  The condition $\cU=\emptyset$ is a very
strong restriction since the locus $\cU$ is {\em open} in $M$. This
condition amounts to vanishing of the spinor projection $\zeta(p)$
arising in the proof of point 4 of the Proposition of Subsection
\ref{sec:spinorproof} for {\em every} point $p$ of $M$; it is also
equivalent with the condition that the image of the map $B$ defined in
\eqref{Bdef} is contained in the {\em frontier} $\partial \fP$ of the
four-dimensional semi-algebraic body $\fP$, rather that in the body
$\fP$ itself.

We also note that the cosmooth generalized distribution ${\hat
  \cD}\eqdef \ker {\hat V}_+\cap \ker {\hat V}_-\cap \ker {\hat V}_3$
defined on ${\hat M}$ may have transverse or non-transverse
intersection with the distribution $\pi_1^\ast(TM)$. This is one
reason why one cannot conclude (as \cite{Palti} does) that the
stabilizer stratification of $M$ would be ``directly inherited'' from that
of ${\hat M}$. As we show in a different publication, the relation
between the stabilizer stratifications of $M$ and ${\hat M}$ is in
fact rather involved, in particular due to the non-transversality
issue mentioned above.

\paragraph{Remark.} 
Loc cit. gives an argument (see the discussion there introducing
equation \cite[(3.9)]{Palti}) according to which \eqref{theta} would
always have to hold.  That argument relies on confusing $\theta$ (a
one-form which exists on ${\hat M}$ by the definition of ${\hat
  M}\eqdef M\times S^1$ and therefore is not a spinor bilinear) with a
combination of one-forms constructed from the canonical lifts to
${\hat M}$ of the supersymmetry generators $\xi_1,\xi_2\in
\Gamma(M,S)$. It is further based on the assumption that $\theta$
would induce, in certain cases, a {\em nowhere-vanishing} vector
field/one-form on $M$. However, the projection of $\theta$ on the
bundle $\pi_1^\ast(T^\ast M)\subset T^\ast {\hat M}$ always vanishes,
hence that projection can never define a non-vanishing one-form on $M$
and thus it can never give a non-trivial singlet for the structure
group of $M$. For these reasons, the argument given in loc. cit.
cannot be used to conclude that $\theta$ would always have to be a
linear combination of $\hV_\pm$ and $\hV_3$.


\end{document}